\documentclass[a4paper,fleqn,usenatbib]{mnras}

\usepackage{newtxtext,newtxmath}

\usepackage[T1]{fontenc}
\usepackage{ae,aecompl}
\usepackage{bm}
\usepackage{xcolor}
\usepackage{scalerel}
\usepackage{tikz}
\usetikzlibrary{svg.path}

\definecolor{orcidlogocol}{HTML}{A6CE39}
\tikzset{
  orcidlogo/.pic={
    \fill[orcidlogocol] svg{M256,128c0,70.7-57.3,128-128,128C57.3,256,0,198.7,0,128C0,57.3,57.3,0,128,0C198.7,0,256,57.3,256,128z};
    \fill[white] svg{M86.3,186.2H70.9V79.1h15.4v48.4V186.2z}
                 svg{M108.9,79.1h41.6c39.6,0,57,28.3,57,53.6c0,27.5-21.5,53.6-56.8,53.6h-41.8V79.1z M124.3,172.4h24.5c34.9,0,42.9-26.5,42.9-39.7c0-21.5-13.7-39.7-43.7-39.7h-23.7V172.4z}
                 svg{M88.7,56.8c0,5.5-4.5,10.1-10.1,10.1c-5.6,0-10.1-4.6-10.1-10.1c0-5.6,4.5-10.1,10.1-10.1C84.2,46.7,88.7,51.3,88.7,56.8z};
  }
}

\newcommand\orcidicon[1]{\href{https://orcid.org/#1}{\mbox{\scalerel*{
\begin{tikzpicture}[yscale=-1,transform shape]
\pic{orcidlogo};
\end{tikzpicture}
}{|}}}}

\usepackage{float}
\usepackage[export]{adjustbox}
\usepackage{graphicx}	

\usepackage{amsmath}	
\usepackage{amssymb}	
\usepackage{multirow}
\usepackage{multicol}
\usepackage{verbatim}   
\usepackage{threeparttable}
\usepackage{upgreek}
\usepackage{rotating}

\usepackage{subcaption}
\usepackage{comment}
\usepackage{longtable,lscape}
\usepackage{caption}
\usepackage{threeparttable}
\newcommand{\lcdm}{$\Lambda$CDM}

\newcommand{\om}{\Omega_{m0}}
\newcommand{\ol}{\Omega_{\Lambda}}
\newcommand{\ok}{\Omega_{k0}}
\newcommand{\FT}[1]{}
\newcommand{\rfe}{${\cal R}_{\rm{Fe\textsc{ii}}}$}
\newcommand{\Feii}{Fe\,\textsc{ii}}
\newcommand{\Mgii}{Mg\,\textsc{ii}}
\newcommand{\Civ}{C\,\textsc{iv}}
\newcommand{\hb}{{\sc{H}}$\beta$\/}
\newcommand{\hii}{H\,\textsc{ii}}

\definecolor{MZ}{RGB}{255,128,0}

\title[\Mgii\ QSO $R-L$ relation study]{Consistency study of high- and low-accreting \Mgii\ quasars: \\
No significant effect of the \Feii\ to \Mgii\ flux ratio on the radius-luminosity relation dispersion}

\author[Khadka et al.]{
Narayan Khadka$^{\orcidicon{0000-0001-5512-2716}1}$\thanks{E-mail: nkhadka@phys.ksu.edu},
Michal Zaja\v{c}ek$^{\orcidicon{0000-0001-6450-1187}2}$\thanks{E-mail: zajacek@mail.muni.cz}, 
Swayamtrupta Panda$^{\orcidicon{0000-0002-5854-7426}{3,4}}$\thanks{E-mail: panda@cft.edu.pl}\thanks{CNPq fellow},
\newauthor \hspace{0.1mm}
Mary Loli Mart\'inez-Aldama$^{\orcidicon{0000-0002-7843-7689}{5}}$\thanks{E-mail: mmartinez@das.uchile.cl},
Bharat Ratra$^{\orcidicon{0000-0002-7307-0726}1}$\thanks{E-mail: ratra@phys.ksu.edu}\\
$^{1}$Department of Physics, Kansas State University, 116 Cardwell Hall, Manhattan, KS 66506, USA\\
$^{2}$Department of Theoretical Physics and Astrophysics, Faculty of Science, Masaryk University, Kotl\'a\v{r}sk\'a 2, 611 37 Brno, Czech Republic\\
$^{3}$Center for Theoretical Physics, Polish Academy of Sciences, Al.\ Lotnik\'{o}w 32/46, 02-668 Warsaw, Poland\\
$^{4}$Laborat\'orio Nacional de Astrof\'isica - MCTIC, R. dos Estados Unidos, 154 - Na\c{c}\~oes, Itajub\'a - MG, 37504-364, Brazil\\
$^{5}$Departamento de Astronomía, Universidad de Chile, Camino del Observatorio 1515, Santiago, Chile\\
}

\date{Accepted XXX. Received YYY; in original form ZZZ}

\pubyear{2022}

\begin{document}
\label{firstpage}
\pagerange{\pageref{firstpage}--\pageref{lastpage}}
\maketitle

\begin{abstract}
We use observations of 66 reverberation-measured \Mgii\ quasars (QSOs) in the redshift range $0.36 \leq z \leq 1.686$ --- a subset of the 78 QSOs we previously studied that also have \rfe\ (flux ratio parameter of UV \Feii\ to \Mgii\ that is used as an accretion-rate proxy) measurements --- to simultaneously constrain cosmological model parameters and QSO 2-parameter and 3-parameter radius-luminosity ($R-L$) relation parameters in six different cosmological models. We find that these QSO $R-L$ relation parameters are independent of the assumed cosmological model and so these QSOs are standardizable through the $R-L$ relations. Also: (1) With the 2-parameter $R-L$ relation, we find that the low-\rfe\ and high-\rfe\ data subsets obey the same $R-L$ relation within the error bars. (2) Extending the 2-parameter $R-L$ relation to a 3-parameter one does not result in the hoped-for reduction in the intrinsic dispersion of the $R-L$ relation. (3) Neither of the 3-parameter $R-L$ relations provide a significantly better fit to the measurements than does the 2-parameter $R-L$ relation. These are promising results for the on-going development of \Mgii\ cosmological probes. The first and third of these results differ significantly from those we found elsewhere from analyses of reverberation-measured \hb\ QSOs.  
\end{abstract}




\section{Introduction}
\label{sec:Introduction}

The spatially-flat $\Lambda$CDM model \citep{Peebles1984} is consistent with most cosmological observations \citep[see, e.g.][]{Scolnicetal2018, Yuetal2018, PlanckCollaboration2020, eBOSSCollaboration2021}. This model assumes that the dark energy responsible for the currently accelerated cosmological expansion is a time-independent cosmological constant $\Lambda$ that contributes $\sim 70\%$ of the current cosmological energy budget, with non-relativistic cold dark matter (CDM) contributing $\sim 25\%$, and with most of the remaining $\sim 5\%$ contributed by non-relativistic baryonic matter. However, observational data do not rule out a little spatial curvature or dynamical dark energy, so we also study such cosmological models in this paper. 

Some measurements (of the same quantity) appear mutually incompatible when considered in the spatially-flat $\Lambda$CDM model \citep[see, e.g.][]{DiValentinoetal2021b, PerivolaropoulosSkara2021, Abdallaetal2022}. It is not clear whether these discrepancies point to new physics beyond the spatially-flat $\Lambda$CDM model or instead are a reflection of as yet unidentified systematic errors in one or both of the mutually incompatible data sets. New alternate cosmological probes might help resolve this issue. Such probes include \hii\ starburst galaxy observations which reach to redshift $z \sim 2.4$ \citep{ManiaRatra2012, Chavezetal2014, GonzalezMoranetal2021, Caoetal2020, Caoetal2021a, Caoetal_2021c, Johnsonetal2022, Mehrabietal2022}, quasar (QSO) angular size observations which reach to $z \sim 2.7$ \citep{Caoetal2017, Ryanetal2019, Caoetal2020, Caoetal2021b, Zhengetal2021, Lianetal2021}, QSO X-ray and UV flux observations which reach to $z \sim 7.5$ \citep{RisalitiLusso2015, RisalitiLusso2019, KhadkaRatra2020a, KhadkaRatra2020b, KhadkaRatra2021a, KhadkaRatra2022, Lussoetal2020, Rezaeietal2022, Luongoetal2021, Dainottietal2022a},\footnote{Although the most recent \cite{Lussoetal2020} QSO flux compilation can only be used to derive significantly lower-$z$ cosmological constraints since it assumes a UV--X-ray correlation model that is invalid above $z \sim 1.5-1.7$ \citep{KhadkaRatra2021a, KhadkaRatra2022}.} and gamma-ray burst observations that reach to $z \sim 8.2$ \citep{Wang_2016, Wangetal2022, Dirirsa2019, Demianskietal_2021, KhadkaRatra2020c, Khadkaetal2021a, Huetal2021, OrlandoMarco2021, CaoKhadkaRatra2022, CaoDainottiRatra2022a, CaoDainottiRatra2022b, Dainottietal2022b}. 

Reverberation-measured QSOs might prove to be another useful cosmological probe. The method is based on the observed broad-line region (hereafter BLR) radius-luminosity ($R-L$) relation or rather correlation \citep{1991ApJ...370L..61K,1993PASP..105..247P,2000ApJ...533..631K,2013ApJ...767..149B} that allows one to convert the measured rest-frame time-delay of the broad emission line with respect to the photoionizing continuum to an absolute monochromatic luminosity of the QSO.\footnote{For early discussions of reverberation-measured QSOs in cosmology, see \citet{watson2011}, \citet{haas2011}, and \citet{czerny2013}; more recent discussions can be traced back through \citet{MartinezAldama2019}, \citet{2019FrASS...6...75P}, \citet{2020mbhe.confE..10M}, \citet{Michal2021}, and \citet{Czerny2021}.} We have shown that the largest available compilation of 78 reverberation-measured \Mgii\ QSOs are standardizable through the observed 2-parameter $R-L$ relation and that the current best \Mgii\ cosmological constraints are weak but consistent with those determined from better-established cosmological probes \citep{khadka2021}.\footnote{In our analyses we simultaneously determine cosmological-model and $R-L$-correlation parameters; this allows us to circumvent the circularity problem. We found that the QSO $R-L$ relation is independent of the assumed cosmological model so these \Mgii\ QSOs are standardizable and, thus, useful for the purpose of constraining cosmological parameters.} 

More recently, we performed similar analyses using the largest available compilation of 118 H$\beta$ QSO sources, where we found that the 2-parameter $R-L$ relation \hb\ cosmological constraints were somewhat ($\sim 2\sigma$) inconsistent with those from better-established cosmological probes \citep{Khadkaetal2021c}. In contrast, we have found that cosmological parameters determined from the largest available data set of 38 high-quality \Civ\ QSOs are consistent with those from better-established BAO+$H(z)$ data as well as with those from \Mgii\ QSOs, which allows us to jointly analyze the \Civ\ and \Mgii\ samples, as well as use the joint \Civ+\Mgii+BAO+$H(z)$ sample to constrain cosmological parameters \citep{2022arXiv220515552C}. Combining the \Civ+\Mgii\ sample with the BAO+$H(z)$ sample alters the cosmological constraints by $\sim 0.1\sigma$ at most in comparison with those from the BAO+$H(z)$ sample alone.

To try to reduce the intrinsic dispersion of the QSO sample, attempts have previously been made to correct for the accretion-rate effect observed in the 2-parameter $R-L$ relation \citep{duwang_2019, Mary2020, Michal2021,2022arXiv220111062P} by extending it to a 3-parameter $R-L$ relation. Partially to determine whether such a procedure might alter the \hb\ cosmological constraints, we also considered a 3-parameter $R-L$ relation, with optical \rfe\ being the third parameter \citep{Khadkaetal2021c}.\footnote{For H$\beta$ quasars, \rfe\ is the ratio of the flux densities of the optical \Feii\ line to the \hb\ broad component, and analogously, for MgII sources, it is defined as the ratio of the flux densities of the UV \Feii\ pseudocontinuum and the MgII broad line. \rfe\ is used as an accretion-rate intensity proxy.} More interestingly, we confirmed that the low- and high-\rfe\ H$\beta$ sources obey significantly different $R-L$ relations, which suggests that it might not be possible to standardize H$\beta$ QSOs using an $R-L$ relation. We also discovered that the intrinsic dispersion did not significantly decrease when going from the 2- to the 3-parameter $R-L$ relation. These findings motivated us to examine here the same issues for the \Mgii\ sources, whose cosmological constraints are compatible with those derived using better-established cosmological probes \citep{khadka2021}.  

In this paper, we also investigate two 3-parameter $R-L$ relations for \Mgii\ QSO sources, one in which the time-delay depends exponentially on \rfe, and the other where it depends on a power of \rfe. Our conclusions are qualitatively similar in the two 3-parameter cases. We find that for \Mgii\ data and the 2-parameter $R-L$ relation, the low-\rfe\ and high-\rfe\ data subsets obey the same $R-L$ relation within the error bars. This differs from what happens in the \hb\ case \citep{Khadkaetal2021c} and is a hopeful result for the on-going development of \Mgii\ cosmological probes. In the 3-parameter $R-L$ relation cases, the inclusion of the third parameter increases the $R-L$ relation parameter differences between the low-\rfe\ and high-\rfe\ data subsets. This increase is not very significant for the exponential \rfe\ case but is a little more significant for the power-law \rfe\ $R-L$ relation. However, we find that extending the 2-parameter $R-L$ relation to a 3-parameter does not reduce the intrinsic dispersion of the $R-L$ relation,\footnote{This is similar to what happens in the \hb\ case \citep{Khadkaetal2021c}.} which was the main motivation for extending the 2-parameter $R-L$ relation. In \citet{Khadkaetal2021c} we found that for the full \hb\ data set the 3-parameter $R-L$ relation provided a significantly better fit to the measurements than did the 2-parameter $R-L$ relation, unlike what we find for \Mgii\ data here. This is because the \hb\ low-\rfe\ and high-\rfe\ data subsets obey very different 2-parameter $R-L$ relations, unlike what we find for \Mgii\ data here.

In Sec.~2 we describe the dark energy cosmological models we use. In Sec.~3 we summarize the \Mgii\ QSO data we use to constrain cosmological-model and $R-L$ correlation parameters. In Sec.~4 we summarize the data analysis methods we use. We present our results in Sec.~5, discuss some of them in more detail in Sec.~6, and conclude in Sec.~7.

\section{Parameters and Cosmological models}
\label{sec:models}

In this study we use \Mgii\ QSOs to simultaneously infer cosmological model parameters $\mathbf{p}$ as well as the parameters of either the 2-parameter or the 3-parameter $R-L$ correlation.

We use three pairs of dark-energy cosmological models with flat and non-flat spatial geometries, hence in total 6 cosmological models.\footnote{For discussions of constraints on spatial curvature see \citet{Chenetal2016}, \citet{Ranaetal2017}, \citet{Oobaetal2018a, Oobaetal2018b}, \citet{ParkRatra2019a, ParkRatra2019b}, \citet{Wei2018}, \citet{DESCollaboration2019}, \citet{Lietal2020}, \citet{EfstathiouGratton2020}, \citet{DiValentinoetal2021a}, \citet{Vagnozzietal2020, Vagnozzietal2021}, \citet{KiDSCollaboration2021}, \citet{ArjonaNesseris2021}, \citet{Dhawanetal2021}, \citet{Renzietal2021}, \citet{Gengetal2022}, \citet{WeiMelia2022}, \citet{MukherjeeBanerjee2022}, and references therein.} The cosmological expansion rate or the Hubble parameter $H(z,\mathbf{p})$ is introduced for each cosmological model below as a function of the cosmological redshift $z$ and a set of cosmological parameters $\mathbf{p}$ specific to each model. $H(z,\mathbf{p})$ is then used to make theoretical predictions for \Mgii\ QSOs, specifically their rest-frame BLR time delays $\tau$ that are related to the BLR radius as $R=c\tau$ where $c$ is the speed of light.

The fundamental observables for each \Mgii\ QSO is its redshift $z$, determined from the shift of its spectral lines, and the monochromatic flux density $F_{3000}$ at 3000\,\AA\, measured in ${\rm erg\,s^{-1}\,cm^{-2}}$. To obtain the $R-L$ relation for each cosmological model, we calculate the monochromatic luminosity at 3000\,\AA\, using
\begin{equation}
    L_{3000}(z,\mathbf{p})=4\pi D_L(z,\mathbf{p})^2 F_{3000}\,,
    \label{eq_L3000}
\end{equation}
where the luminosity distance $D_{\rm L}(z,\mathbf{p})$ (expressed in cm) is a function of $z$ and of the cosmological-model parameters $\mathbf{p}$. Depending on the considered cosmological model and its current spatial curvature density parameter $\Omega_{k0}$, the luminosity distance for a given \Mgii\ QSO at redshift $z$ is  
\begin{equation}
  \label{eq:DL}
\resizebox{0.475\textwidth}{!}{%
    $D_L(z,\mathbf{p}) = 
    \begin{cases}
    \frac{c(1+z)}{H_0\sqrt{\Omega_{\rm k0}}}\sinh\left[\frac{H_0\sqrt{\Omega_{\rm k0}}}{c}D_C(z,\mathbf{p})\right] & \text{if}\ \Omega_{\rm k0} > 0, \\
    \vspace{1mm}
    (1+z)D_C(z,\mathbf{p}) & \text{if}\ \Omega_{\rm k0} = 0,\\
    \vspace{1mm}
    \frac{c(1+z)}{H_0\sqrt{|\Omega_{\rm k0}|}}\sin\left[\frac{H_0\sqrt{|\Omega_{\rm k0}|}}{c}D_C(z,\mathbf{p})\right] & \text{if}\ \Omega_{\rm k0} < 0,
    \end{cases}$%
    }
\end{equation}
where $H_0$ is the Hubble constant and $D_C(z,\mathbf{p})$ is the comoving distance defined for a given cosmological model as
\begin{equation}
\label{eq:gz}
   D_C(z,\mathbf{p}) = c\int^z_0 \frac{dz'}{H(z',\mathbf{p})}\,.
\end{equation}

Observations indicate the existence of the $R-L$ relation for \Mgii\, QSOs at known redshifts $z$ \citep{2019ApJ...880...46C,Michal2020,Homayouni2020,Michal2021,Zhefu2021} that can generally be expressed as \citep{Mary2020}
\begin{equation}
\label{eq:corr}
   \log \left({\frac{\tau} {\rm day}}\right) = \beta + \gamma \log\left[{\frac{L_{3000}(z,\mathbf{p})}{10^{44}\,{\rm erg\,s^{-1}}}}\right]+kq\,,
\end{equation}
where the intercept $\beta$ and the slope $\gamma$ are treated as free parameters. For the standard 2-parameter $R-L$ relation, $k=0$, while for the extended 3-parameter $R-L$ relation, $k$ is another free parameter and the quantity $q$ is the observationally determined \rfe\ or $\log{}$\rfe. The $R-L$ relation free parameters $\beta$, $\gamma$, and $k$ and the cosmological-model parameters $\mathbf{p}$ that enter the $R-L$ relation via $L_{3000}(z,\mathbf{p})$, see eq.~\eqref{eq_L3000}, are fitted simultaneously as is explained in detail in Sec.~\ref{sec:methods}.

In the $\Lambda$CDM model the cosmological expansion rate
\begin{equation}
\label{eq:friedLCDM}
    H(z,\mathbf{p}) = H_0\sqrt{\Omega_{m0}(1+z)^3 + \Omega_{k0}(1+z)^2 + \Omega_{\Lambda}}\,.
\end{equation}
The parameters $\Omega_{m0}$, $\Omega_{k0}$, and $\Omega_{\Lambda}$, that are constrained through the energy budget equation $\Omega_{m0}$ + $\Omega_{k0}$ + $\Omega_{\Lambda}$ = 1, are the current values of the non-relativistic matter density parameter, the spatial curvature density parameter, and the cosmological constant density parameter, respectively. In the spatially non-flat $\Lambda$CDM model it is  conventional to choose $\Omega_{m0}$, $\Omega_{k0}$, and $H_0$ to be the free parameters. In the spatially-flat $\Lambda$CDM model we use the same set of free parameters but with $\Omega_{k0}=0$ to account for the flat geometry.

In the XCDM dynamical dark energy parametrization the cosmological expansion rate
\begin{equation}
\label{eq:XCDM}
    H(z,\mathbf{p}) = H_0\sqrt{\Omega_{m0}(1+z)^3 + \Omega_{k0}(1+z)^2 + \Omega_{X0}(1+z)^{3(1+\omega_X)}},
\end{equation}
where $\Omega_{X0}$ is the current value of the $X$-fluid dark energy density parameter and is constrained with $\Omega_{m0}$ and $\Omega_{k0}$ through the energy budget equation $\Omega_{m0}$ + $\Omega_{k0}$ + $\Omega_{X0}$ = 1. $\omega_X$ is the equation of state parameter of the $X$-fluid and is the ratio of the pressure to the energy density of the $X$-fluid. In the spatially non-flat XCDM parametrization it is conventional to choose $\Omega_{m0}$, $\Omega_{k0}$, $\omega_X$, and $H_0$ to be the free parameters. In the spatially-flat XCDM parametrization we use the same set of free parameters but with $\Omega_{k0}=0$ to account for the flat geometry. The XCDM parametrization with $\omega_X = -1$ corresponds to the $\Lambda$CDM model. 

The $\phi$CDM model is a physically-complete dynamical dark energy model with the scalar field $\phi$ being the dynamical dark energy \citep{PeeblesRatra1988, RatraPeebles1988, Pavlovetal2013}.\footnote{For discussions of constraints on the $\phi$CDM model see \citet{Zhaietal2017}, \citet{Oobaetal2018c, Oobaetal2019}, \citet{ParkRatra2018, ParkRatra2019c, ParkRatra2020}, \citet{SolaPercaulaetal2019}, \citet{Singhetal2019}, \citet{UrenaLopezRoy2020}, \citet{SinhaBanerjee2021}, \citet{Xuetal2021}, \citet{deCruzetal2021}, \citet{Jesusetal2021}, \citet{CaoRatra2022}, and references therein.} In this model, the scalar field dark energy density parameter $\Omega_{\phi}(z, \alpha)$ is determined by the scalar field potential energy density $V(\phi)$ that we assume to be an inverse power law of $\phi$,
\begin{equation}
\label{eq:phiCDMV}
    V(\phi) = \frac{1}{2}\kappa m_{p}^2 \phi^{-\alpha}.
\end{equation}
Here $m_{p}$ and $\alpha$ are the Planck mass and a positive parameter respectively, and $\kappa$ is a constant whose value is determined using the shooting method to ensure that the current energy budget equation $\Omega_{m0} + \Omega_{k0} + \Omega_{\phi}(z = 0, \alpha) = 1$ is obeyed.

With this potential energy density, the dynamics of a spatially homogeneous scalar field and the cosmological scale factor $a$ are determined by the scalar field equation of motion and the Friedmann equation. These equations are 
\begin{align}
\label{eq:field}
   & \ddot{\phi}  + 3\frac{\dot{a}}{a}\dot\phi - \frac{1}{2}\alpha \kappa m_{p}^2 \phi^{-\alpha - 1} = 0, \\
\label{eq:friedpCDM}
   & \left(\frac{\dot{a}}{a}\right)^2 = \frac{8 \pi}{3 m_{p}^2}\left(\rho_m + \rho_{\phi}\right) - \frac{k}{a^2}.
\end{align}
Here an overdot denotes a derivative with respect to time, $k$ is positive, zero, and negative for closed, flat, and open spatial hypersurfaces (corresponding to $\Omega_{k0} < 0, =0, {\rm and} >0$), $\rho_m$ is the non-relativistic matter energy density, and $\rho_{\phi}$ is the scalar field energy density given by
\begin{equation}
    \rho_{\phi} = \frac{m^2_p}{32\pi}\left[\dot{\phi}^2 + \kappa m^2_p \phi^{-\alpha}\right].
\end{equation}
The numerical solution of the coupled differential equations (\ref{eq:field}) and (\ref{eq:friedpCDM}) is used to compute $\rho_{\phi}$ and then $\Omega_{\phi}(z, \alpha)$ is determined from the definition
\begin{equation}
    \Omega_{\phi}(z, \alpha) = \frac{8 \pi \rho_{\phi}}{3 m^2_p H^2_0}.
\end{equation}

The cosmological expansion rate in the $\phi$CDM model is
\begin{equation}
    H(z,\mathbf{p}) = H_0\sqrt{\Omega_{m0}(1+z)^3 + \Omega_{k0}(1+z)^2 + \Omega_{\phi}\left(z, \alpha\right)}.
\end{equation}
In the spatially non-flat $\phi$CDM model it is conventional to choose $\Omega_{m0}$, $\Omega_{k0}$, $\alpha$, and $H_0$ to be the free parameters. In the spatially-flat $\phi$CDM model we use the same set of free parameters but with $\Omega_{k0}=0$ to account for the flat geometry. The $\phi$CDM model with $\alpha = 0$ corresponds to the $\Lambda$CDM model.

In the BAO + $H(z)$ data analyses, whose results we use for comparison with \Mgii\ QSO cosmological constraints, instead of $\Omega_{m0}$ we use the present values of the physical CDM and baryonic matter density parameters, $\Omega_c h^2$ and $\Omega_b h^2$, as free parameters. Here $\Omega_{m0} = \Omega_c + \Omega_b$ and $h$ is the Hubble constant in units of 100 km s$^{-1}$ Mpc$^{-1}$.

\begin{figure*}
    \centering
    \includegraphics[width=0.49\textwidth]{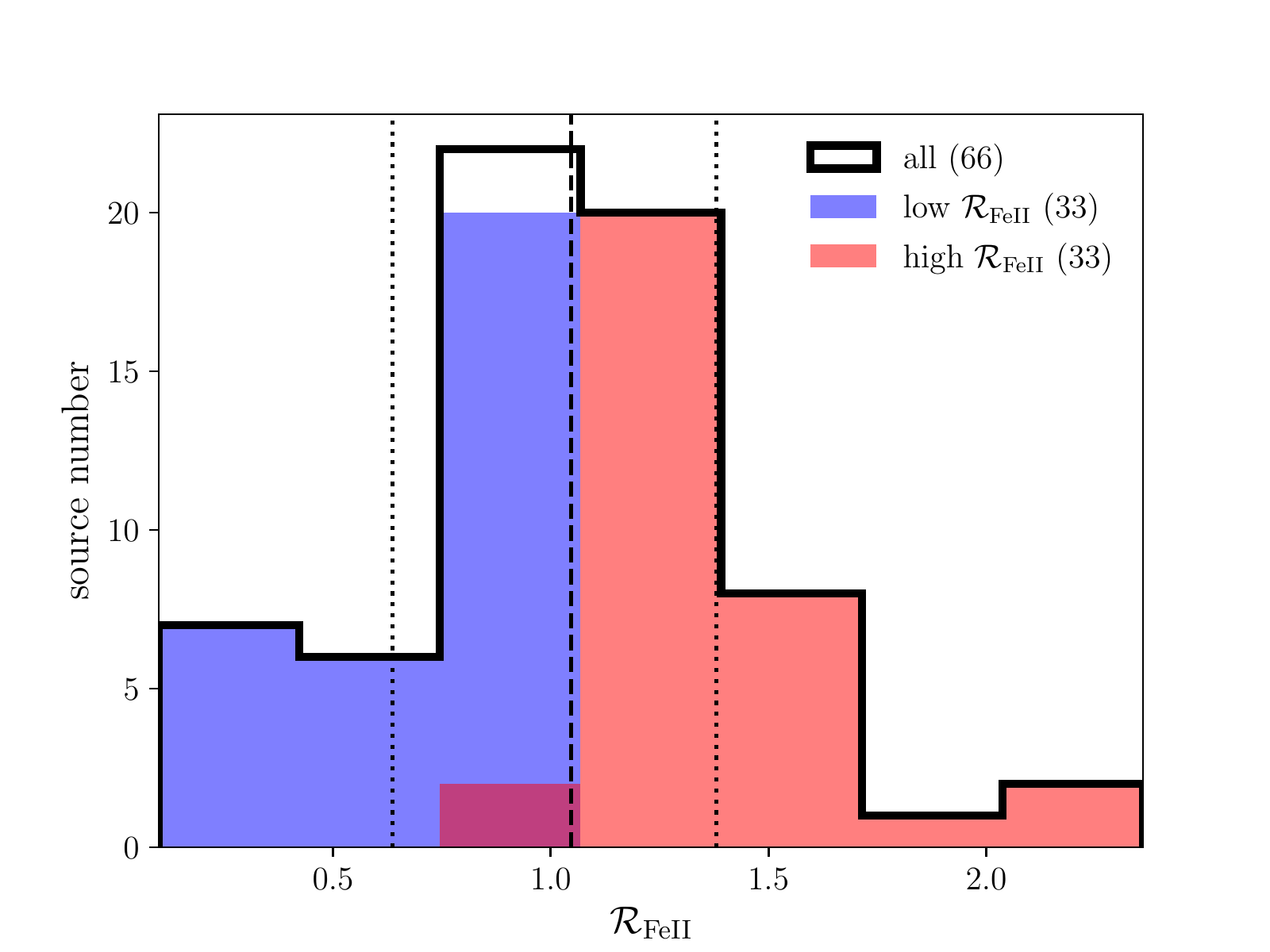}
    \includegraphics[width=0.49\textwidth]{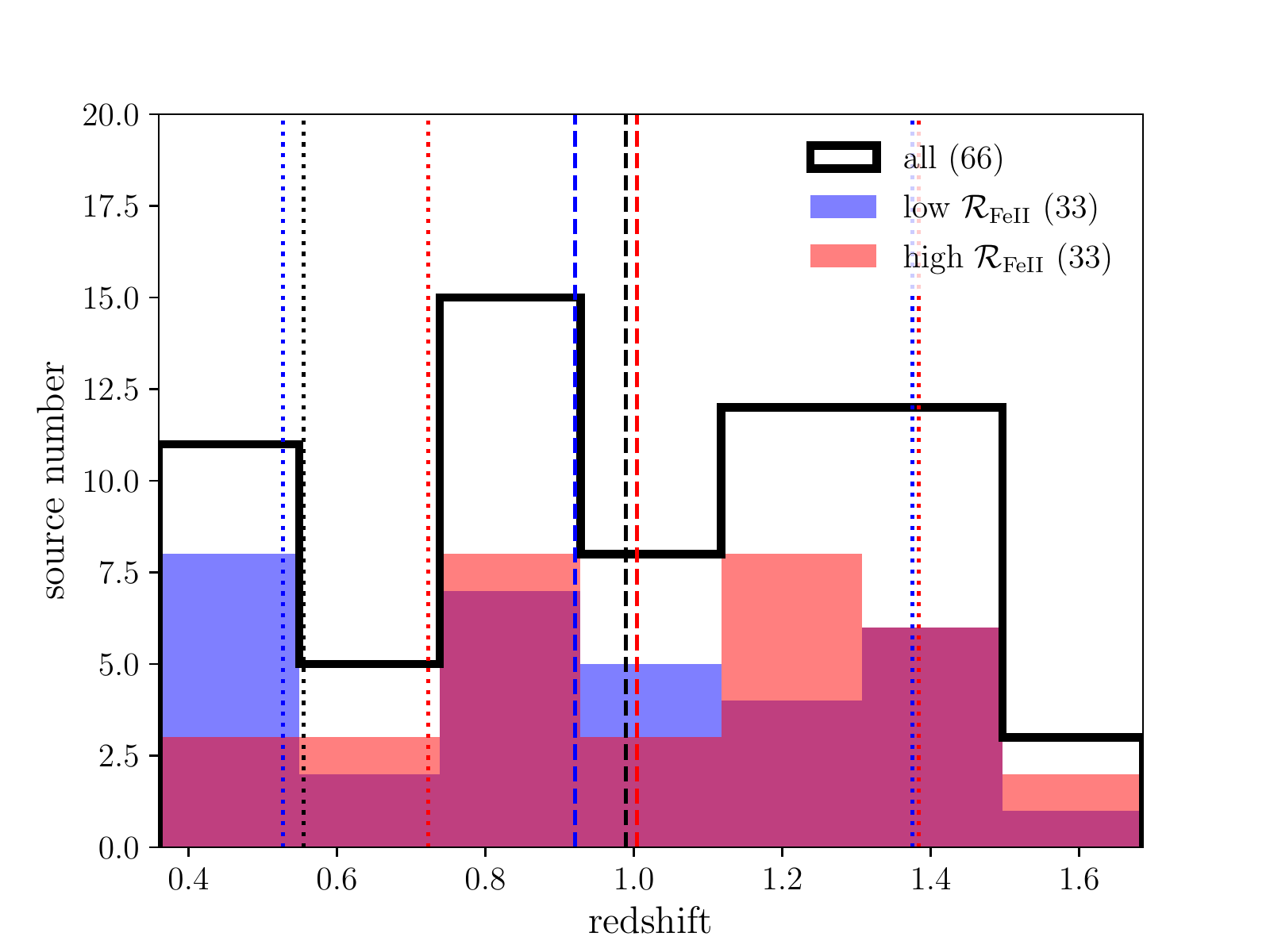}
    \caption{The relative \Feii\ intensity and the redshift distributions of 66 MgII QSOs. Left panel: The \rfe\ distribution for the whole sample (black histogram), low-\rfe\ subsample (blue histogram), and the high-\rfe\ subsample (pink histogram). The dashed vertical line denotes the median \rfe\ value (1.0467), while the dotted vertical lines stand for the 16\% and 84\% percentile values (0.64 and 1.38, respectively). The bin size of the histogram is set by the Sturges' and the Doane's choice, $\Delta$\rfe$\simeq 0.32$, which both give 7 bins for the \rfe\, range of $(0.1, 2.36)$. Right panel: The redshift distribution of 66 QSOs for the whole sample (black solid line), and low- and high-\rfe\ subsamples (33 sources in each subsample) that are depicted by blue and pink bars, respectively. The dashed vertical lines (black, blue, and red) stand for the corresponding median values for the whole, low-\rfe\, and high-\rfe\ subsamples, respectively. The two sets of dotted vertical lines stand for the corresponding 16\%- and 84\%-percentiles for each sample. The bin size of the histogram is detemined via the Sturges's and the Doane's choice, $\Delta z\simeq 0.19$, which both yield 7 bins in the redshift range of $(0.36, 1.686)$.}
    \label{fig_redshift_dist}
\end{figure*}

\section{Data}
\label{sec:data}

In this paper, we use 66 QSOs, with reverberation-measured time-delays of the broad \Mgii\ line with respect to the 3000\AA\, continuum, and with measurements of the intensity of the UV \Feii\ pseudocontinuum given in terms of the ratio parameter \rfe\ $=F(\mathrm{Fe}\textsc{ii}_{2250-2650 \AA})/F(\mathrm{Mg}\textsc{ii})$. These data are listed in Table~\ref{tab:MgQSOdata}, including their redshifts, 3000\AA\, continuum flux densities, rest-frame time-delays of the MgII line, the UV \rfe\ parameter, and the original publication reference. 

This \Mgii\ dataset spans the redshift range of $0.36 \leq z \leq 1.686$ and is a subset of our earlier 78 \Mgii\ QSO compilation \citep{khadka2021} for which the \rfe\ parameter is reliably inferred. The \rfe\ range for the whole sample is $(0.1, 2.36)$, or $(-1.0, 0.37)$ when using the log-scale. We also consider two low- and high-\rfe\ subsets of these 66 QSOs, divided at the median \rfe\ = 1.0467 value, each containing 33 QSOs. The \rfe\ distribution for 66 QSOs is shown in Fig.~\ref{fig_redshift_dist} (left panel), including the median as well as 16\%- and 84\%-percentile values of the distribution. The \rfe{}-subsample flag is included in Table~\ref{tab:MgQSOdata} in the last column where (1) denotes a low-\rfe\ source and (2) corresponds to a high-\rfe\ source. The low-\rfe\ subset spans the redshift range $0.36 \leq z \leq 1.587$ and the high-\rfe\ subset spans $0.4253 \leq z \leq 1.686$. The redshift distribution of all 66 QSOs, as well as low- and high-\rfe\ subsamples, is displayed in Fig.~\ref{fig_redshift_dist} (right panel). For all the sources, the median redshift value is $z=0.990$ (0.556 and 1.384 for 16\%- and 84\% percentiles, respectively), while it is $z=0.921$ for low-\rfe\ (16\% percentile: 0.527; 84\% percentile: 1.376) and $z=1.004$ for high-\rfe\ sources (16\% percentile: 0.723; 84\% percentile: 1.384). The redshift distribution of the \Mgii\ QSOs is quite different from that of the H$\beta$ QSOs, which peaks at the lowest-redshift bin between $z=0$ and $0.1$ for the whole H$\beta$ dataset as well as for low- and high-\rfe\ subsamples, see Fig.~1 in \citet{Khadkaetal2021c}.

\begin{figure}
    \centering
    \includegraphics[width=0.48\textwidth]{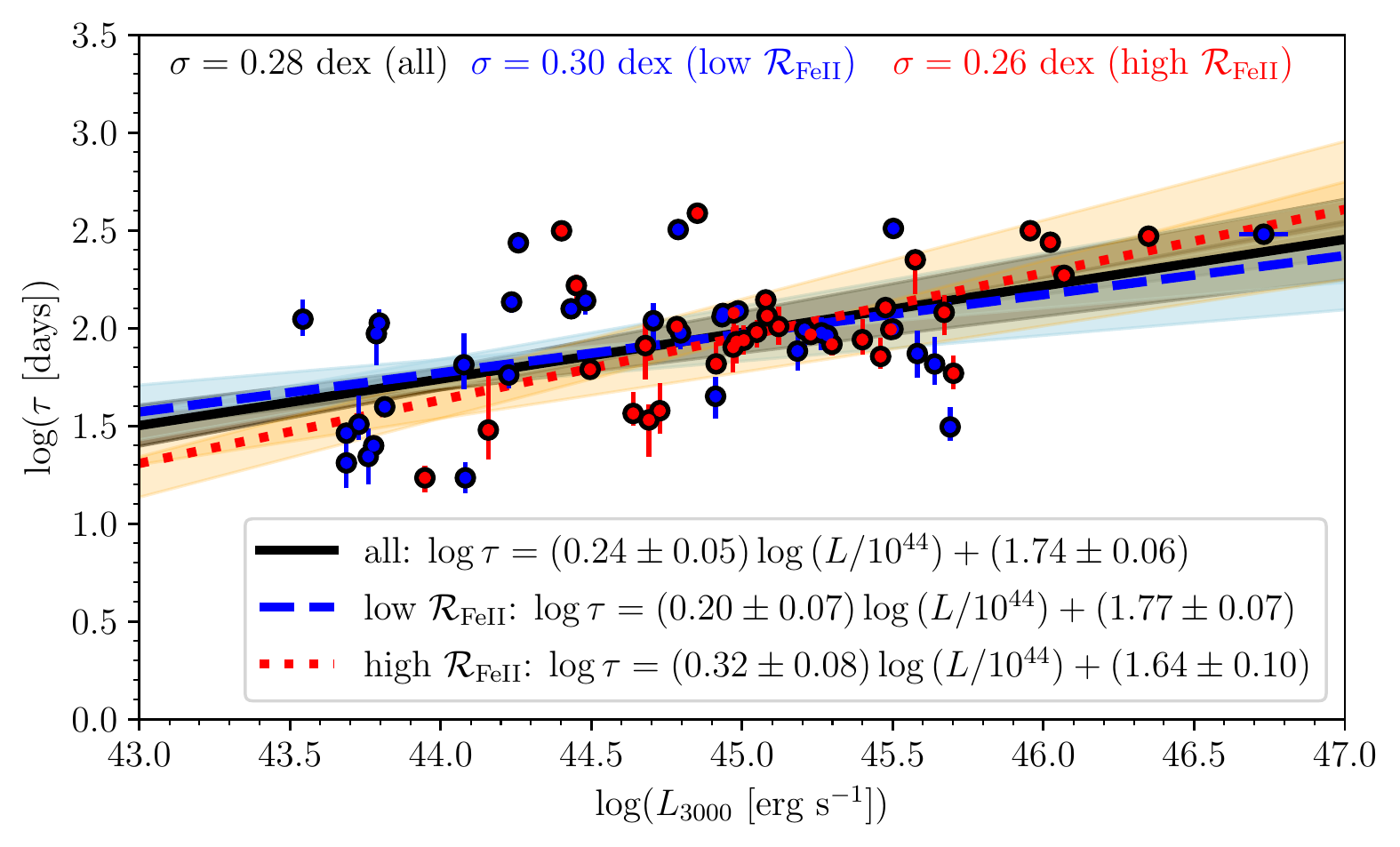}
    \caption{The \Mgii-based $R-L$ relation ( in a spatially flat $\Lambda$CDM model with $H_0=70\,{\rm km\,s^{-1}\,Mpc^{-1}}$, $\Omega_{m0}=0.3$, and $\Omega_{\Lambda}=0.7$) for 66 QSOs with the inferred relative \Feii\ intensities. The low-\rfe\ sources are depicted with blue circles, while high-\rfe\ sources are represented by red circles. The black solid, blue dashed, and red dotted lines correspond to the best-fit $R-L$ relations for the whole sample, the low-, and the high-\rfe\ subsamples, respectively. The best-fit parameters are listed in the legend. The shaded areas around each line correspond to $1\sigma$ confidence intervals. The intrinsic rms scatter for each sample is listed along the top part of the plot.}
    \label{fig_RL_relation}
\end{figure}

The \Mgii\ $R-L$ relation shows a significant positive correlation for the whole dataset of 66 sources as well as for the low-\rfe\ and the high-\rfe\ subsamples (33 sources in each subsample), see Fig.~\ref{fig_RL_relation} for the graphical depiction of the $R-L$ relation including the corresponding best-fit relations for all three datasets and the intrinsic scatters. For Fig.~\ref{fig_RL_relation}, to obtain the monochromatic luminosity at $3000$\,\AA, we assume a spatially-flat $\Lambda\text{CDM}$ model with $H_0=70\,{\rm km\,s^{-1}\,Mpc^{-1}}$, $\Omega_{m0}=0.3$, and $\Omega_{\Lambda}=0.7$. It is clear that the best-fit $R-L$ relations are consistent with each other within 1$\sigma$ confidence intervals,
\begin{align}
    \text{All:} \log{\tau} &= (0.24 \pm 0.05)\log{L_{44}}+(1.74 \pm 0.06)\,\notag\,,\\
    \text{Low-\rfe:} \log{\tau} &= (0.20 \pm 0.07)\log{L_{44}}+(1.77 \pm 0.07)\,\notag\,,\\
    \text{High-\rfe:} \log{\tau} &= (0.32 \pm 0.08)\log{L_{44}}+(1.64 \pm 0.10)\,\,,
    \label{eqs_RL_flat}
\end{align}
where $\tau$ is the source-frame time delay expressed in days, $L_{44}=L_{3000}/10^{44}\,{\rm erg\,s^{-1}}$ corresponds to the monochromatic 3000\,\AA\ luminosity $L_{3000}$ (expressed in ${\rm erg\,s^{-1}}$) normalized to $10^{44}\,{\rm erg\,s^{-1}}$, and $\log \equiv \log_{10}$. The intrinsic scatter of $\sigma_{\rm ext}=0.28$, $0.30$, and $0.26$ dex for the whole, low-\rfe, and high-\rfe\ samples are comparable, though it appears to be lower for the high-\rfe\ subsample. For the whole sample of 66 sources, the Pearson correlation coefficient is $r=0.51$ ($p=1.01\times 10^{-5}$) and the Spearman rank-order correlation coefficient is $s=0.40$ ($p=9.67 \times 10^{-4}$). For the low-\rfe\ subsample, we obtain a Pearson correlation coefficient of $r=0.46$ ($p=0.00738$) and a Spearman rank-order correlation coefficient of $s=0.37$ ($p=0.0354$). For the high-\rfe\ subsample, the Pearson correlation coefficient is $r=0.57$ ($p=5.79\times 10^{-4}$) and the Spearman rank-order correlation coefficient is $s=0.48$ ($p=4.91\times 10^{-3}$). Hence, for the high-\rfe\ subsample the correlation is stronger and more significant than for the low-\rfe\ subsample.

\begin{figure}
    \centering
    \includegraphics[width=\columnwidth]{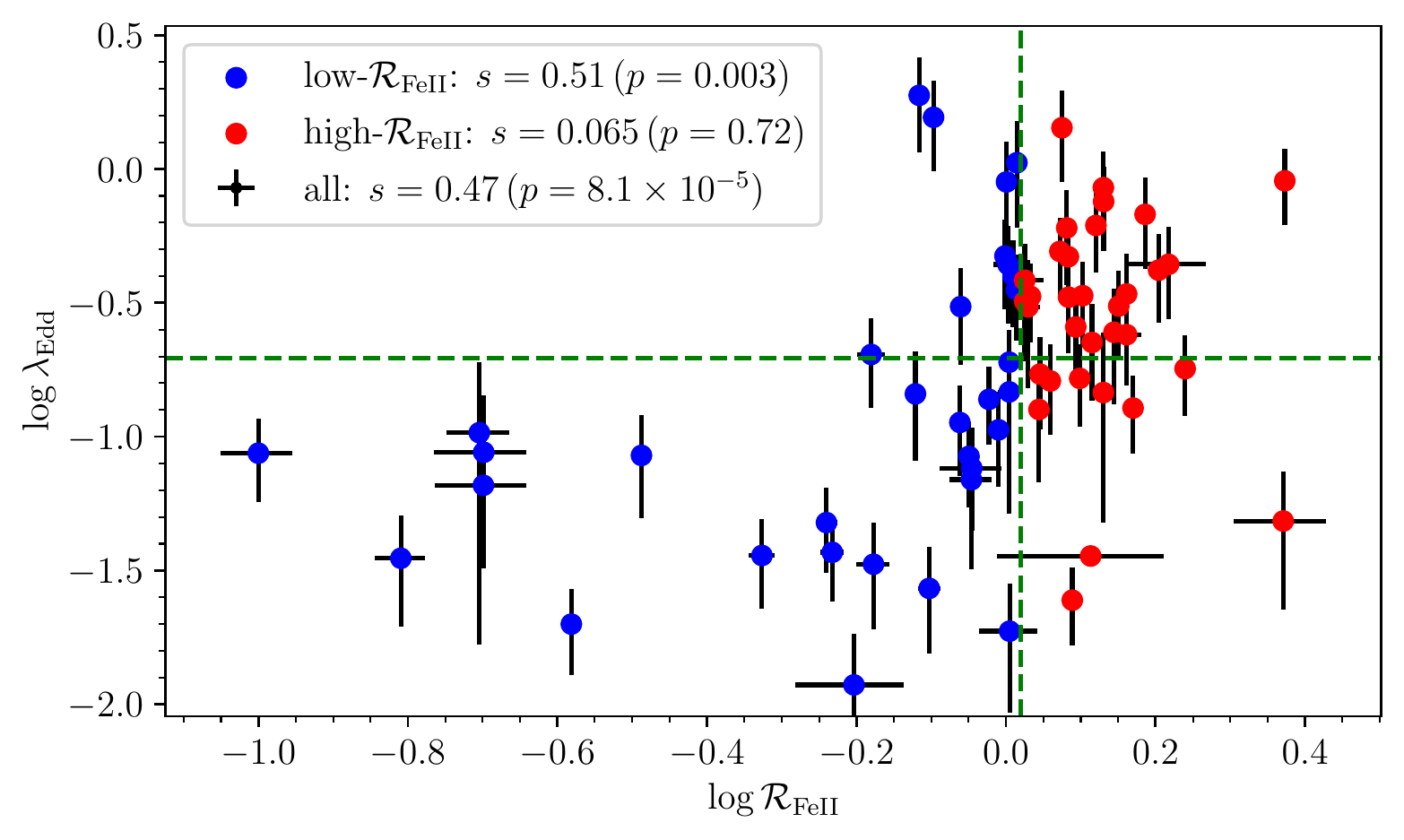}
    \includegraphics[width=\columnwidth]{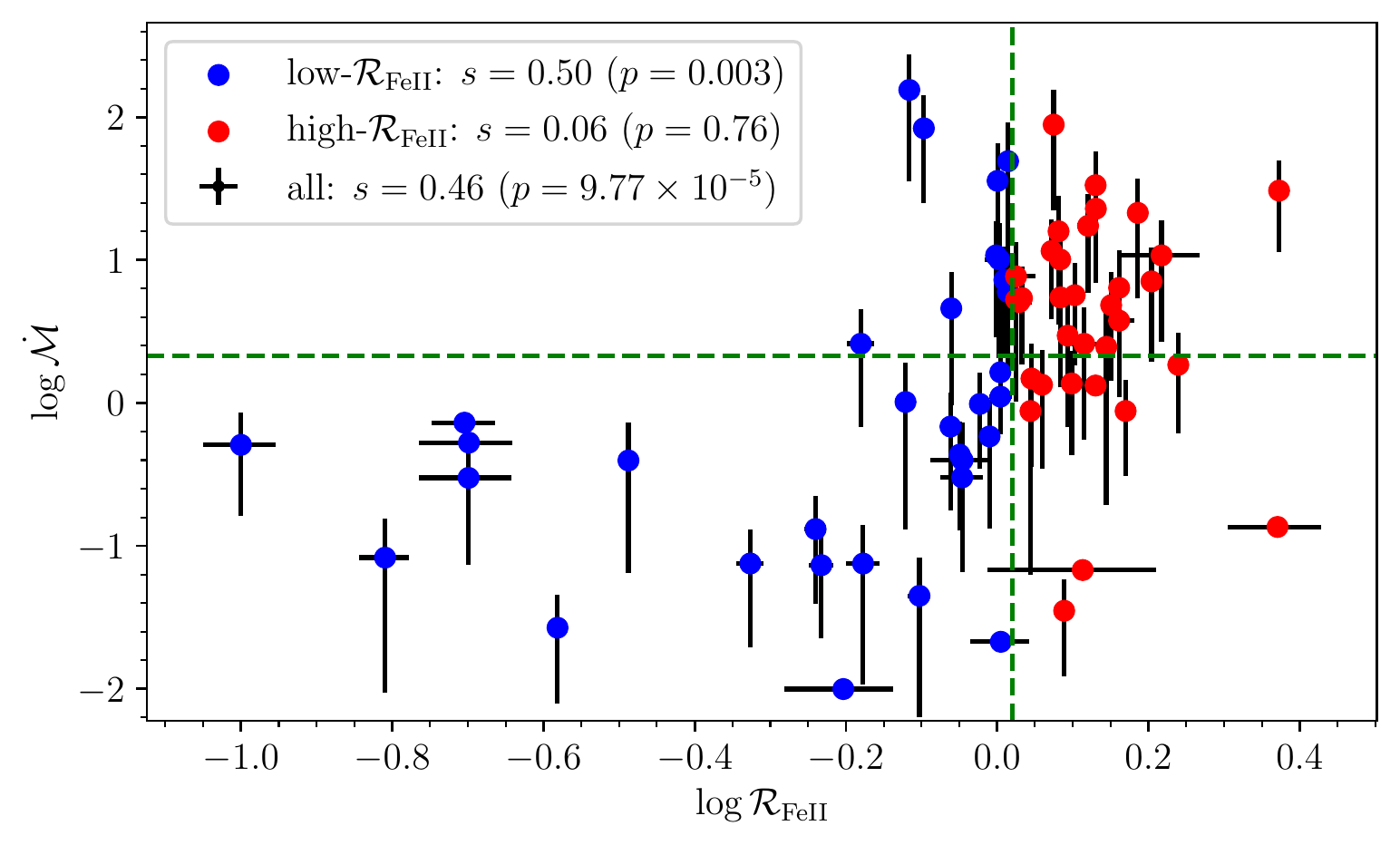}
    \caption{Relation between the relative \Feii\ strength expressed by the ratio \rfe\, and the Eddington ratio $\lambda_{\rm Edd}$ (top panel) and the dimensionless accretion rate $\dot{\mathcal{M}}$ (bottom panel). The dashed green lines mark the median values of the corresponding quantities. For the whole sample, \rfe\, shows a significant positive correlation with respect to both quantities as measured by the Spearman correlation coefficient, see the legend. However, when the sample is divided the significance of the correlation drops considerably, especially for the high-\rfe\ subsample, which exhibits no significant correlation with the quantities of interest.}
    \label{fig_correlation}
\end{figure}

The main motivation for including the \rfe\ parameter is that it is expected to be correlated with the Eddington ratio $\lambda_{\rm Edd}$\footnote{The Eddington ratio is $\lambda_{\rm Edd}\equiv L_{\rm bol}/L_{\rm Edd}$, where $L_{\rm bol}$ is the bolometric luminosity calculated as $L_{\rm bol}=\kappa_{\rm bol}L_{3000}$, where $\kappa_{\rm bol}$ is the bolometric correction factor taken from \citet{netzer2019}. $L_{\rm Edd}$ stands for the Eddington luminosity, where we adopt the standard relation for stationary spherical accretion. The black hole mass is calculated using the reveberation method, $M_{\bullet}=f_{\rm vir}c\tau \text{FWHM}^2/G$, where FWHM is the full width at half maximum in ${\rm km\,s^{-1}}$ for the \Mgii\ line, and $f_{\rm vir}$ is the virial factor calculated using the FWHM-dependent relation for \Mgii\ \citep{2018NatAs...2...63M}.} or the dimensionless accretion-rate $\dot{\mathcal{M}}$ \citep{2019ApJ...882...79P,2020ApJ...900...64S}. \footnote{The dimensionless accretion rate is $\dot{\mathcal{M}}=26.2 (L_{44}/\cos{\theta})^{3/2} m_7^{-2}$ \citep{2014ApJ...793..108W}, where $\theta$ is the accretion-disk angle with respect to the line of sight and we fix $\cos{\theta}=0.75$ according to the mean angle for type I AGN taking into account the covering factor of the dusty molecular torus \citep{2010ApJ...714..561L,2015ApJ...803...57I}. $m_7$ is the reverberation-measured black hole mass scaled to $10^7\,M_{\odot}$, i.e. $m_7\equiv M_{\bullet}/(10^7\,M_{\odot})$.} Since both $\lambda_{\rm Edd}$ and $\dot{\mathcal{M}}$ depend on the \Mgii\ QSO black-hole mass that is determined via reverberation mapping, they are intrinsically correlated with $\tau$ as well as $L_{3000}$. Their inclusion would bias the extended $R-L$ relation, which is avoided by using the independent observable \rfe. Since the scatter about the $R-L$ relation appears to be driven by the accretion rate \citep{Mary2020}, the inclusion of \rfe\ as an independent observable could in principle decrease the scatter. We show the correlation between \rfe\ and the two accretion-rate quantities in Fig.~\ref{fig_correlation}, computed in the fixed flat $\Lambda$CDM model of the previous paragraph. For the whole sample of 66 sources, \rfe\ exhibits a significant positive correlation with both $\lambda_{\rm Edd}$ and $\dot{\mathcal{M}}$ as measured by the Spearman rank-order correlation coefficient, $s=0.47$ ($p=8.07\times 10^{-5}$) and $s=0.46$ ($p=9.77 \times 10^{-5}$), respectively. However, when the sample is divided in two with respect to the median of \rfe, the correlation significance drops considerably, especially for the high-\rfe\ subsample, which does not show a significant correlation.

In this study, we also use 11 BAO measurements listed in Table~1 of \cite{KhadkaRatra2021a} and 31 $H(z)$ observations listed in Table~2 of \cite{Ryanetal2018}. We use the better-established joint BAO+$H(z)$ sample cosmological constraints for comparison with the cosmological constraints based on \Mgii\ QSOs, in all 6 cosmological models considered in this study.

\section{Methods and Model comparison}
\label{sec:methods}

First, we consider the 2-parameter $R-L$ relation using eq.~\eqref{eq:corr} with $k=0$; see also \citet{khadka2021} for the analysis using 78 \Mgii\, QSOs and the 2-parameter $R-L$ relation. When we analyse \Mgii\ data assuming the 2-parameter $R-L$ relation, we denote the three data sets without a prime (\Mgii\ QSO-66, \Mgii\ low-\rfe, and \Mgii\ high-\rfe).

Attempts have been made to correct for the accretion rate effect observed in the 2-parameter $R-L$ relation \citep{duwang_2019,Mary2020, Michal2021} by extending it to a 3-parameter $R-L$ relation; the hope is that this will reduce the intrinsic dispersion and result in tighter cosmological constraints. In our \hb\ QSO work \citep{Khadkaetal2021c} we did not find evidence for this, however, the \hb\ QSO cosmological constraints were somewhat inconsistent with those from better-established cosmological probes. Here we examine this issue for \Mgii\ QSOs whose cosmological constraints are consistent with those from better-established cosmological probes \citep{khadka2021}.

For this study we consider two 3-parameter $R-L$ relations, the first being eq.~\eqref{eq:corr} with $q=$\rfe\, where $k$ is the third free parameter associated with the intensity of the UV \Feii\  flux ratio parameter \rfe\ and must be determined from data. When we analyse \Mgii\ data assuming this 3-parameter $R-L$ relation we denote the three data sets with a single prime (\Mgii$^{\prime}$ QSO-66, \Mgii$^{\prime}$ low-\rfe, and \Mgii$^{\prime}$ high-\rfe).

The second 3-parameter $R-L$ relation we use assumes a power-law dependency of $\tau$ on \rfe\, hence $q=\log$\rfe\, in eq.~\eqref{eq:corr}. When we analyse \Mgii\ data assuming this 3-parameter $R-L$ relation we denote the three data sets with a double prime (\Mgii$^{\prime\prime}$ QSO-66, \Mgii$^{\prime\prime}$ low-\rfe, and \Mgii$^{\prime\prime}$ high-\rfe).

The rest-frame time-delay of a QSO at known redshift in a given cosmological model can be theoretically predicted using eq.~\eqref{eq:corr}, along with eqs.\ (\ref{eq_L3000}) and (\ref{eq:DL}). The log likelihood function that compares predicted time-delays with the corresponding observed time-delays, and so allows us to determine the constraints on cosmological-model and QSO-correlation parameters, is \citep{Dago2005}
\begin{equation}
\label{eq:chi2}
    \ln({\rm LF}) = -\frac{1}{2}\sum^{N}_{i = 1} \left[\frac{[\log(\tau^{\rm obs}_{X,i}) - \log(\tau^{\rm th}_{X,i})]^2}{s^2_i} + \ln(2\pi s^2_i)\right].
\end{equation}
Here $\ln$ = $\log_e$, $\tau^{\rm th}_{X,i}(\textbf{\emph{p}})$ and $\tau^{\rm obs}_{X,i}$ are the predicted and observed time-delays at measured redshift $z_i$. In the 2-parameter $R-L$ relation case,  $s^2_i = \sigma^2_{\log{\tau_{\rm obs},i}} + \gamma^2 \sigma^2_{\log{F_{3000},i}} + \sigma_{\rm ext}^2$,  while in the linear 3-parameter $R-L$ relation case,  $s^2_i = \sigma^2_{\log{\tau_{\rm obs},i}} + \gamma^2 \sigma^2_{\log{F_{3000},i}} + k^2 \sigma^2_{{\cal R}_{\text{Fe{\sc II}}},i} + \sigma_{\rm ext}^2$, where $\sigma_{\log{\tau_{\rm obs},i}}$, $\sigma_{\log{F_{3000},i}}$, and $\sigma_{{\cal R}_{\text{Fe{\sc II}}},i}$ are the measurement errors on the observed time-delay ($\tau^{\rm obs}_{X,i}$), measured flux ($F_{3000,i}$), and measured \rfe$_{,i}$ respectively, and in the log 3-parameter $R-L$ relation case,  $s^2_i = \sigma^2_{\log{\tau_{\rm obs},i}} + \gamma^2 \sigma^2_{\log{F_{3000},i}} + k^2 \sigma^2_{\log{{\cal R}_{\text{Fe{\sc II}}}},i} + \sigma_{\rm ext}^2$, where $\sigma_{\log{{\cal R}_{\text{Fe{\sc II}}}},i}$ is the measurement error on the measured log \rfe$_{,i}$. $\sigma_{\rm ext}$ is the intrinsic dispersion of the $R-L$ relation.

The BAO + $H(z)$ cosmological constraints used here are from \cite{KhadkaRatra2021a}; see that paper for the derivation and description of these constraints. The better-established BAO+$H(z)$ data cosmological constraints are used for comparison with the \Mgii\ QSO cosmological constraints.

We perform Markov chain Monte Carlo (MCMC) sampling, as implemented in the {\sc MontePython} code \citep{Brinckmann2019}, to maximize the log likelihood function given in eq.\ (\ref{eq:chi2}). We use top hat priors for each free parameter, see Table \ref{tab:prior}. \Mgii\ data cannot constrain $H_0$ because of the degeneracy between $\beta$ and $H_0$, so in QSO data analyses we set $H_0$ to $70$ ${\rm km}\hspace{1mm}{\rm s}^{-1}{\rm Mpc}^{-1}$. Each MCMC chain satisfies the Gelman-Rubin convergence criterion, $R-1 < 0.05$. The MCMC sampling chains are analysed using the {\sc Python} package {\sc Getdist} \citep{Lewis_2019}. 

\begin{table}
	\centering
	\caption{Non-zero flat prior parameter ranges.}
	\label{tab:prior}
	\begin{threeparttable}
	\begin{tabular}{l|c}
	\hline
	Parameter & Prior range \\
	\hline
	$\Omega_bh^2$ & $[0, 1]$ \\
	$\Omega_ch^2$ & $[0, 1]$ \\
    $\Omega_{m0}$ & $[0, 1]$ \\
    $\Omega_{k0}$ & $[-2, 2]$ \\
    $\omega_{X}$ & $[-5, 0.33]$ \\
    $\alpha$ & $[0, 10]$ \\
    $\sigma_{\rm ext}$ & $[0, 5]$ \\
    $\beta$ & $[0, 10]$ \\
    $\gamma$ & $[0, 5]$ \\
    $k$ & $[-10, 10]$ \\
	\hline
	\end{tabular}
    \end{threeparttable}
\end{table}

We compute the Akaike and the Bayesian information criterion ($AIC$ and $BIC$) values and use them for the comparison of different $R-L$ relations. These are
\begin{align}
\label{eq:AIC}
    AIC =& -2 \ln({\rm LF}_{\rm max}) + 2d,\\
\label{eq:BIC}
    BIC =& -2 \ln({\rm LF}_{\rm max}) + d\ln{N}\, ,
\end{align}
where $\rm LF_{\rm max}$ is the maximum likelihood value, $d$ is the number of free parameters, and $N$ is the number of data points, with the degrees of freedom $dof = N - d$. We compute $\Delta AIC$ and $\Delta BIC$ differences of the 3-parameter $R-L$ relations with respect to the corresponding 2-parameter $R-L$ reference model. Positive (negative) values of $\Delta AIC$ or $\Delta BIC$ indicate that the 3-parameter correlation relation under investigation fit the measurements worse (better) than the 2-parameter reference one. $\Delta AIC(BIC) \in [0, 2]$ is weak evidence in favor of the 2-parameter reference relation, $\Delta AIC(BIC) \in(2, 6]$ is positive evidence for the reference relation, $\Delta AIC(BIC)>6$ is strong evidence for the reference relation, and $\Delta AIC(BIC)>10$ is very strong evidence for the reference reference.

\section{Results}
\label{sec:QSO}

\begin{figure*}
\begin{multicols}{2}
    \includegraphics[width=\linewidth,height=7cm]{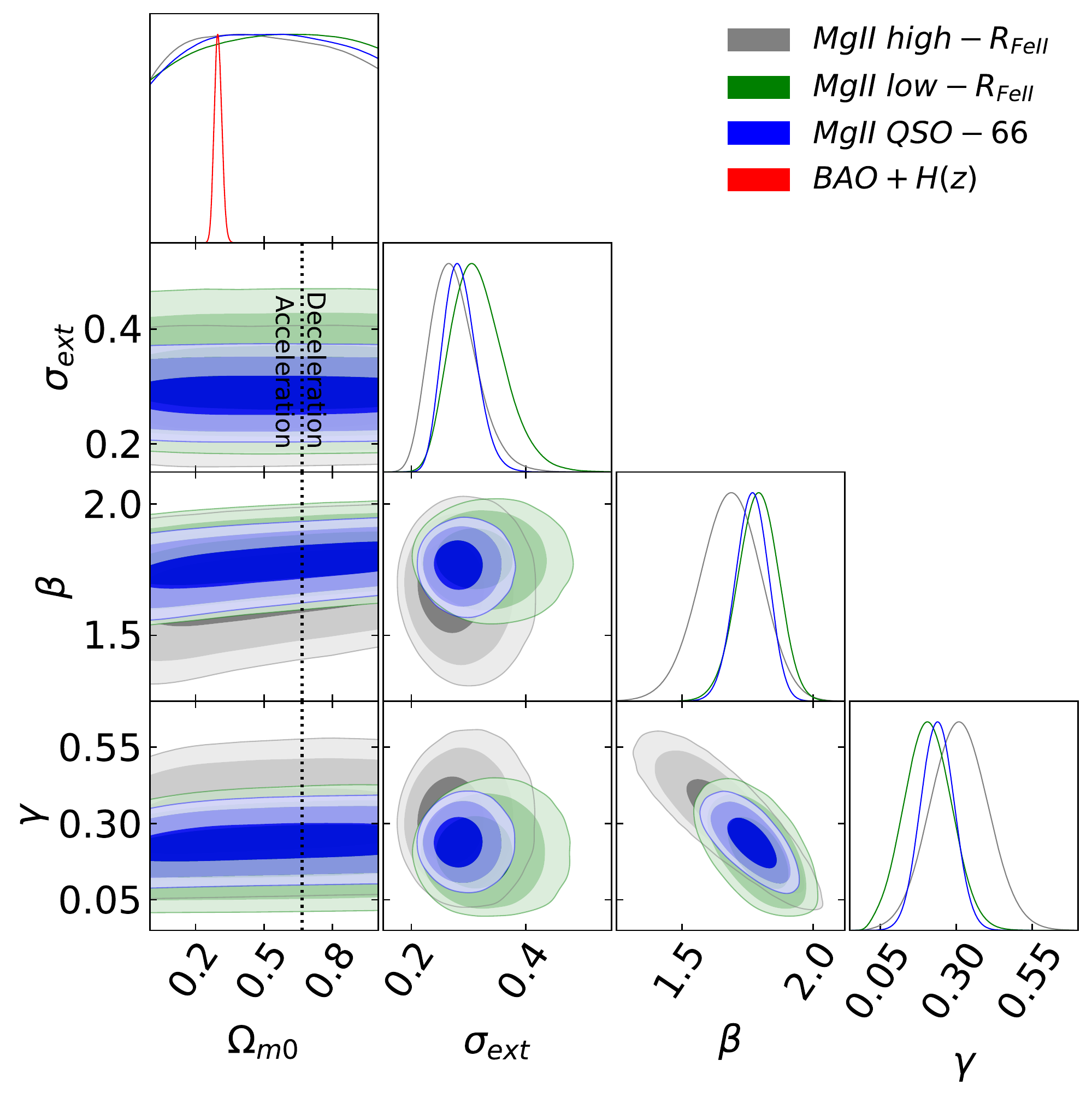}\par
    \includegraphics[width=\linewidth,height=7cm]{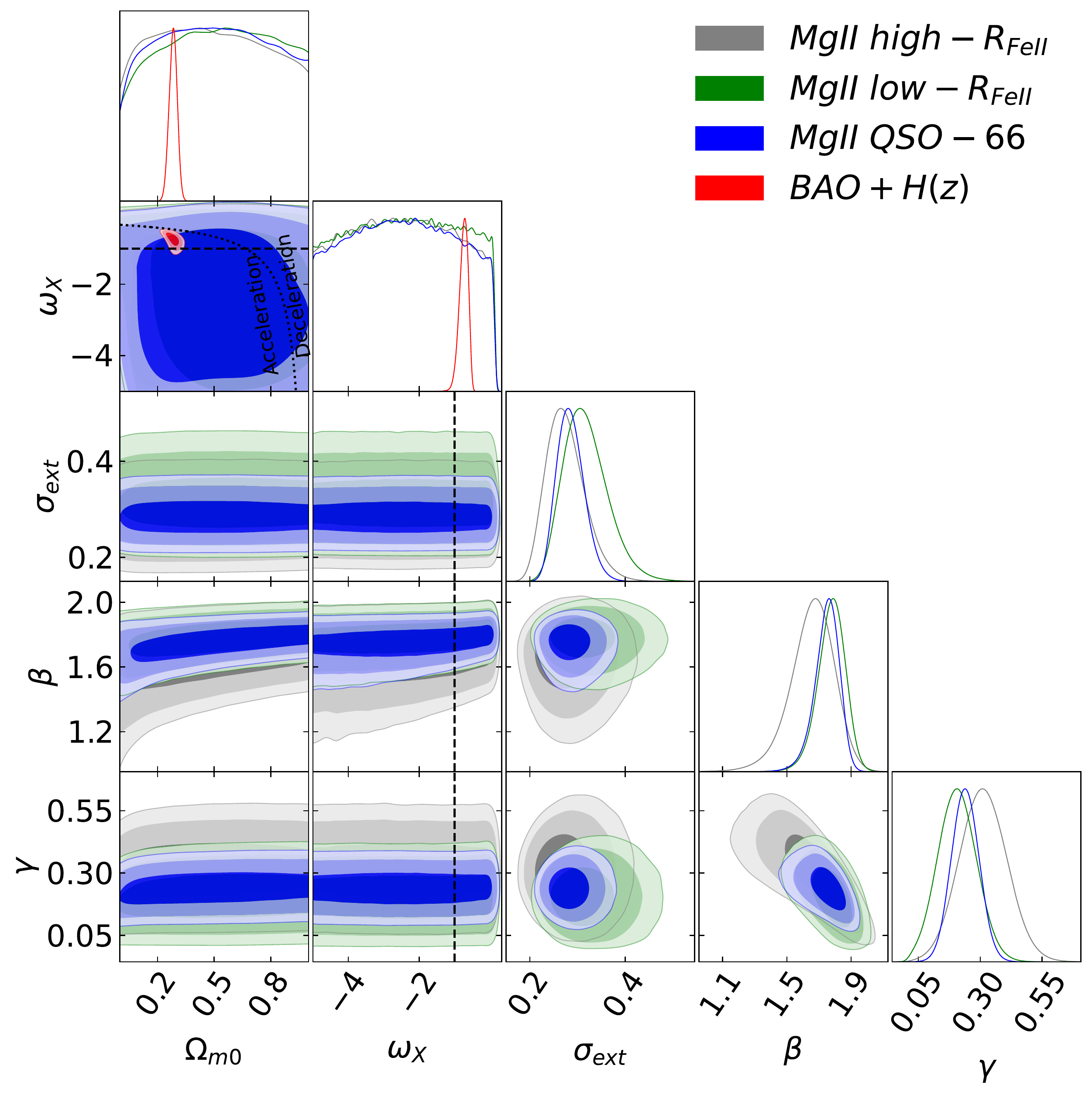}\par
    \includegraphics[width=\linewidth,height=7cm]{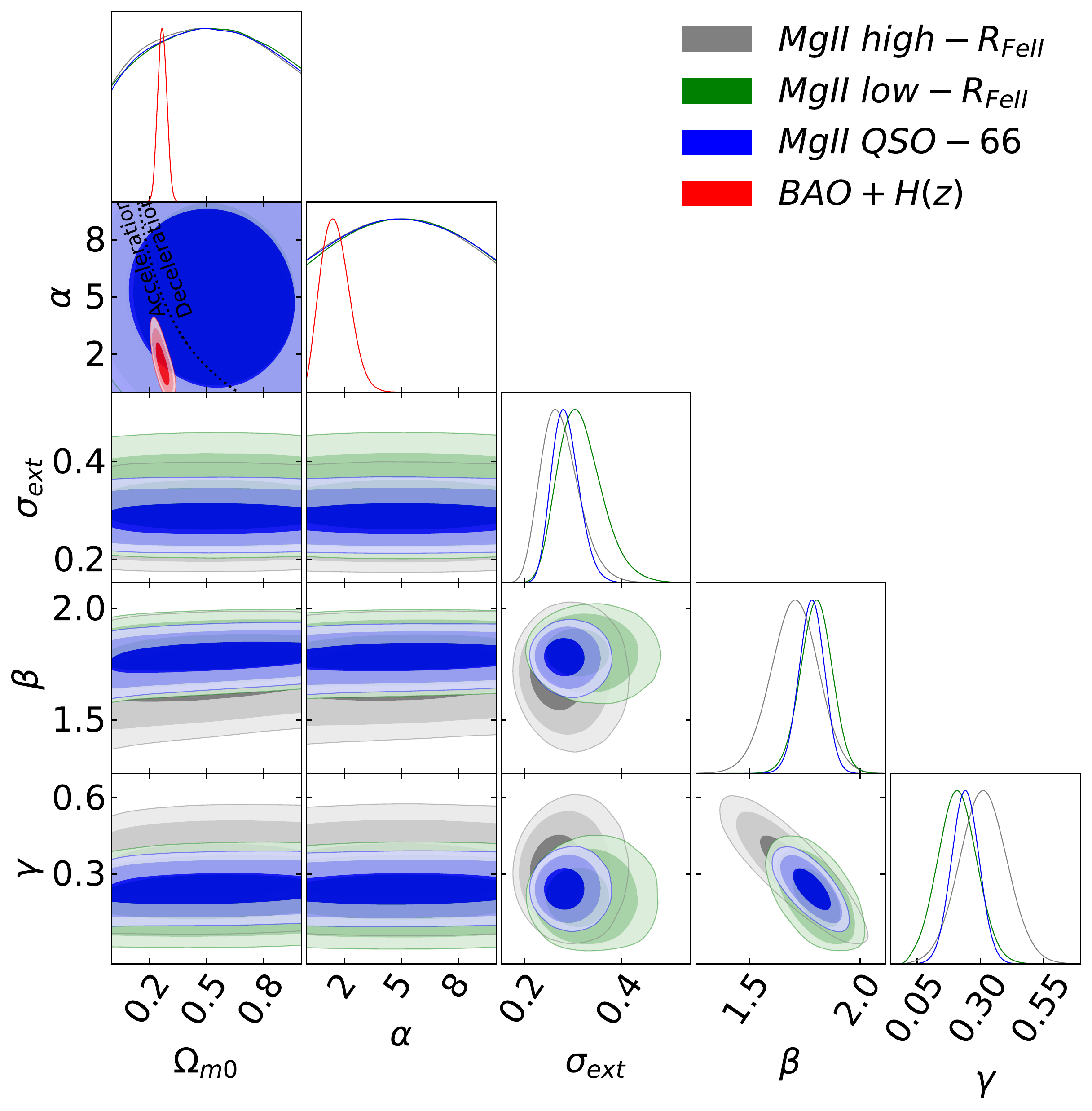}\par
    \includegraphics[width=\linewidth,height=7cm]{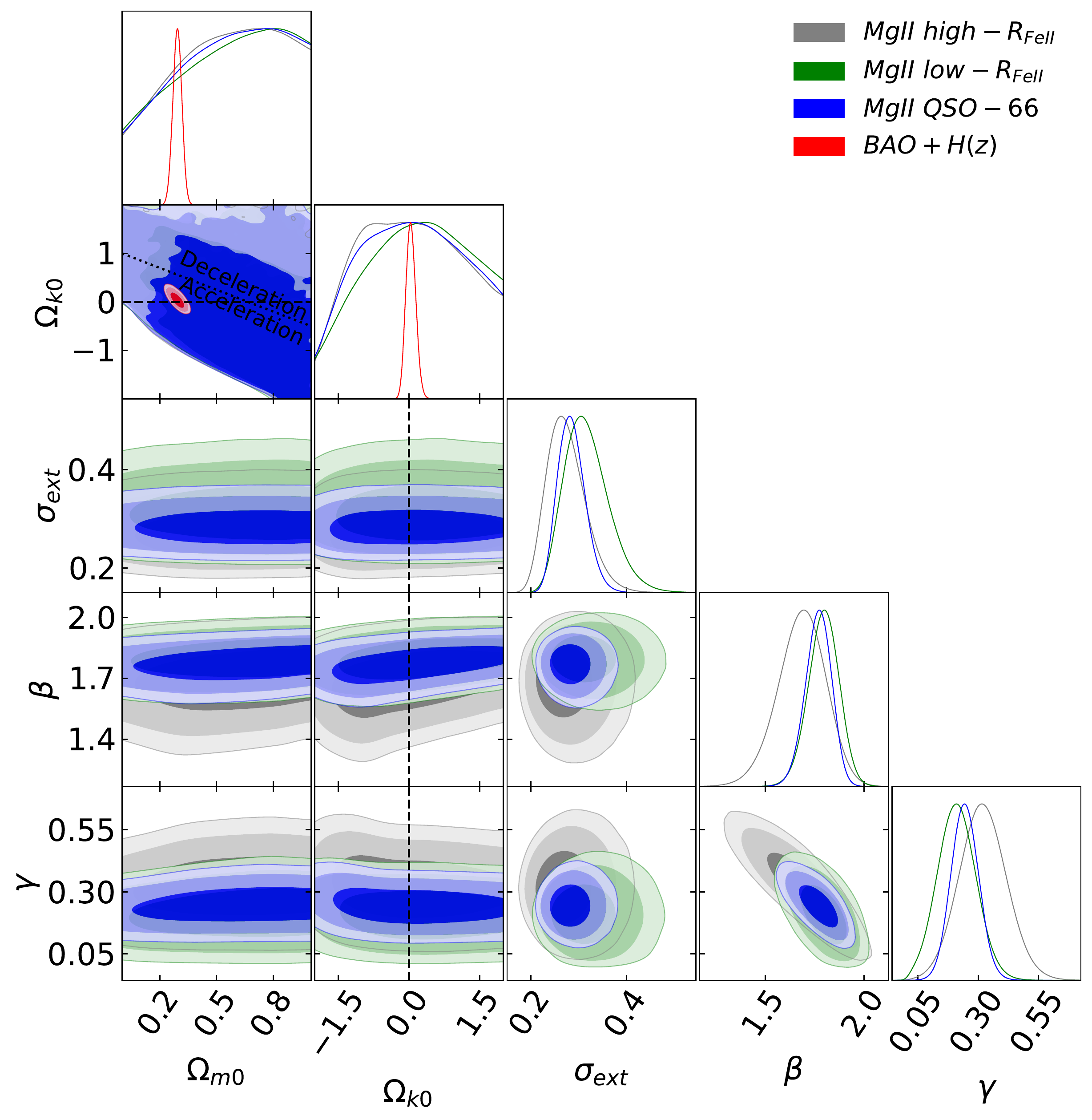}\par
    \includegraphics[width=\linewidth,height=7cm]{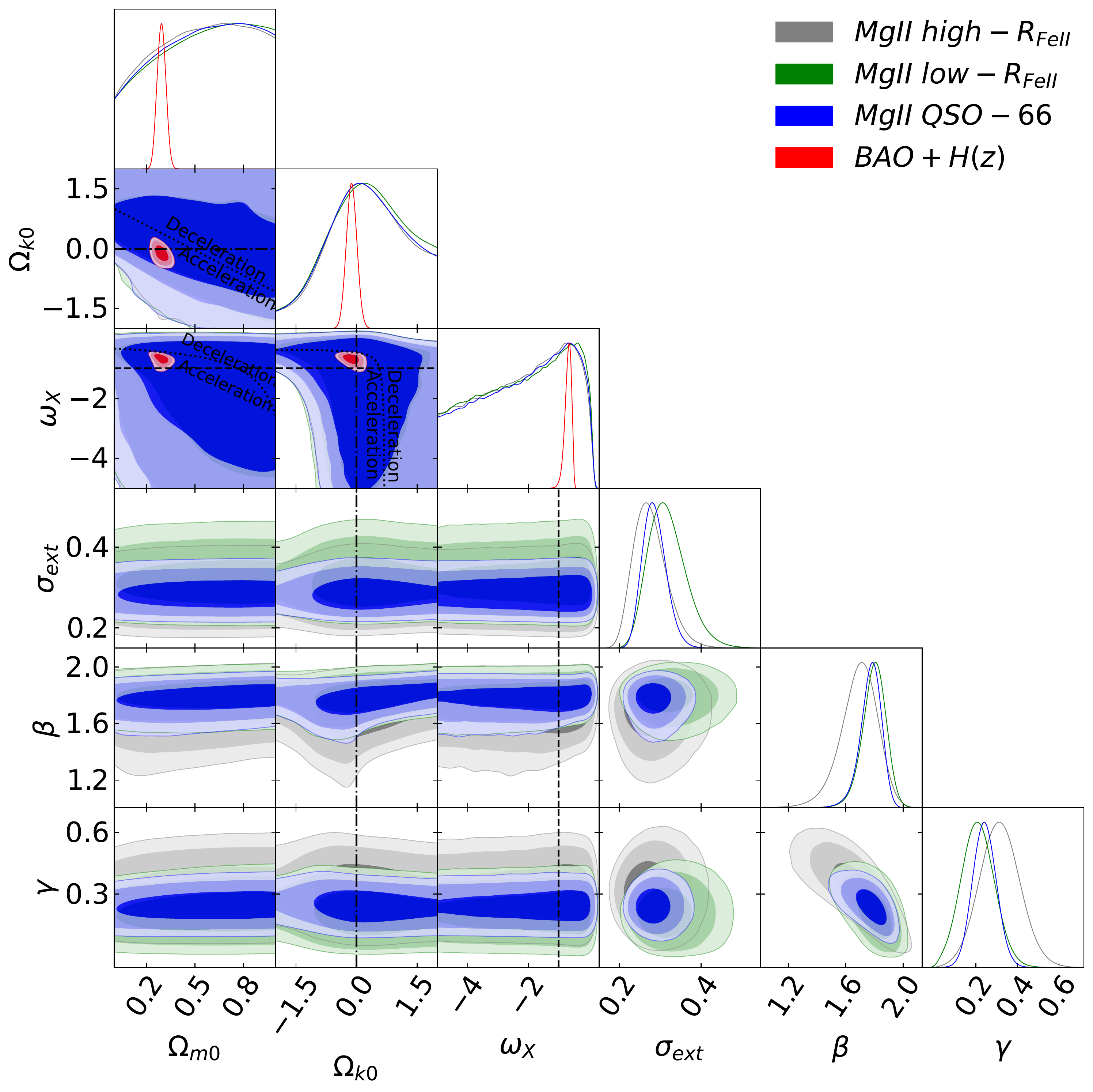}\par
    \includegraphics[width=\linewidth,height=7cm]{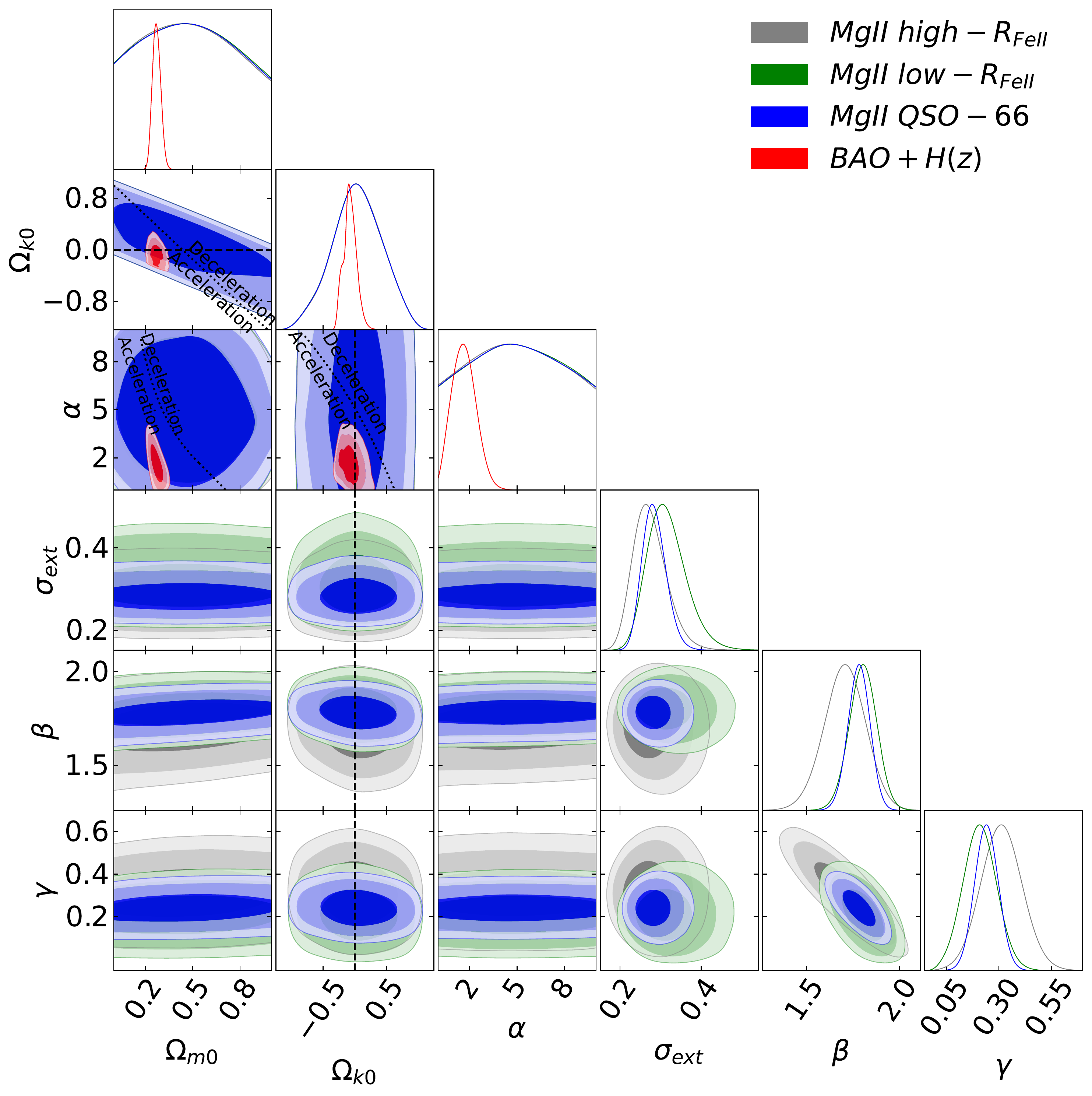}\par
\end{multicols}
\caption{One-dimensional likelihood distributions and two-dimensional likelihood contours at 1$\sigma$, 2$\sigma$, and 3$\sigma$ confidence levels using \Mgii\ high-\rfe\ (gray), \Mgii\ low-\rfe\ (green), \Mgii\ QSO-66 (blue), and BAO + $H(z)$ (red) data for all free parameters. Left column shows the flat $\Lambda$CDM model, flat XCDM parametrization, and flat $\phi$CDM model respectively. The black dotted lines in all plots are the zero acceleration lines. The black dashed lines in the flat XCDM parametrization plots are the $\omega_X=-1$ lines. Right column shows the non-flat $\Lambda$CDM model, non-flat XCDM parametrization, and non-flat $\phi$CDM model respectively. Black dotted lines in all plots are the zero acceleration lines. Black dashed lines in the non-flat $\Lambda$CDM and $\phi$CDM model plots and black dotted-dashed lines in the non-flat XCDM parametrization plots correspond to $\Omega_{k0} = 0$. The black dashed lines in the non-flat XCDM parametrization plots are the $\omega_X=-1$ lines.}
\label{fig:1}
\end{figure*}

\begin{figure*}
\begin{multicols}{2}
    \includegraphics[width=\linewidth,height=7cm]{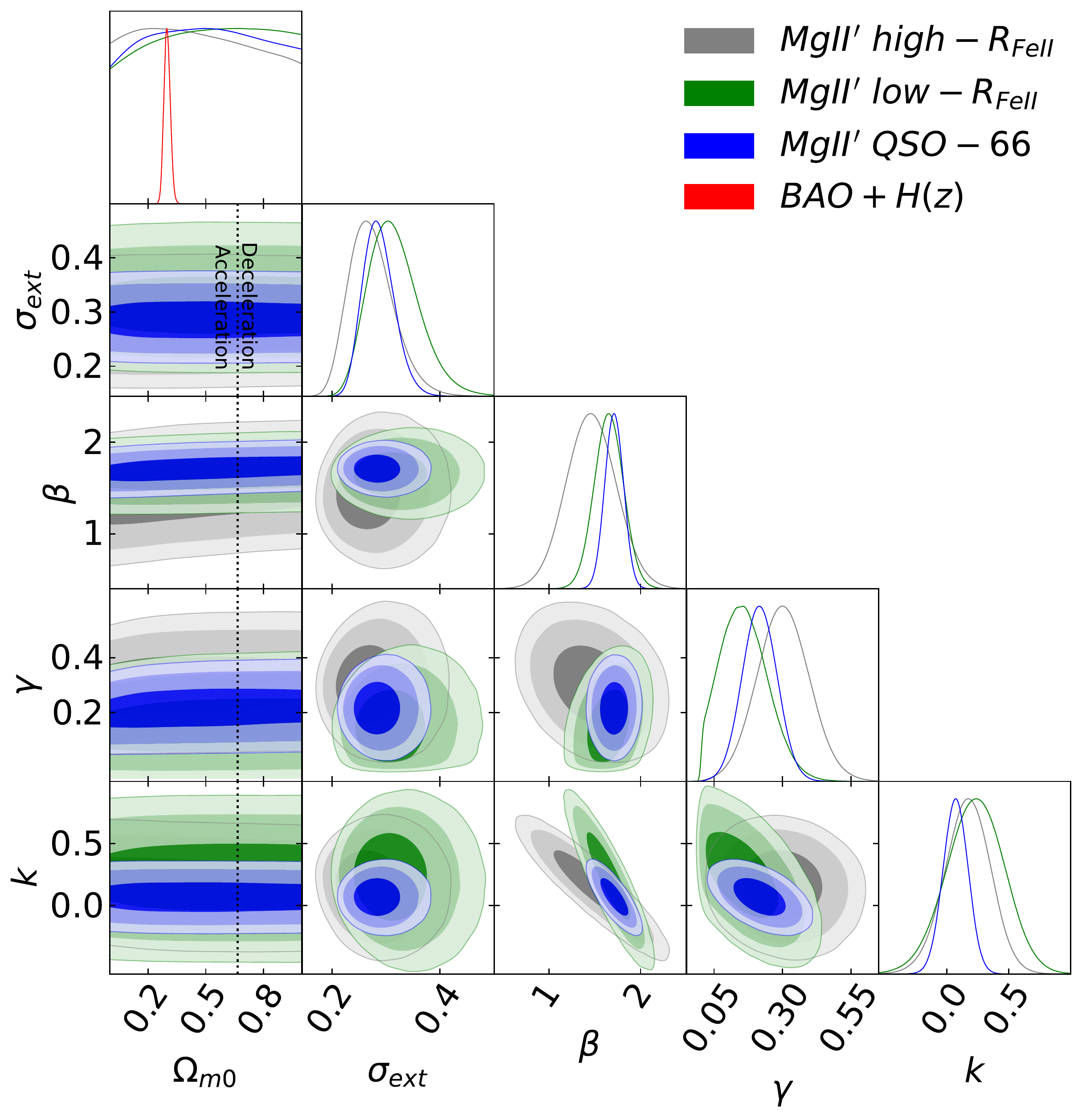}\par
    \includegraphics[width=\linewidth,height=7cm]{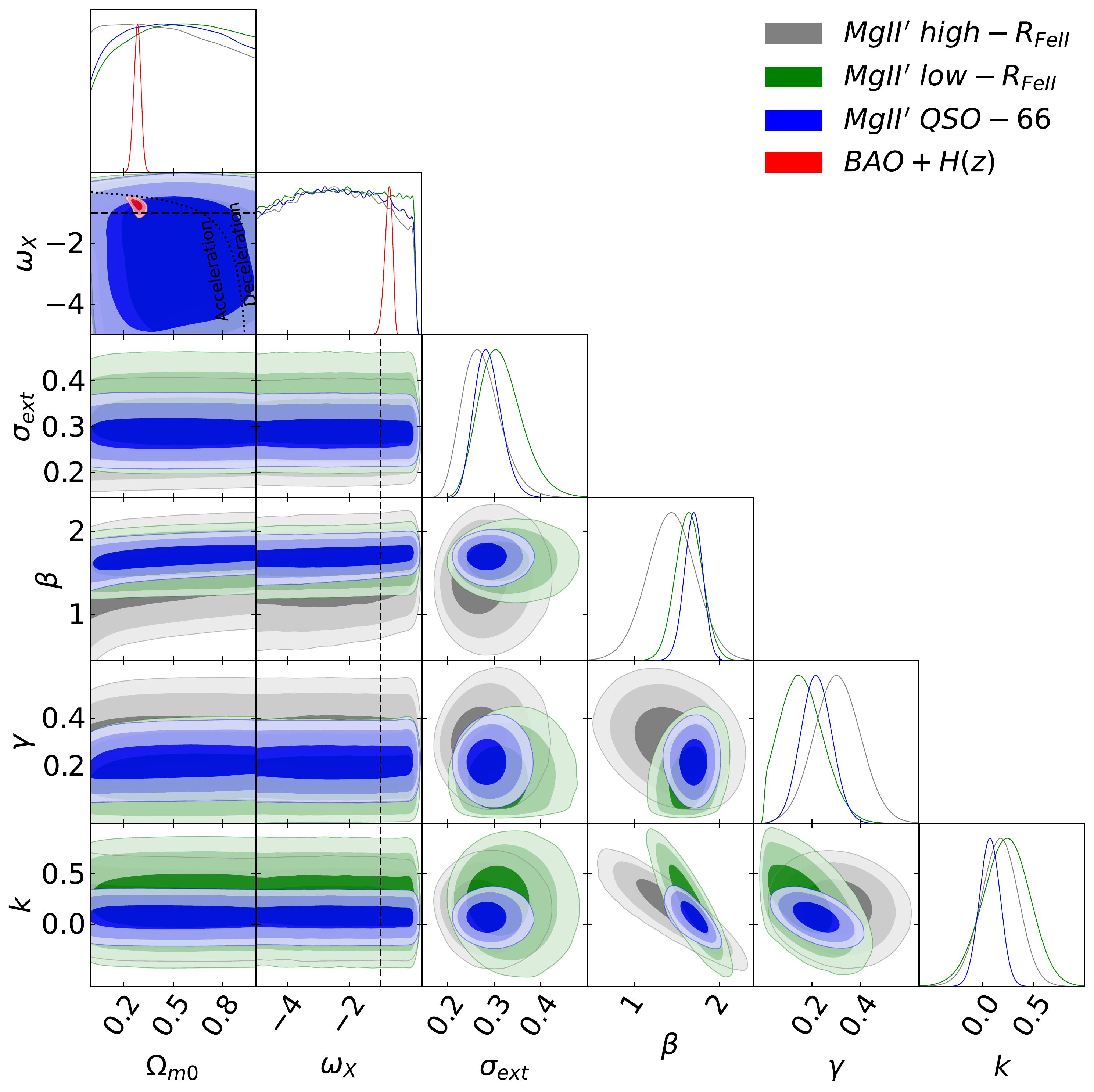}\par
    \includegraphics[width=\linewidth,height=7cm]{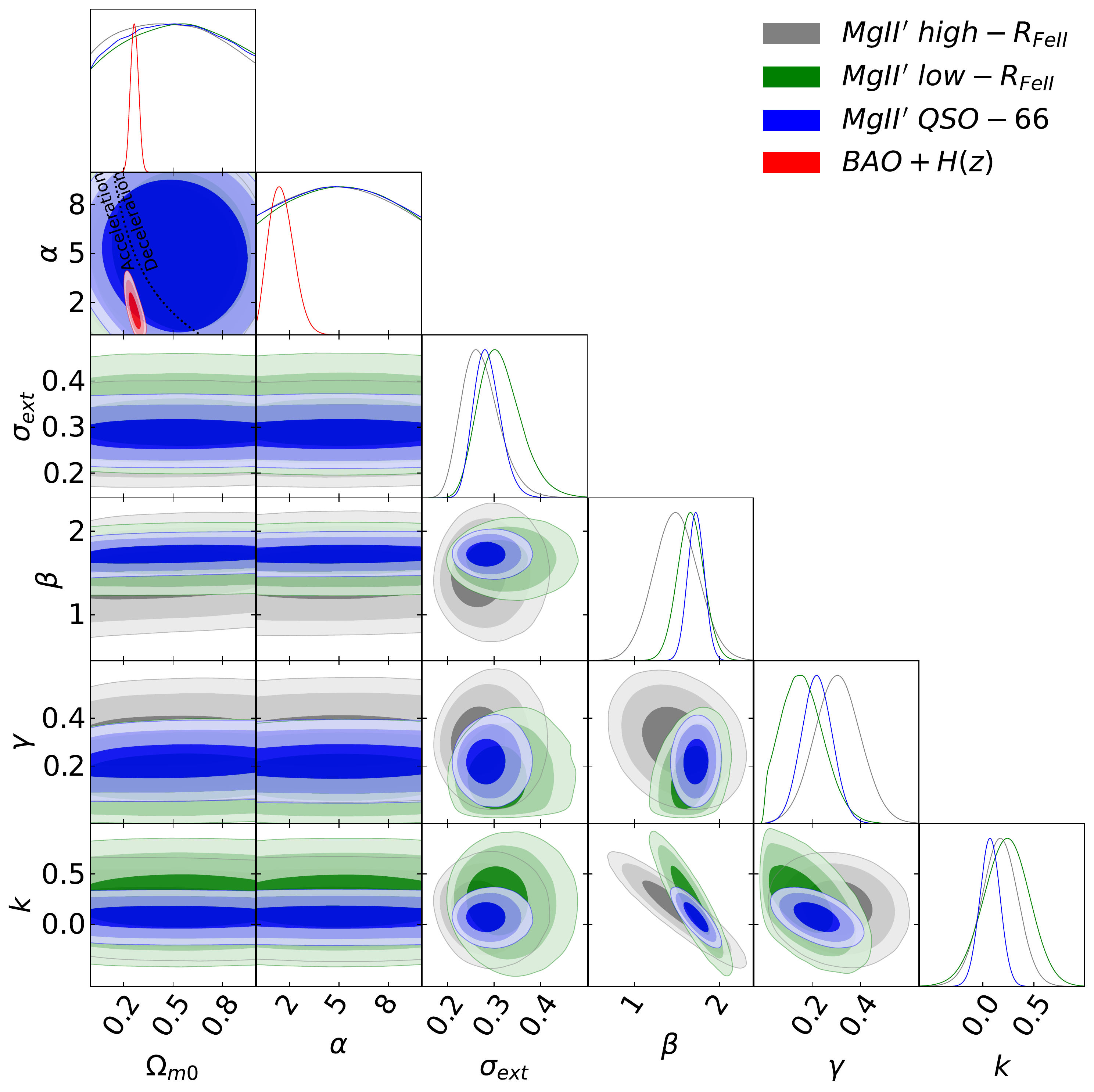}\par
    \includegraphics[width=\linewidth,height=7cm]{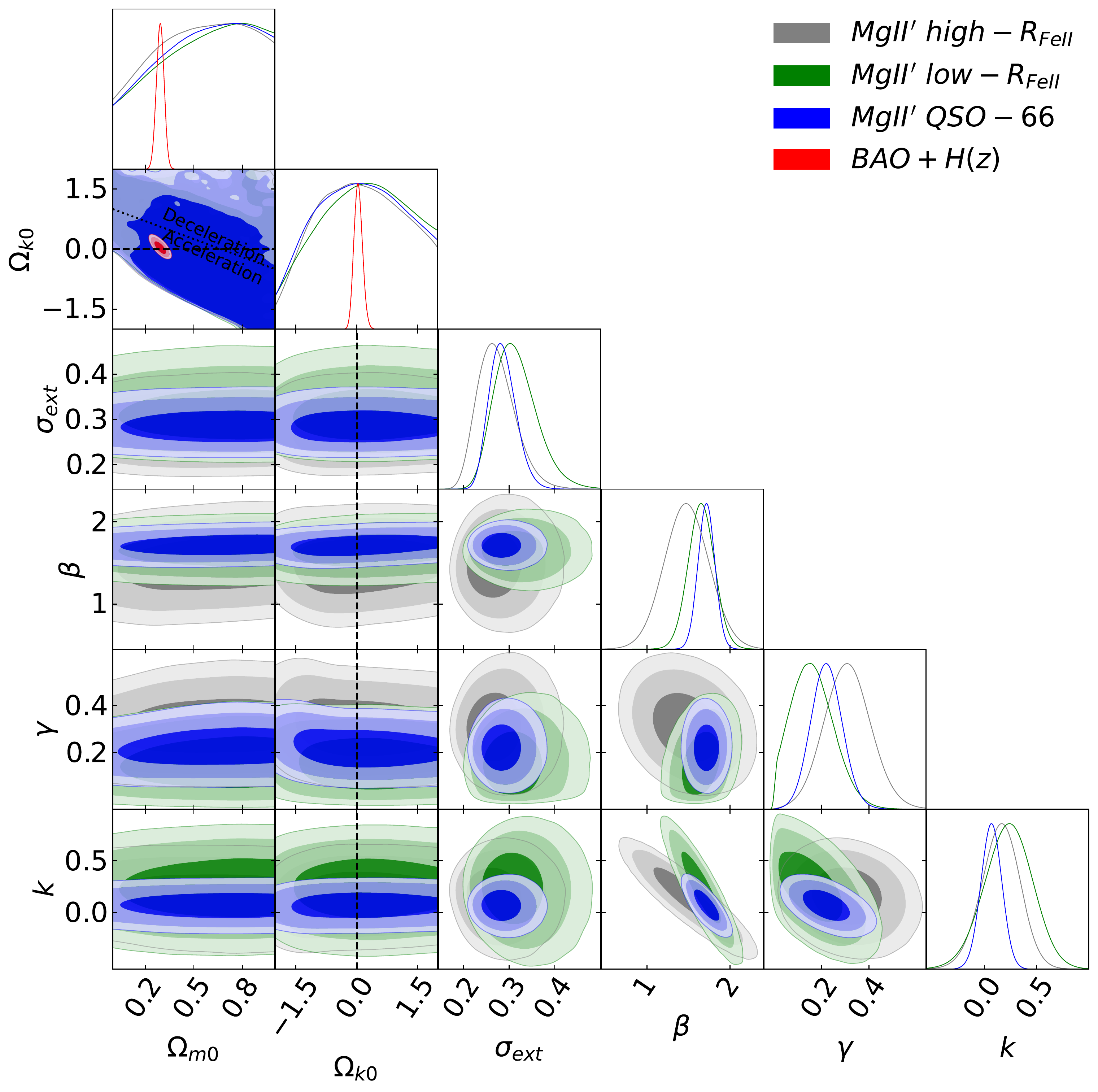}\par
    \includegraphics[width=\linewidth,height=7cm]{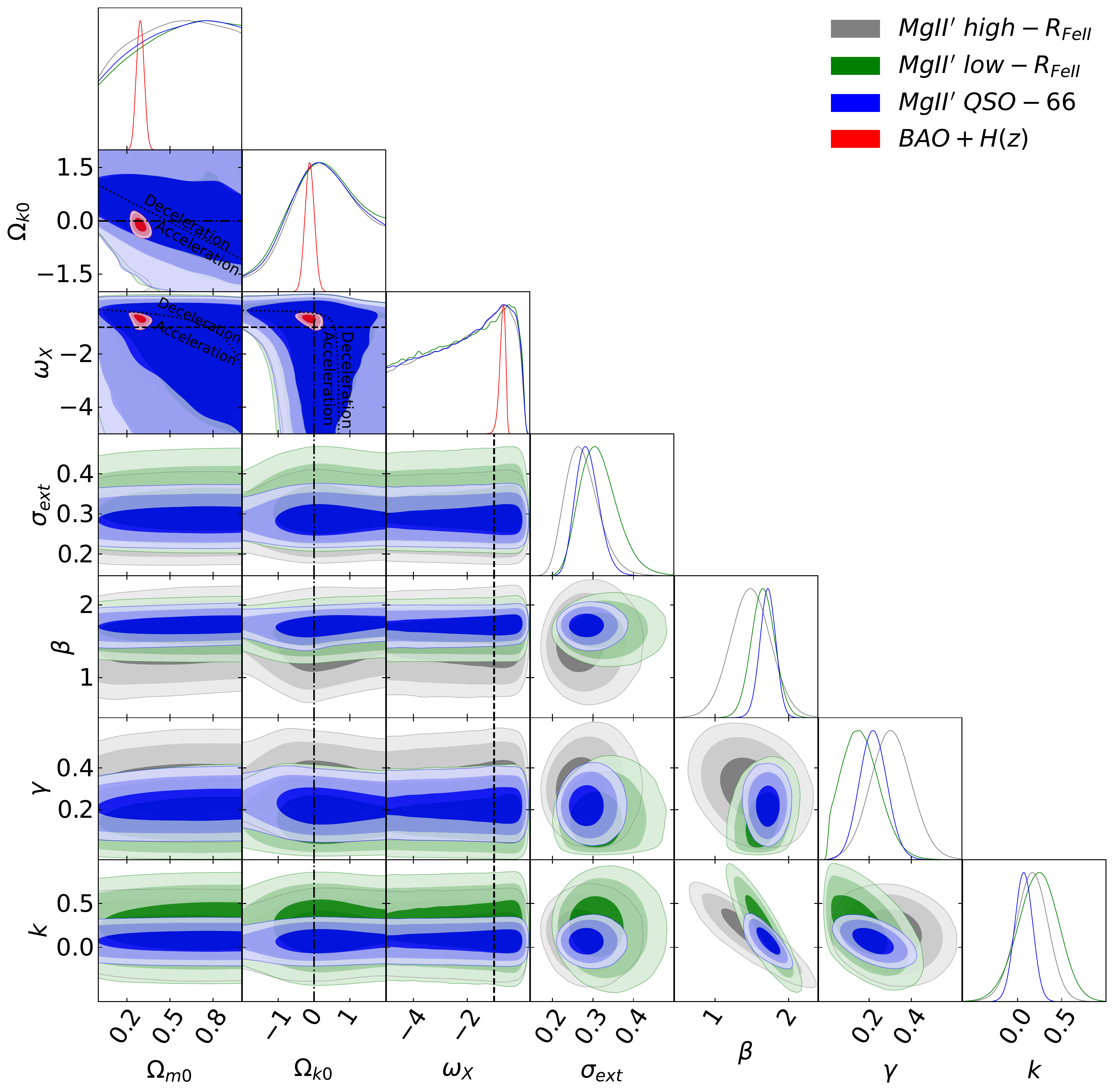}\par
    \includegraphics[width=\linewidth,height=7cm]{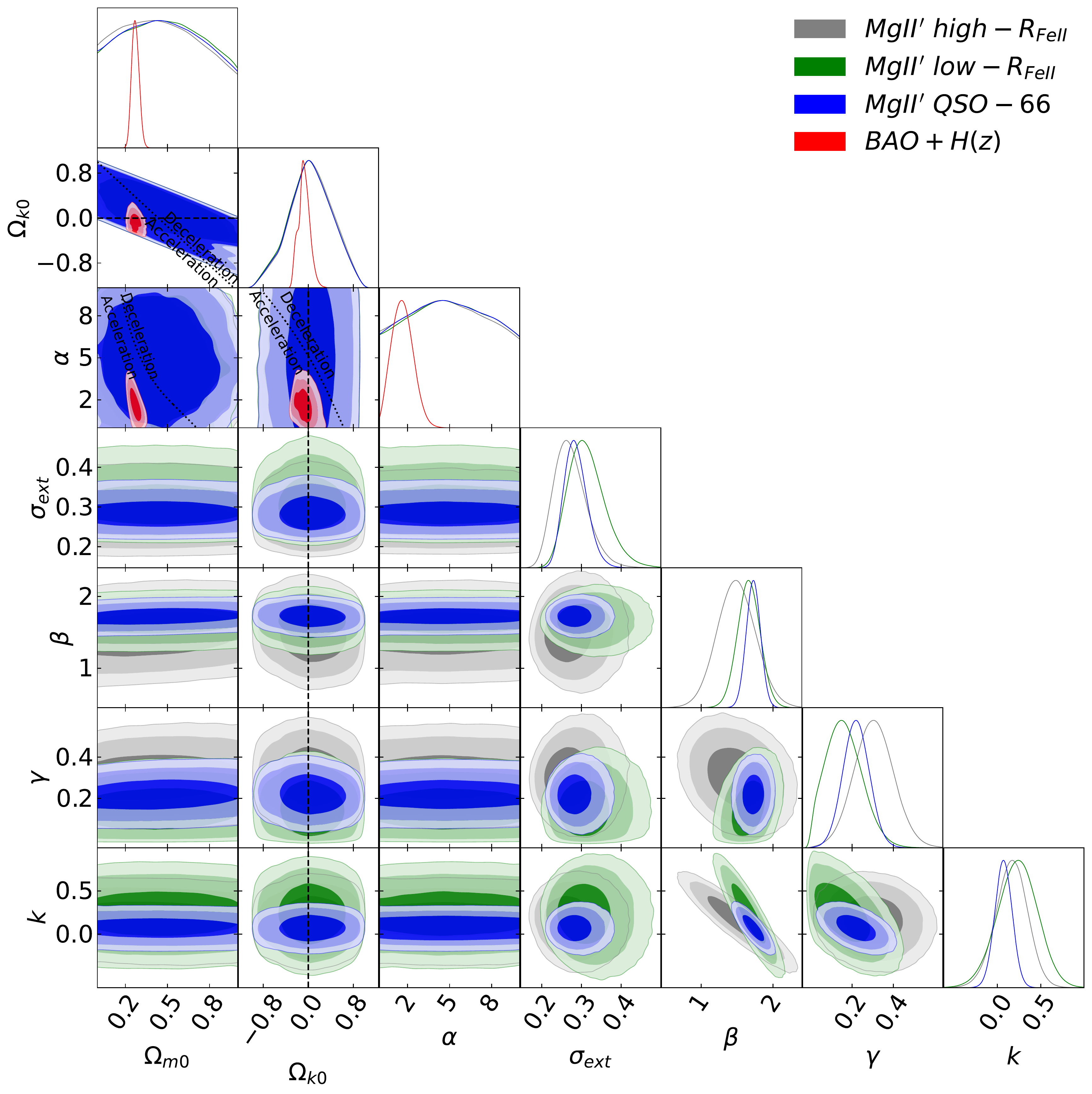}\par
\end{multicols}
\caption{One-dimensional likelihood distributions and two-dimensional likelihood contours at 1$\sigma$, 2$\sigma$, and 3$\sigma$ confidence levels using \Mgii$^{\prime}$ high-\rfe\ (gray), \Mgii$^{\prime}$ low-\rfe\ (green), \Mgii$^{\prime}$ QSO-66 (blue), and BAO + $H(z)$ (red) data for all free parameters. Left column shows the flat $\Lambda$CDM model, flat XCDM parametrization, and flat $\phi$CDM model respectively. The black dotted lines in all plots are the zero acceleration lines. The black dashed lines in the flat XCDM parametrization plots are the $\omega_X=-1$ lines. Right column shows the non-flat $\Lambda$CDM model, non-flat XCDM parametrization, and non-flat $\phi$CDM model respectively. Black dotted lines in all plots are the zero acceleration lines. Black dashed lines in the non-flat $\Lambda$CDM and $\phi$CDM model plots and black dotted-dashed lines in the non-flat XCDM parametrization plots correspond to $\Omega_{k0} = 0$. The black dashed lines in the non-flat XCDM parametrization plots are the $\omega_X=-1$ lines.}
\label{fig:2}
\end{figure*}

\begin{figure*}
\begin{multicols}{2}
    \includegraphics[width=\linewidth,height=7cm]{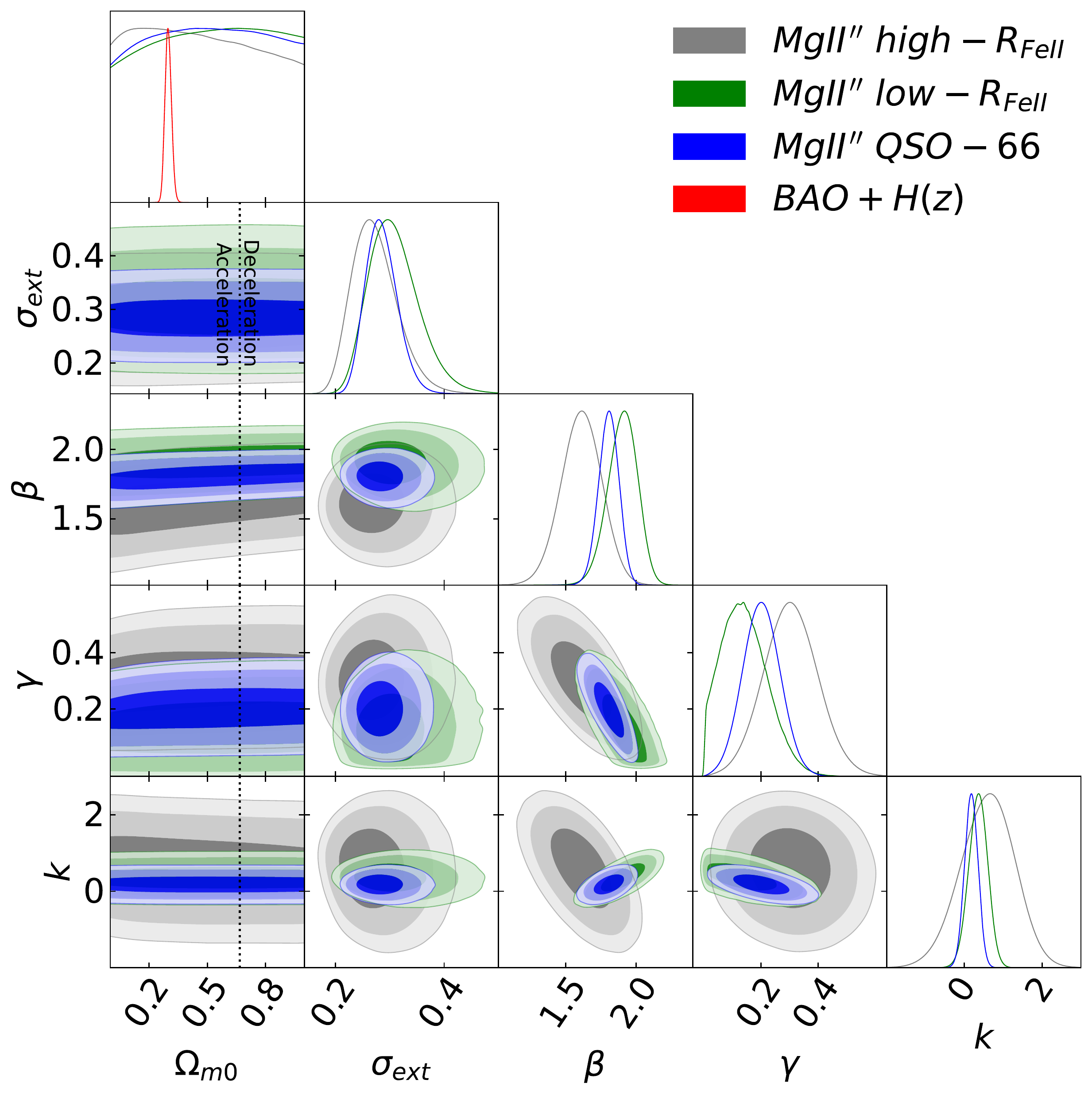}\par
    \includegraphics[width=\linewidth,height=7cm]{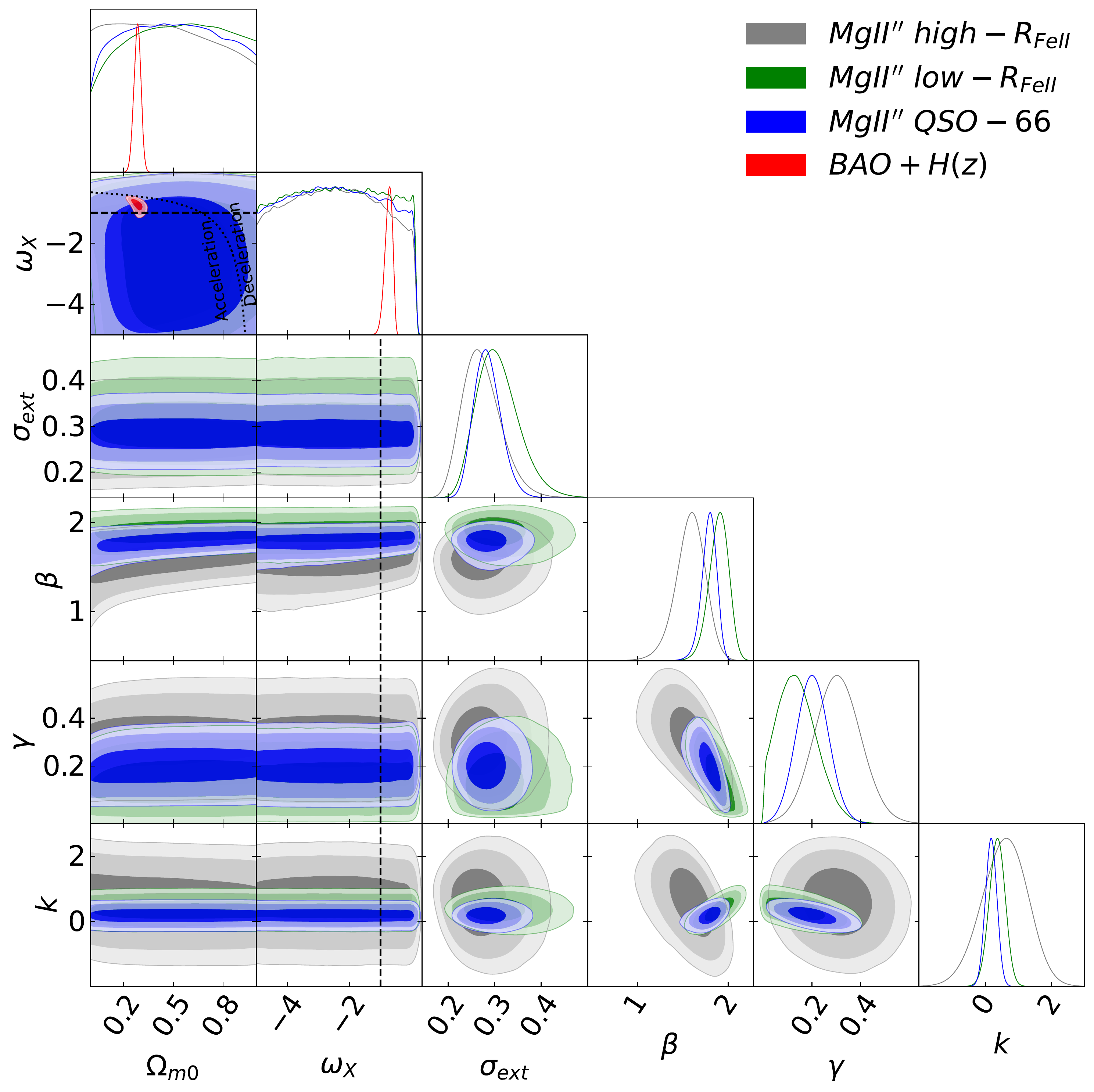}\par
    \includegraphics[width=\linewidth,height=7cm]{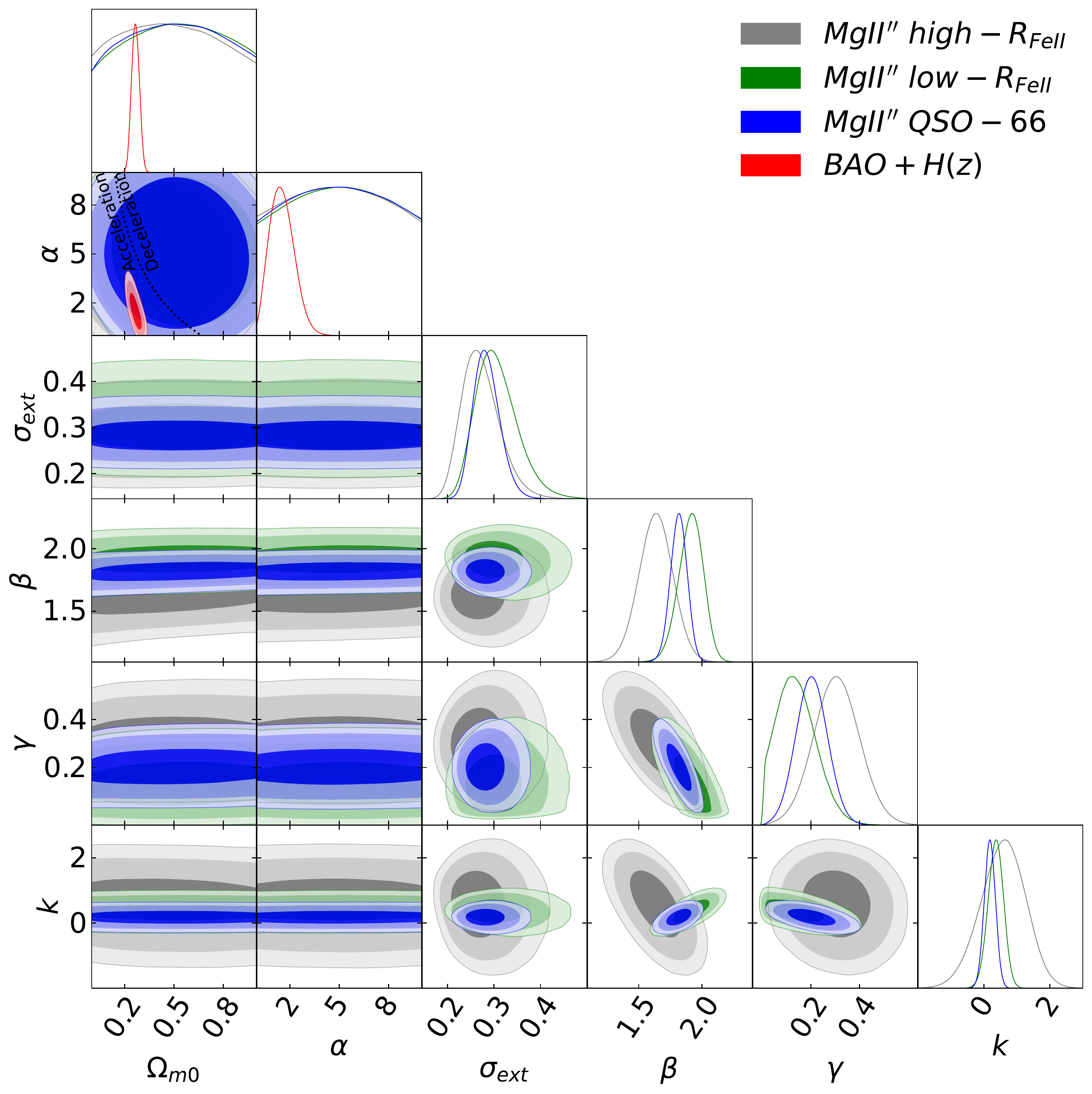}\par
    \includegraphics[width=\linewidth,height=7cm]{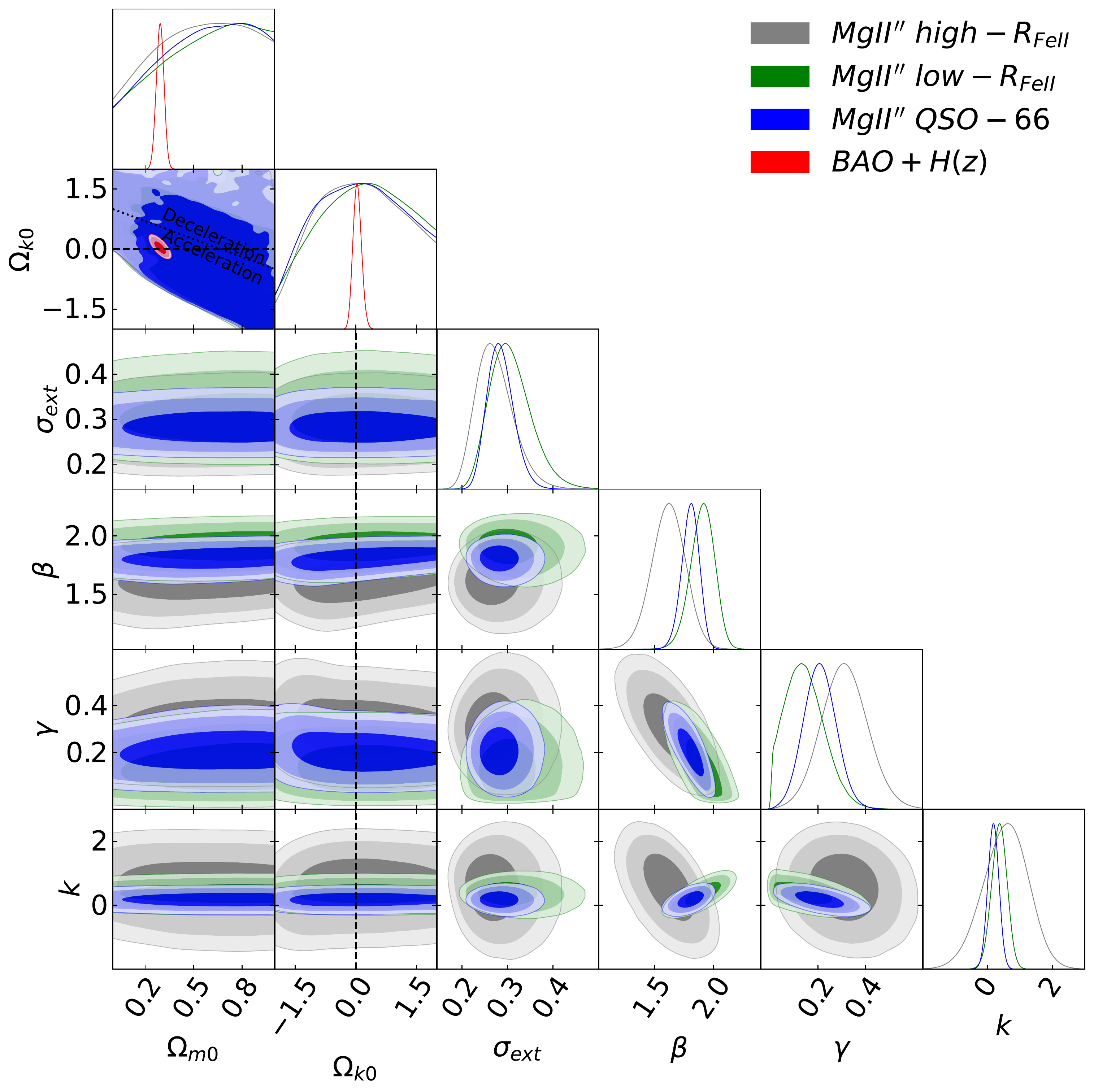}\par
    \includegraphics[width=\linewidth,height=7cm]{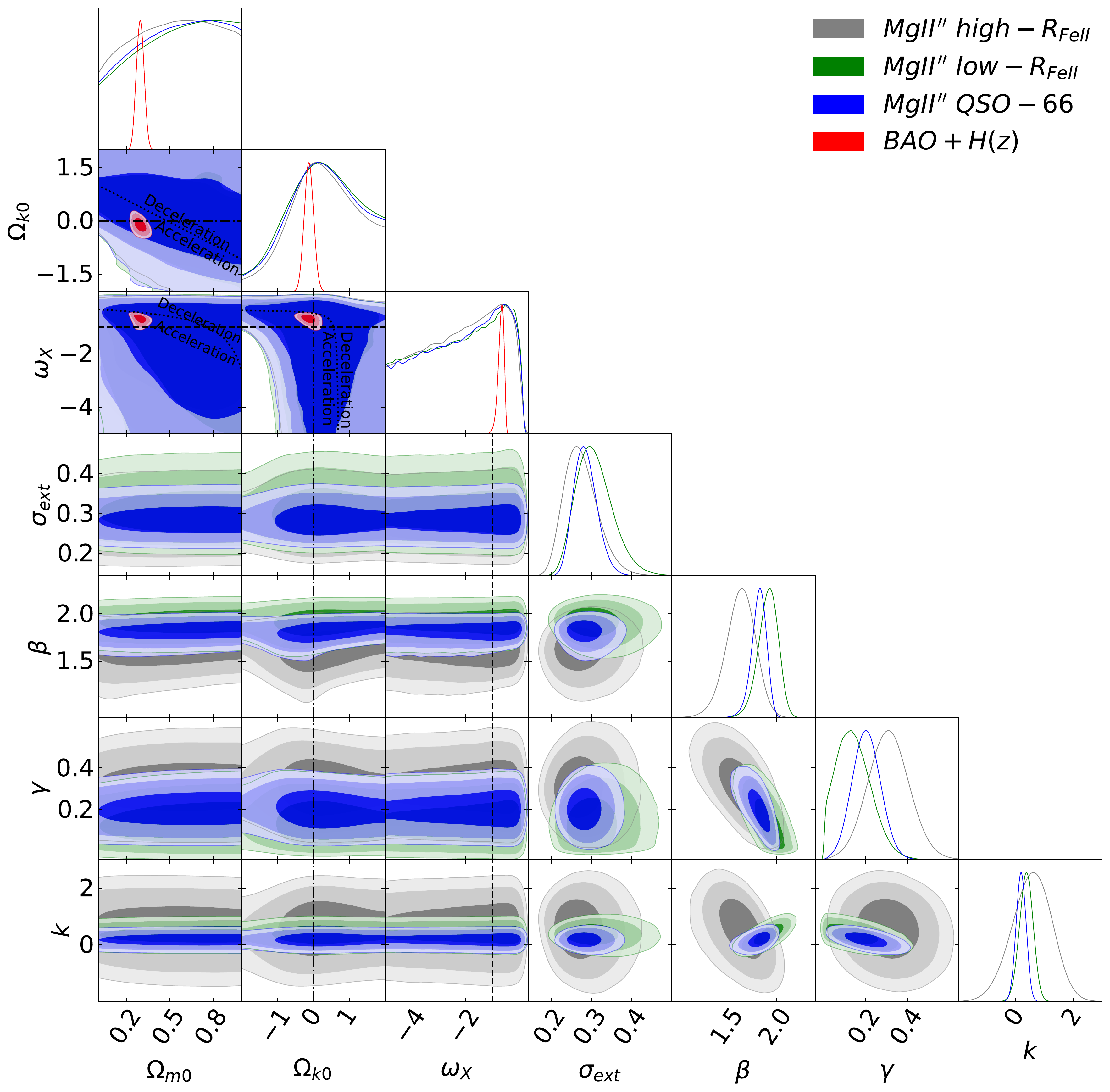}\par
    \includegraphics[width=\linewidth,height=7cm]{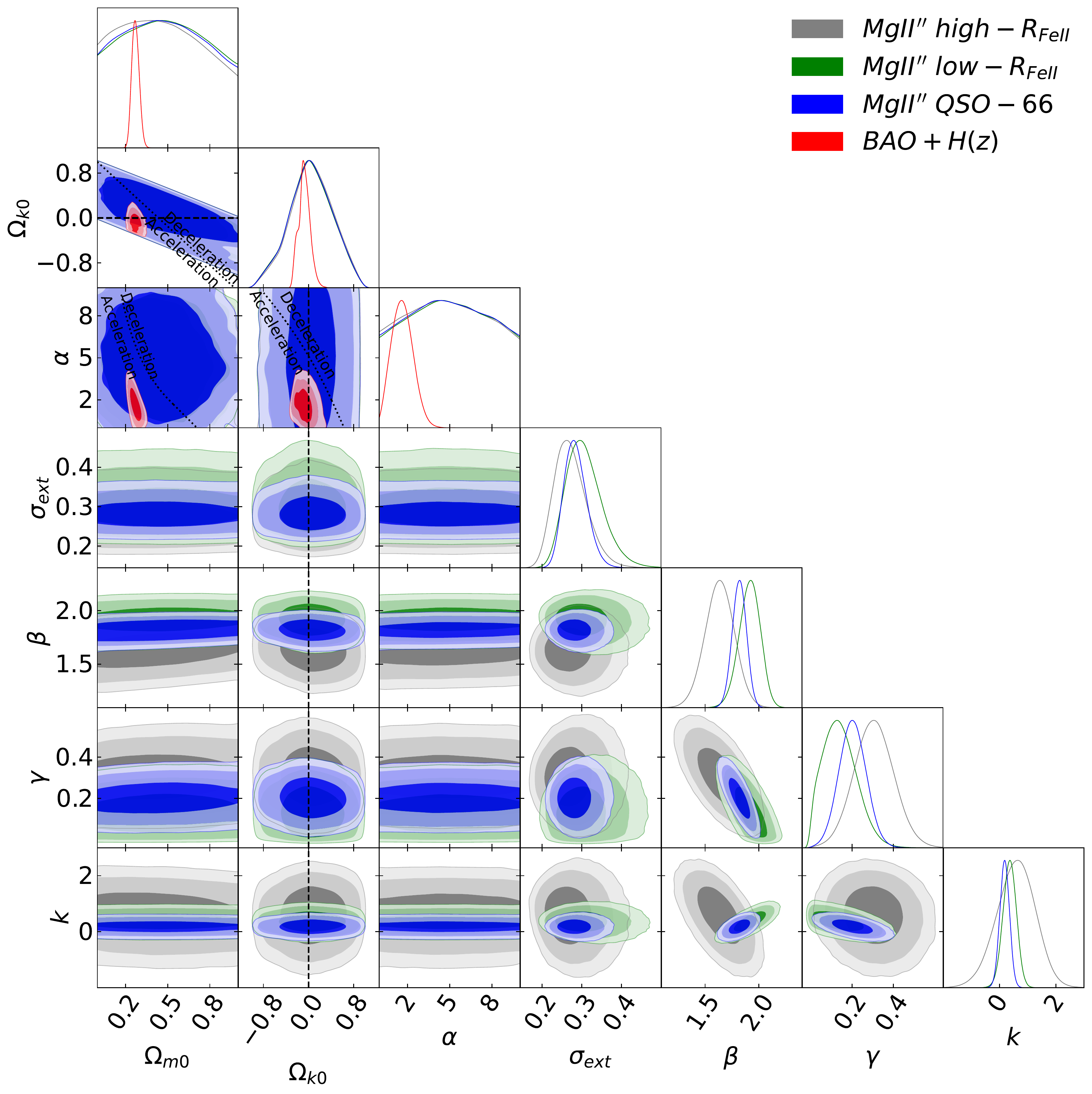}\par
\end{multicols}
\caption{One-dimensional likelihood distributions and two-dimensional likelihood contours at 1$\sigma$, 2$\sigma$, and 3$\sigma$ confidence levels using using \Mgii$^{\prime \prime}$ high-\rfe\ (gray), \Mgii$^{\prime \prime}$ low-\rfe\ (green), \Mgii$^{\prime \prime}$ QSO-66 (blue), and BAO + $H(z)$ (red) data for all free parameters. Left column shows the flat $\Lambda$CDM model, flat XCDM parametrization, and flat $\phi$CDM model respectively. The black dotted lines in all plots are the zero acceleration lines. The black dashed lines in the flat XCDM parametrization plots are the $\omega_X=-1$ lines. Right column shows the non-flat $\Lambda$CDM model, non-flat XCDM parametrization, and non-flat $\phi$CDM model respectively. Black dotted lines in all plots are the zero acceleration lines. Black dashed lines in the non-flat $\Lambda$CDM and $\phi$CDM model plots and black dotted-dashed lines in the non-flat XCDM parametrization plots correspond to $\Omega_{k0} = 0$. The black dashed lines in the non-flat XCDM parametrization plots are the $\omega_X=-1$ lines.}
\label{fig:3}
\end{figure*}

\begin{figure*}
\begin{multicols}{2}
    \includegraphics[width=\linewidth,height=7cm]{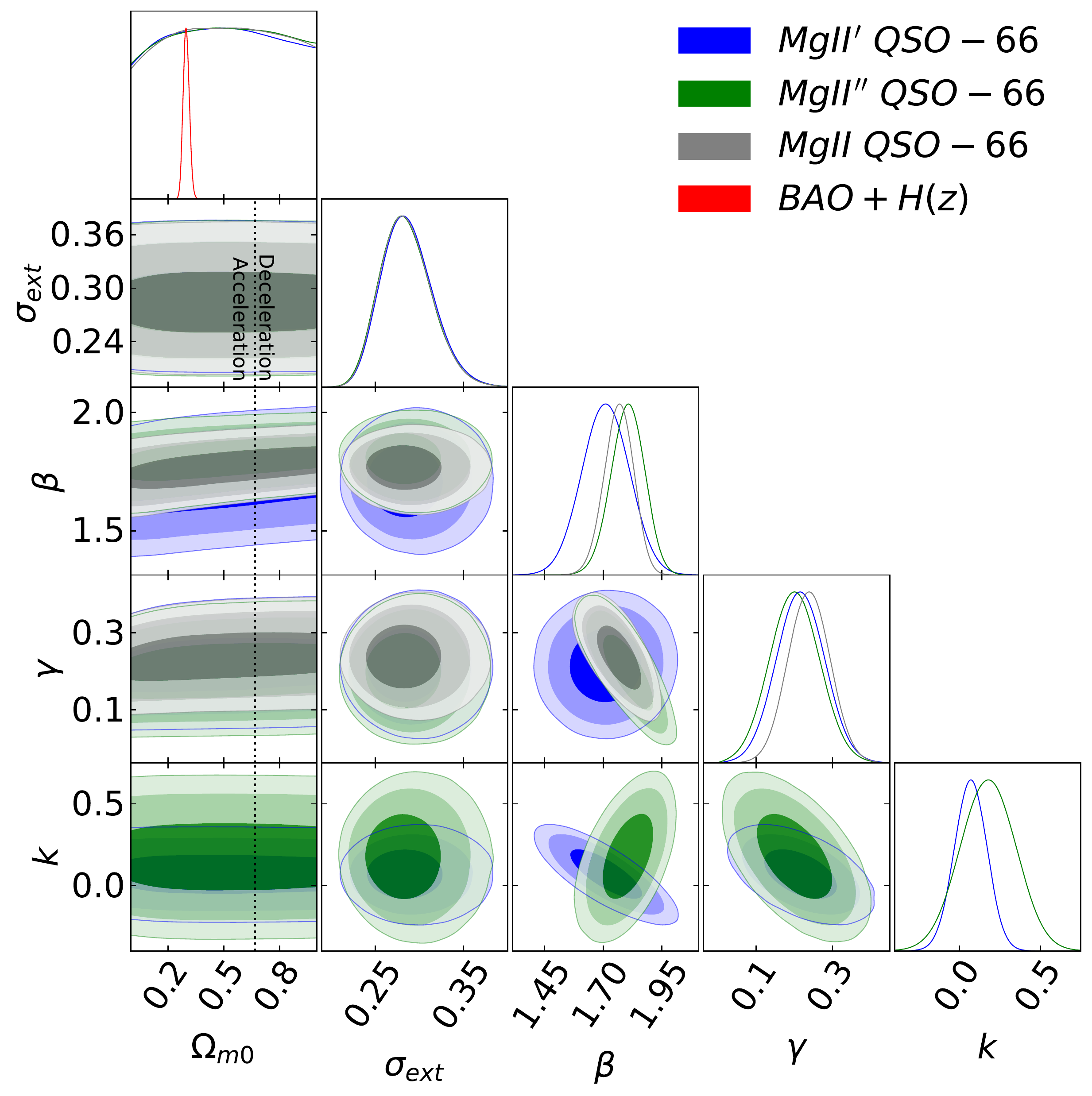}\par
    \includegraphics[width=\linewidth,height=7cm]{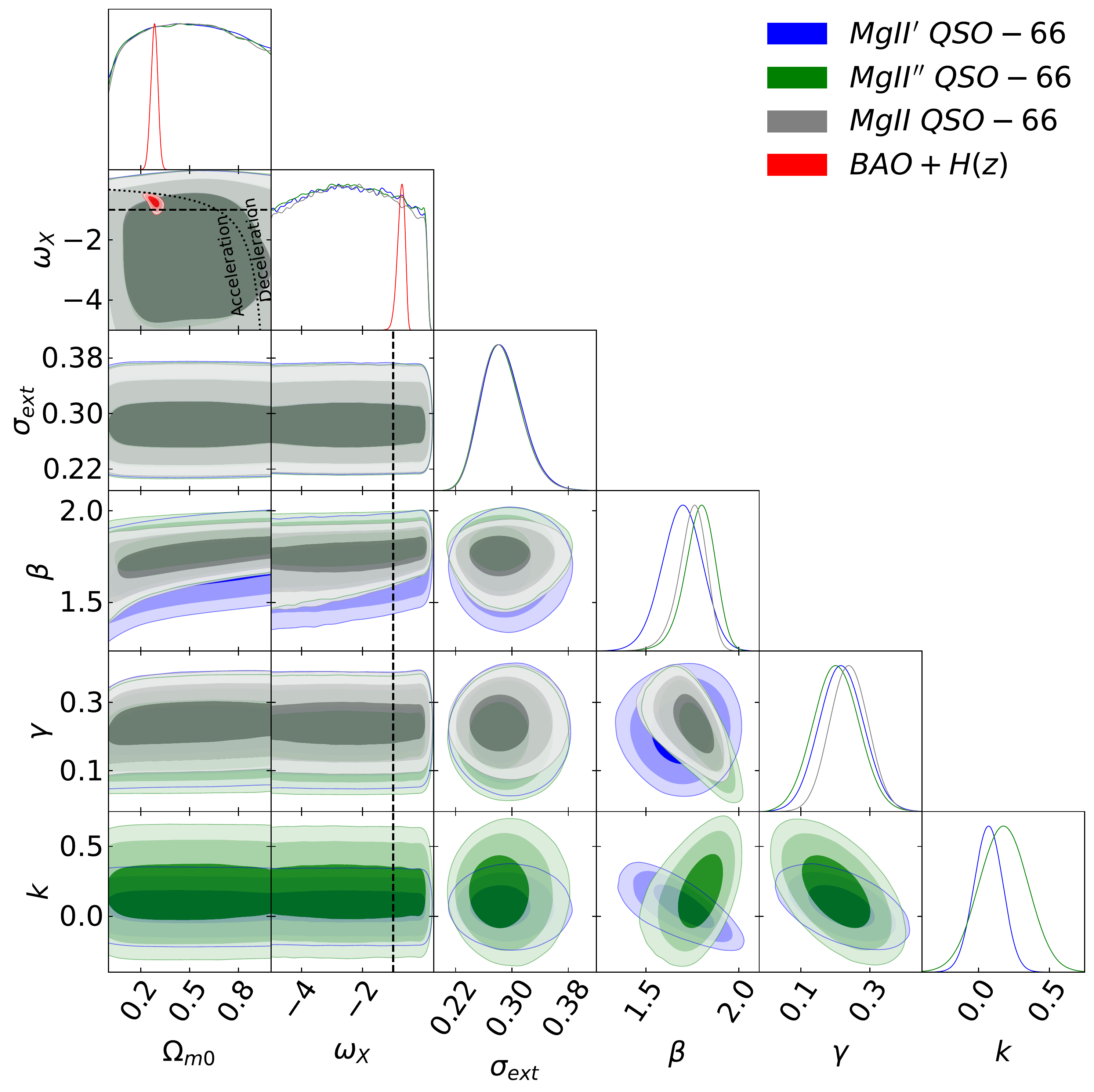}\par
    \includegraphics[width=\linewidth,height=7cm]{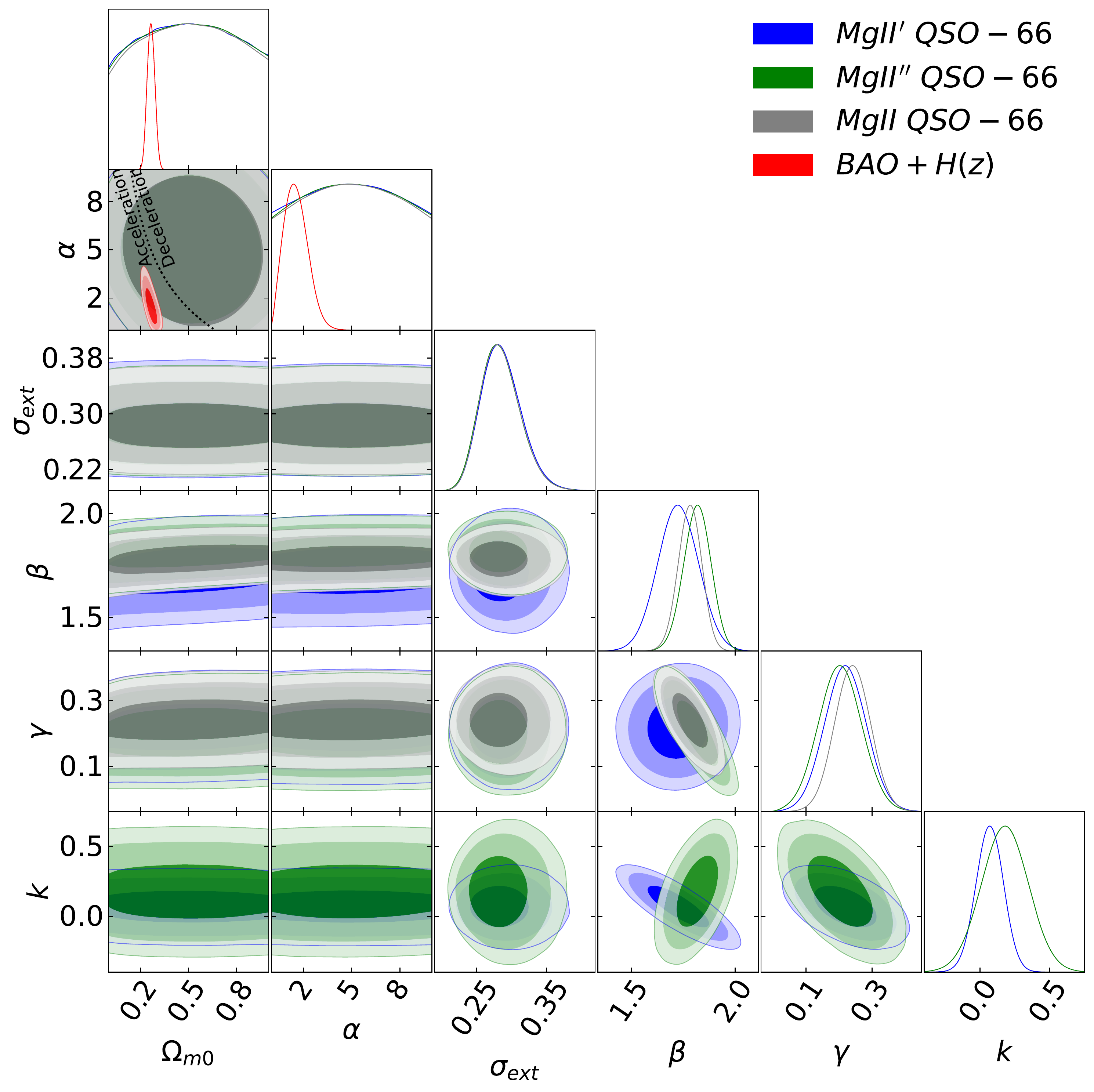}\par
    \includegraphics[width=\linewidth,height=7cm]{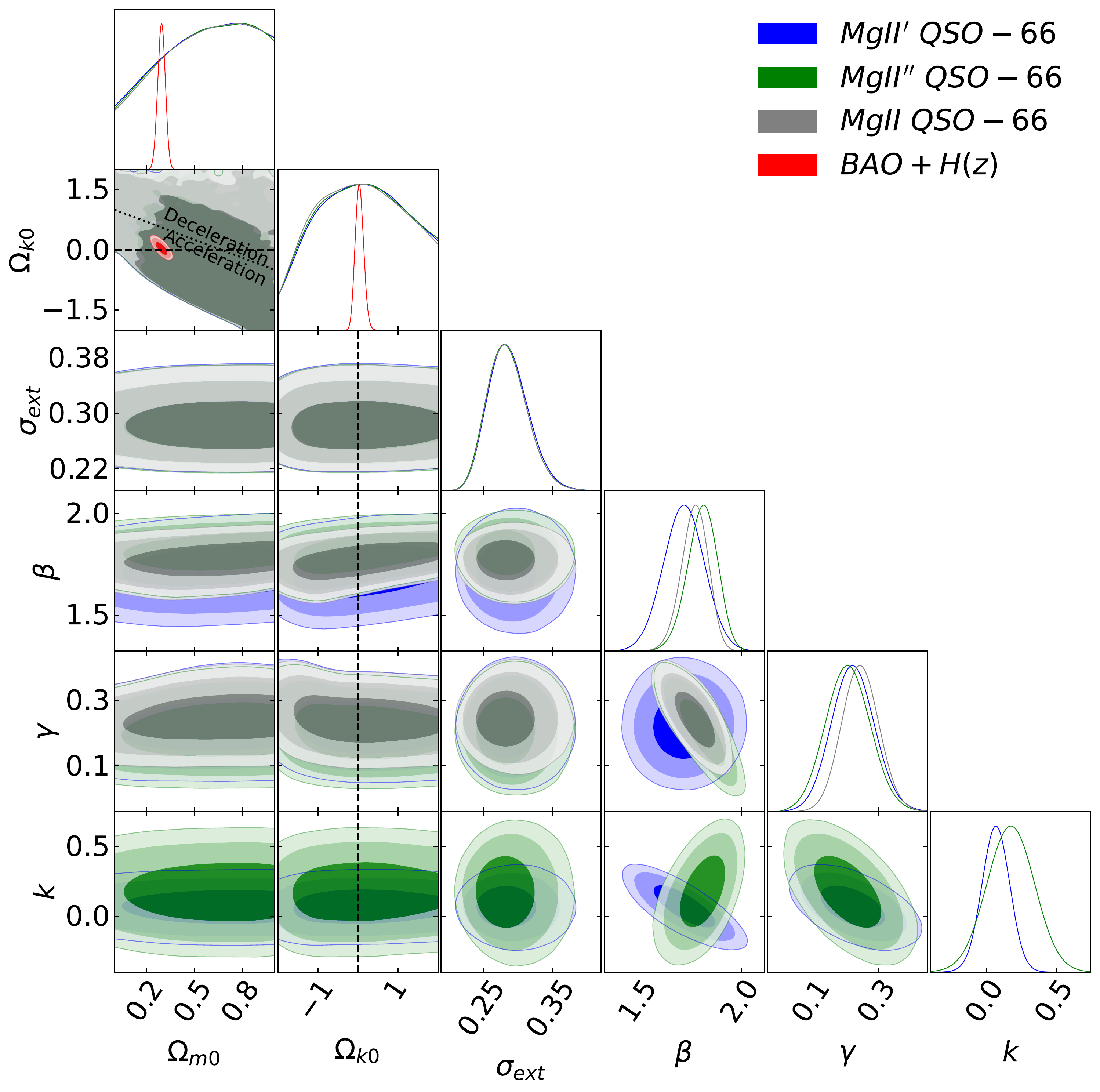}\par
    \includegraphics[width=\linewidth,height=7cm]{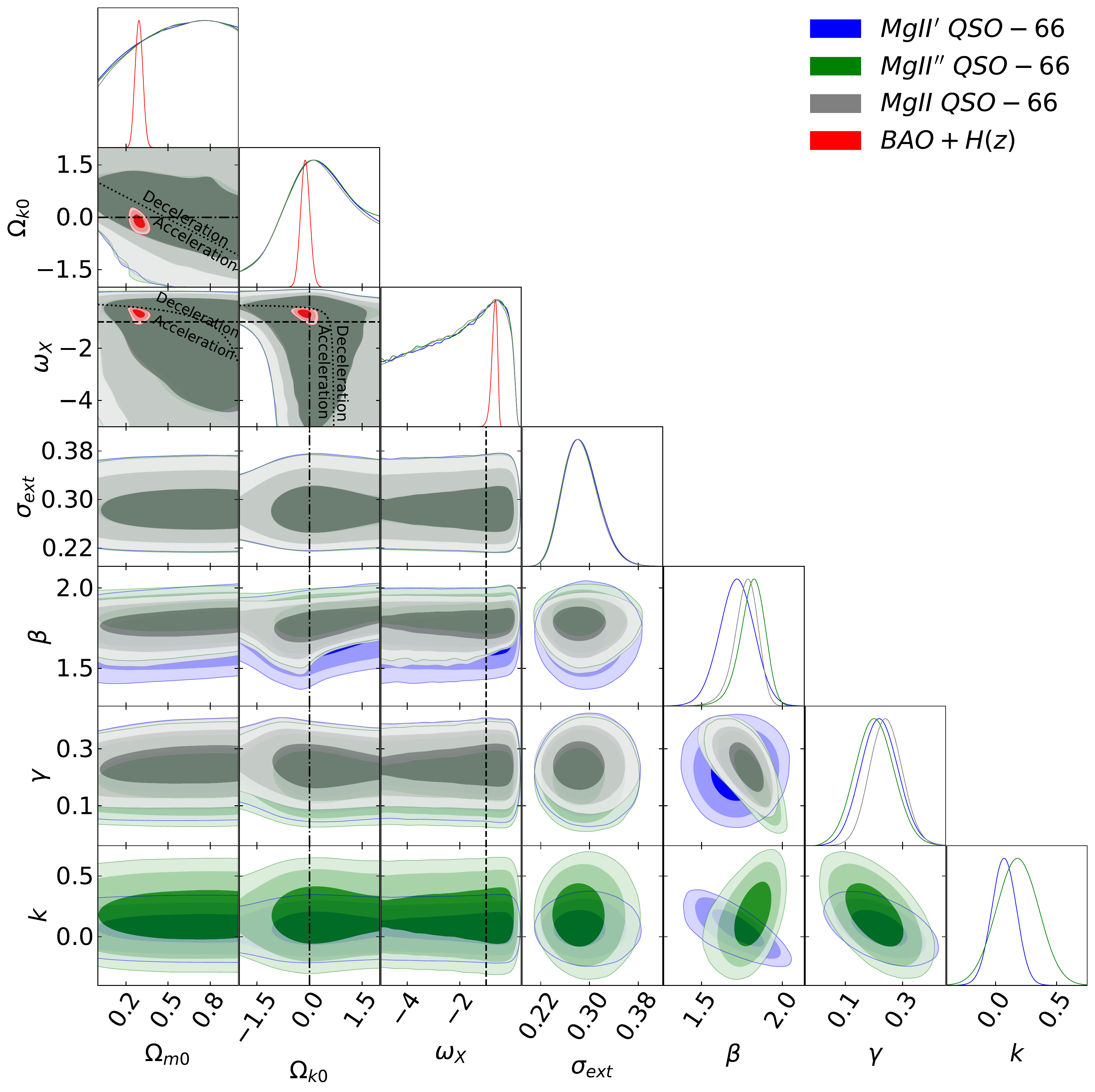}\par
    \includegraphics[width=\linewidth,height=7cm]{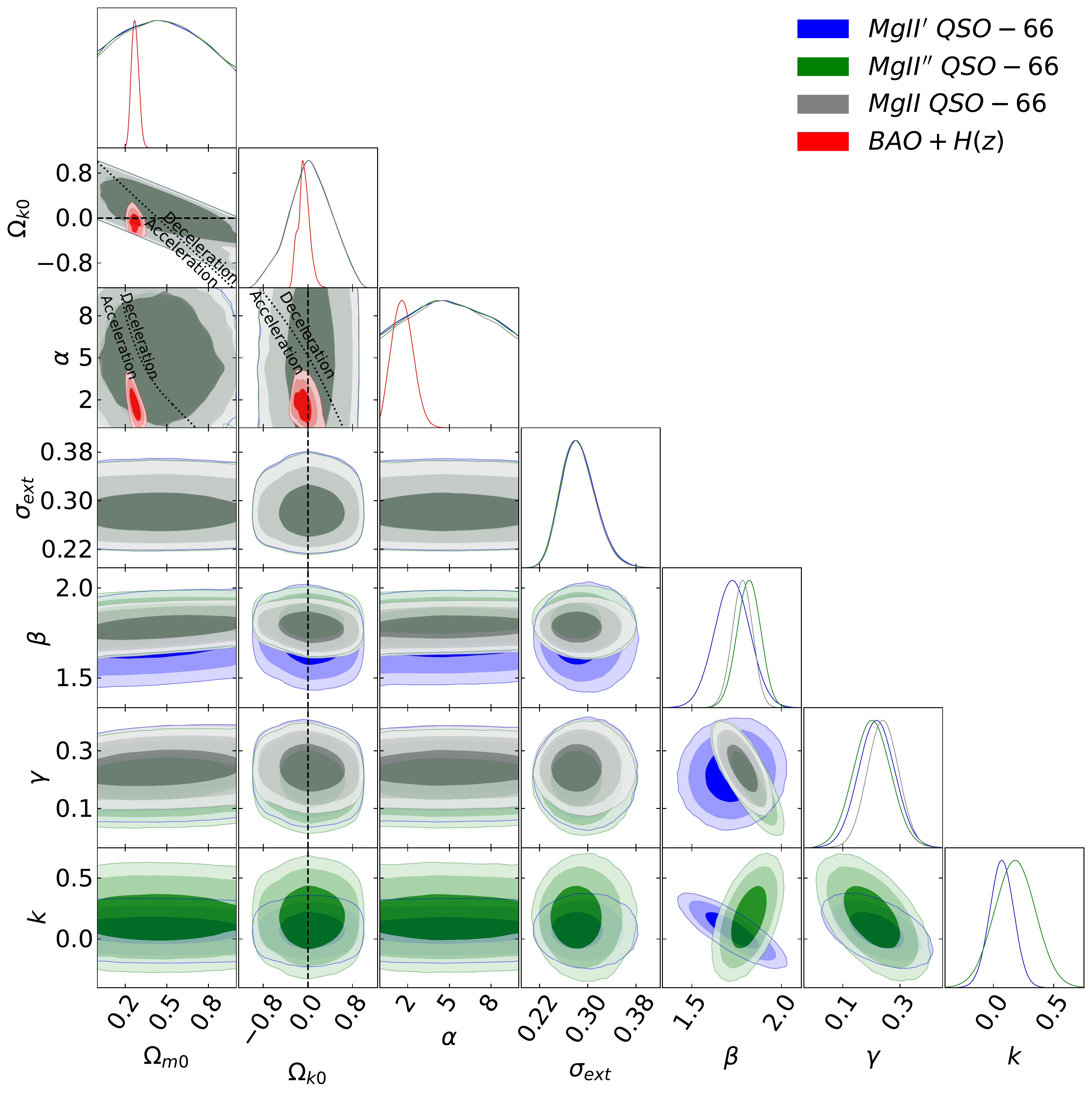}\par
\end{multicols}
\caption{One-dimensional likelihood distributions and two-dimensional likelihood contours at 1$\sigma$, 2$\sigma$, and 3$\sigma$ confidence levels using using \Mgii$^{\prime}$ QSO-66 (blue), \Mgii$^{\prime \prime}$ QSO-66 (green), \Mgii\ QSO-66 (gray), and BAO + $H(z)$ (red) data for all free parameters. Left column shows the flat $\Lambda$CDM model, flat XCDM parametrization, and flat $\phi$CDM model respectively. The black dotted lines in all plots are the zero acceleration lines. The black dashed lines in the flat XCDM parametrization plots are the $\omega_X=-1$ lines. Right column shows the non-flat $\Lambda$CDM model, non-flat XCDM parametrization, and non-flat $\phi$CDM model respectively. Black dotted lines in all plots are the zero acceleration lines. Black dashed lines in the non-flat $\Lambda$CDM and $\phi$CDM model plots and black dotted-dashed lines in the non-flat XCDM parametrization plots correspond to $\Omega_{k0} = 0$. The black dashed lines in the non-flat XCDM parametrization plots are the $\omega_X=-1$ lines.}
\label{fig:4}
\end{figure*}

\begin{figure*}
\begin{multicols}{2}
    \includegraphics[width=\linewidth,height=7cm]{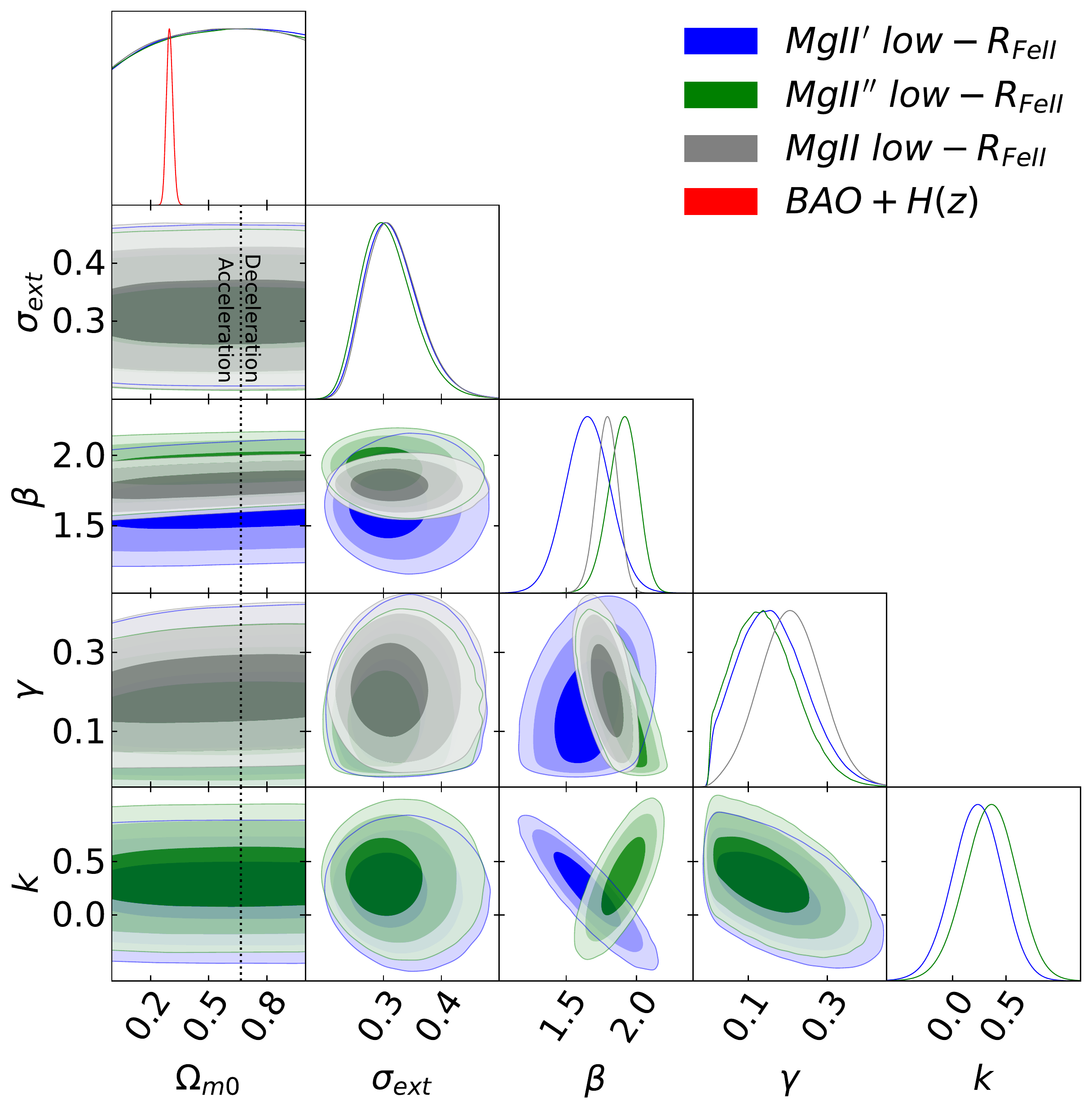}\par
    \includegraphics[width=\linewidth,height=7cm]{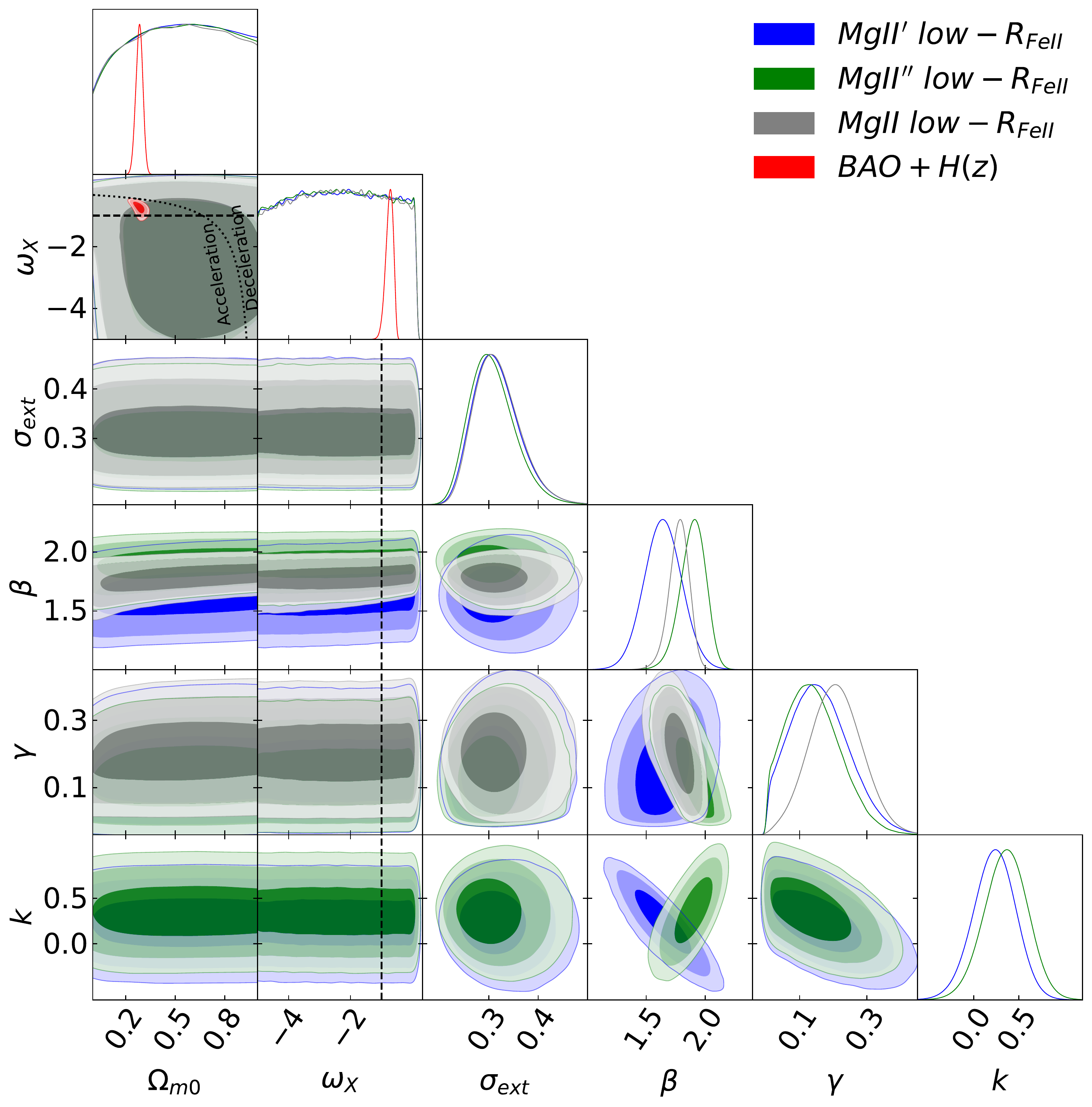}\par
    \includegraphics[width=\linewidth,height=7cm]{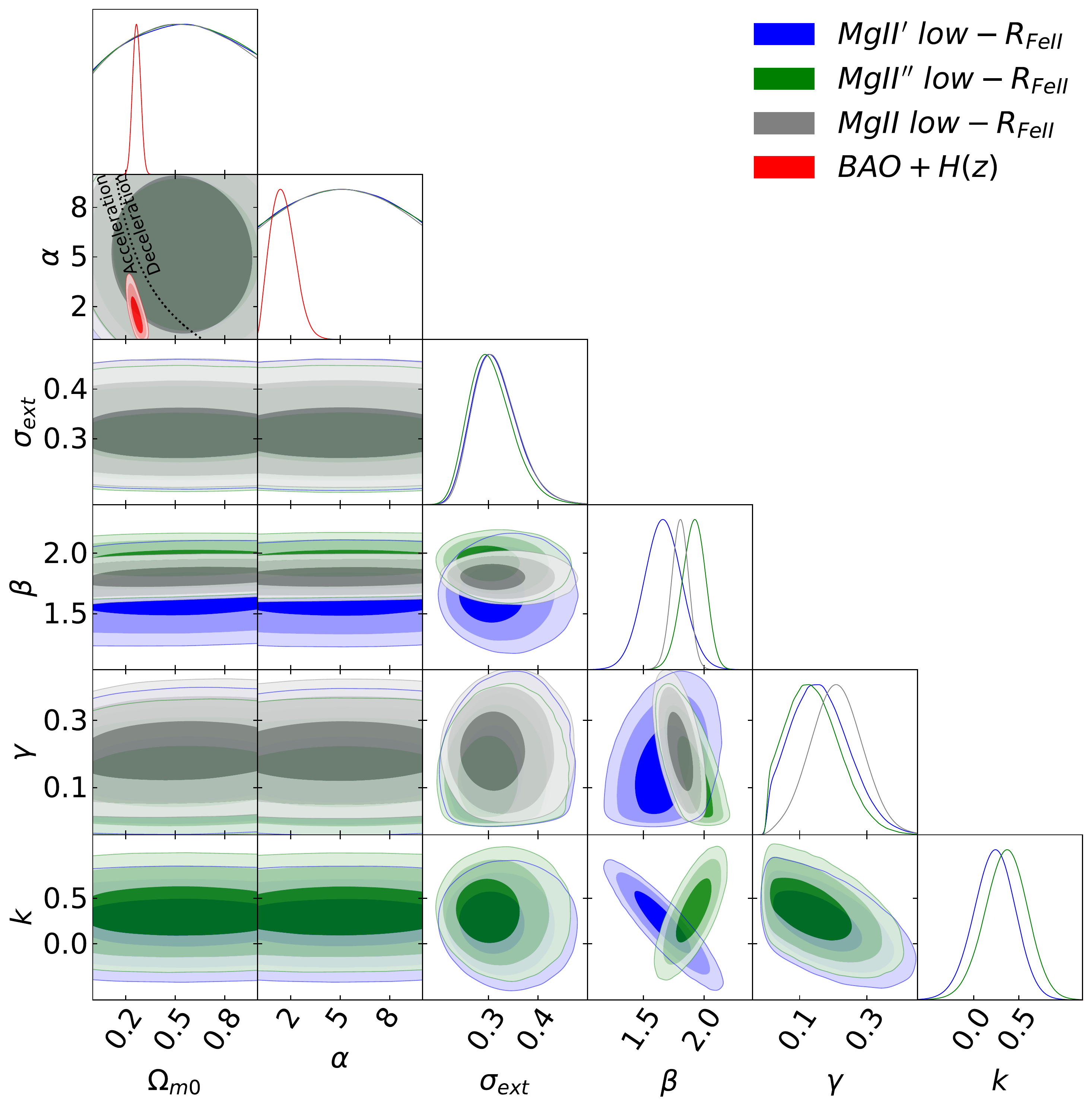}\par
    \includegraphics[width=\linewidth,height=7cm]{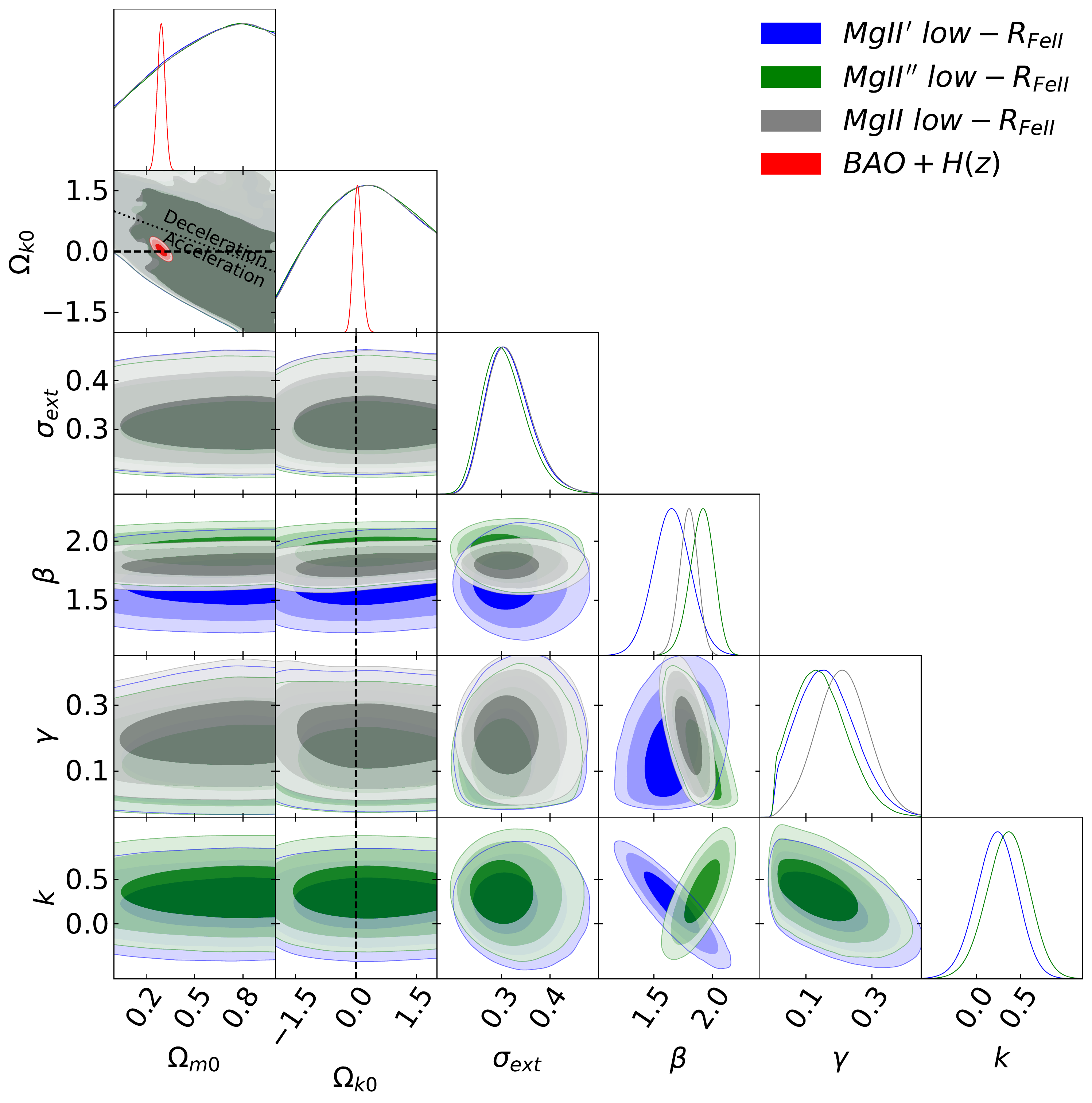}\par
    \includegraphics[width=\linewidth,height=7cm]{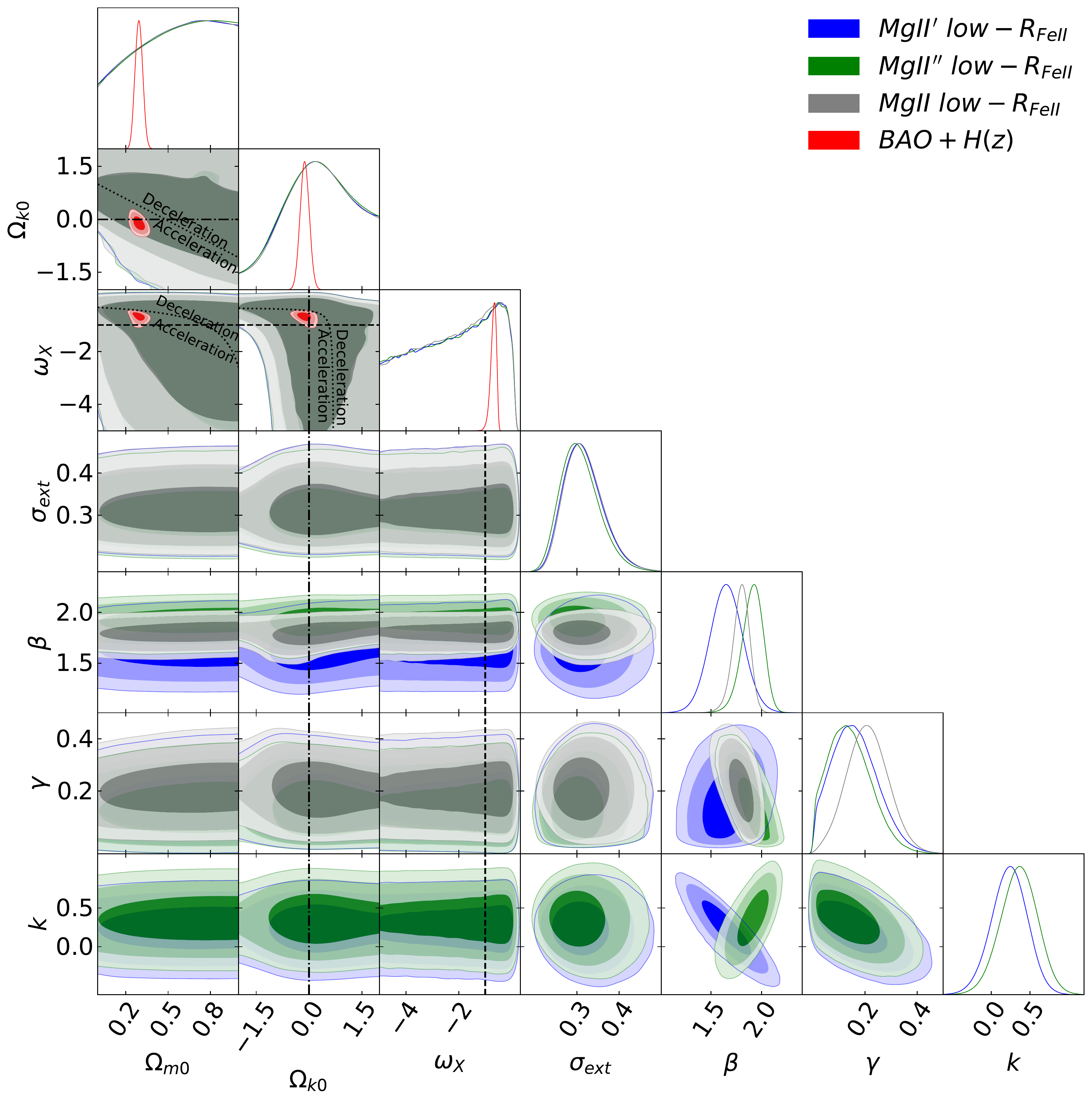}\par
    \includegraphics[width=\linewidth,height=7cm]{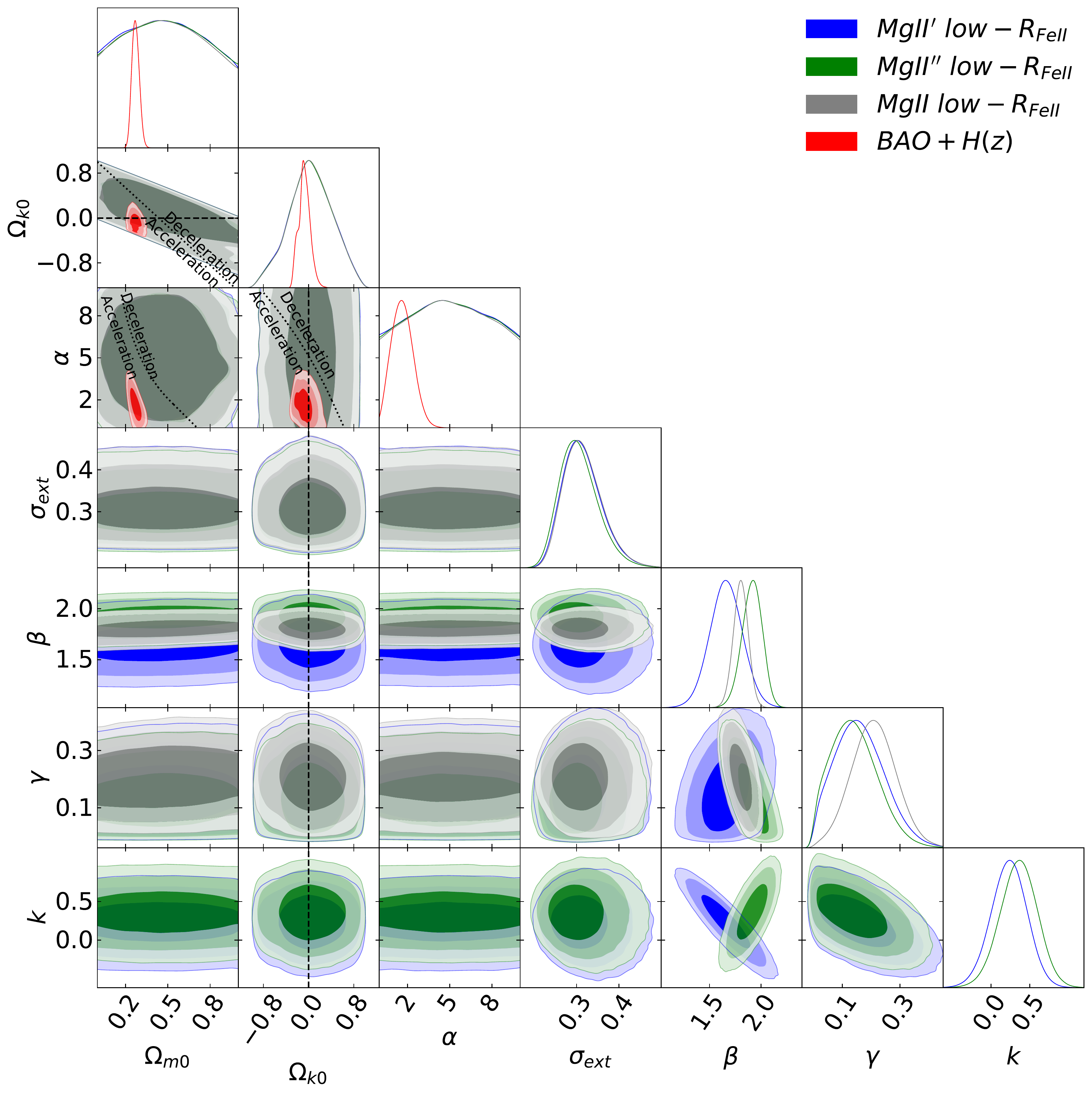}\par
\end{multicols}
\caption{One-dimensional likelihood distributions and two-dimensional likelihood contours at 1$\sigma$, 2$\sigma$, and 3$\sigma$ confidence levels using using \Mgii$^{\prime}$ low-\rfe\ (blue), \Mgii$^{\prime \prime}$ low-\rfe\ (green),  \Mgii\ low-\rfe\ (gray), and BAO + $H(z)$ (red) data for all free parameters. Left column shows the flat $\Lambda$CDM model, flat XCDM parametrization, and flat $\phi$CDM model respectively. The black dotted lines in all plots are the zero acceleration lines. The black dashed lines in the flat XCDM parametrization plots are the $\omega_X=-1$ lines. Right column shows the non-flat $\Lambda$CDM model, non-flat XCDM parametrization, and non-flat $\phi$CDM model respectively. Black dotted lines in all plots are the zero acceleration lines. Black dashed lines in the non-flat $\Lambda$CDM and $\phi$CDM model plots and black dotted-dashed lines in the non-flat XCDM parametrization plots correspond to $\Omega_{k0} = 0$. The black dashed lines in the non-flat XCDM parametrization plots are the $\omega_X=-1$ lines.}
\label{fig:5}
\end{figure*}

\begin{figure*}
\begin{multicols}{2}
    \includegraphics[width=\linewidth,height=7cm]{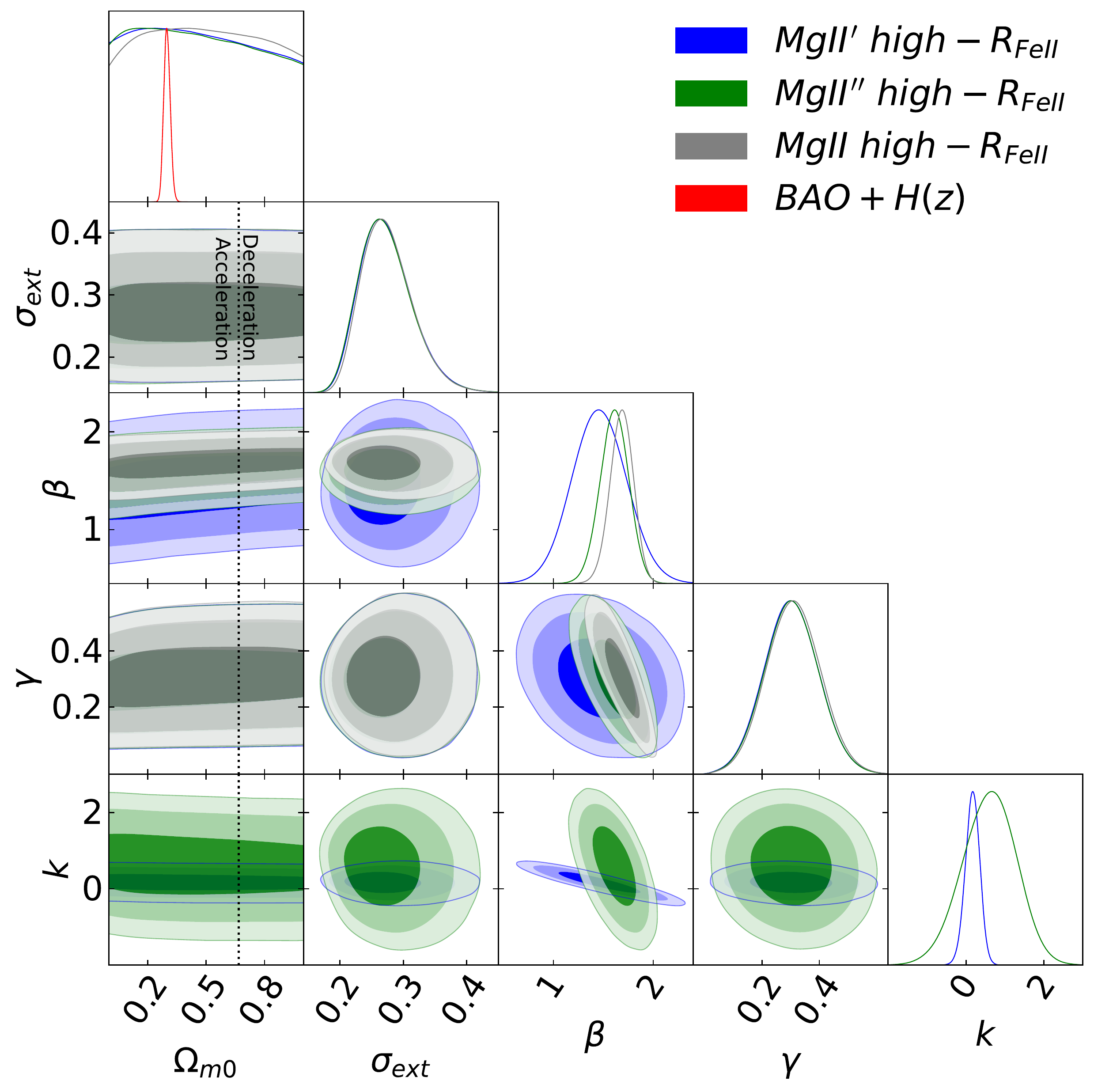}\par
    \includegraphics[width=\linewidth,height=7cm]{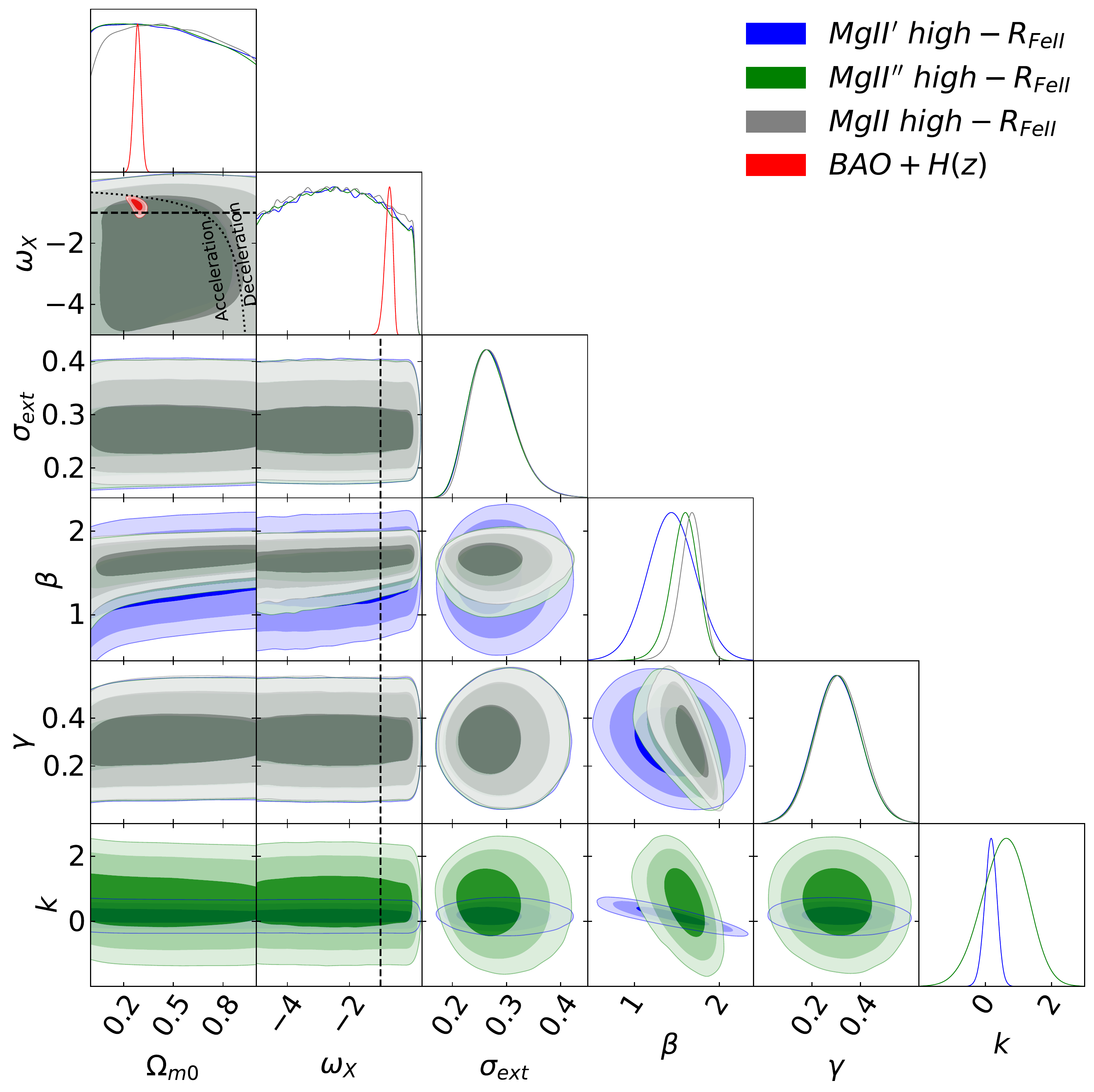}\par
    \includegraphics[width=\linewidth,height=7cm]{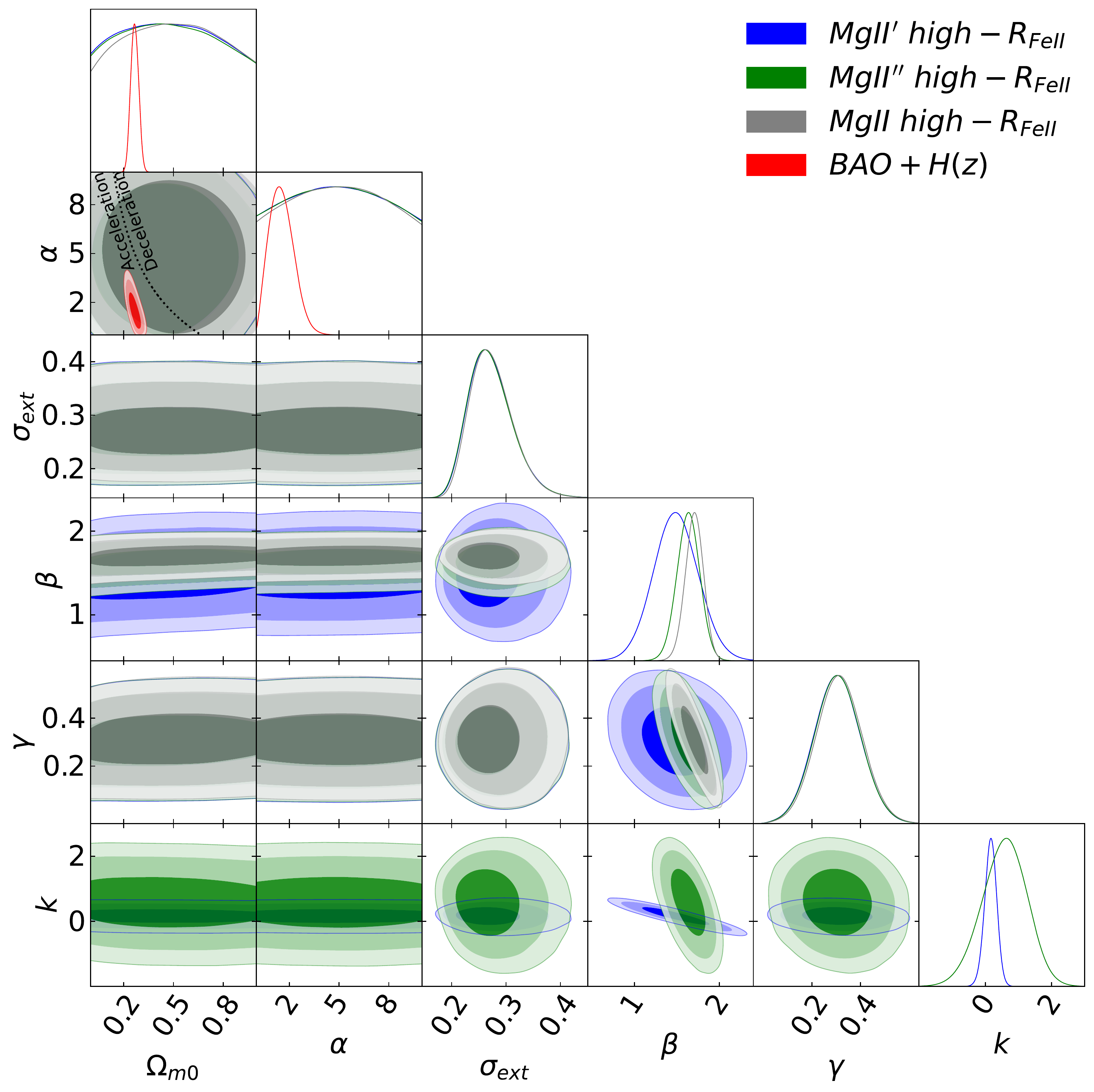}\par
    \includegraphics[width=\linewidth,height=7cm]{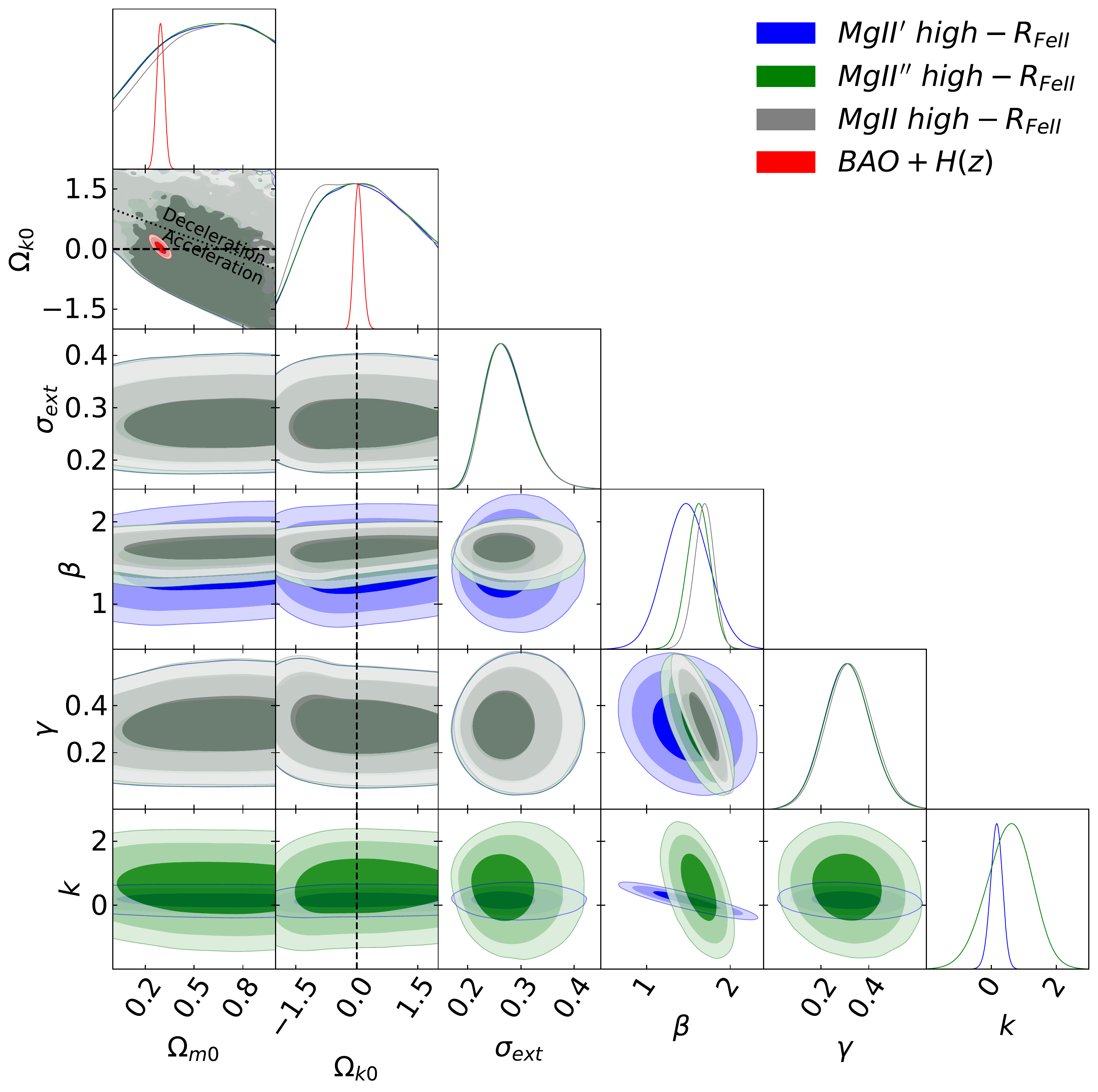}\par
    \includegraphics[width=\linewidth,height=7cm]{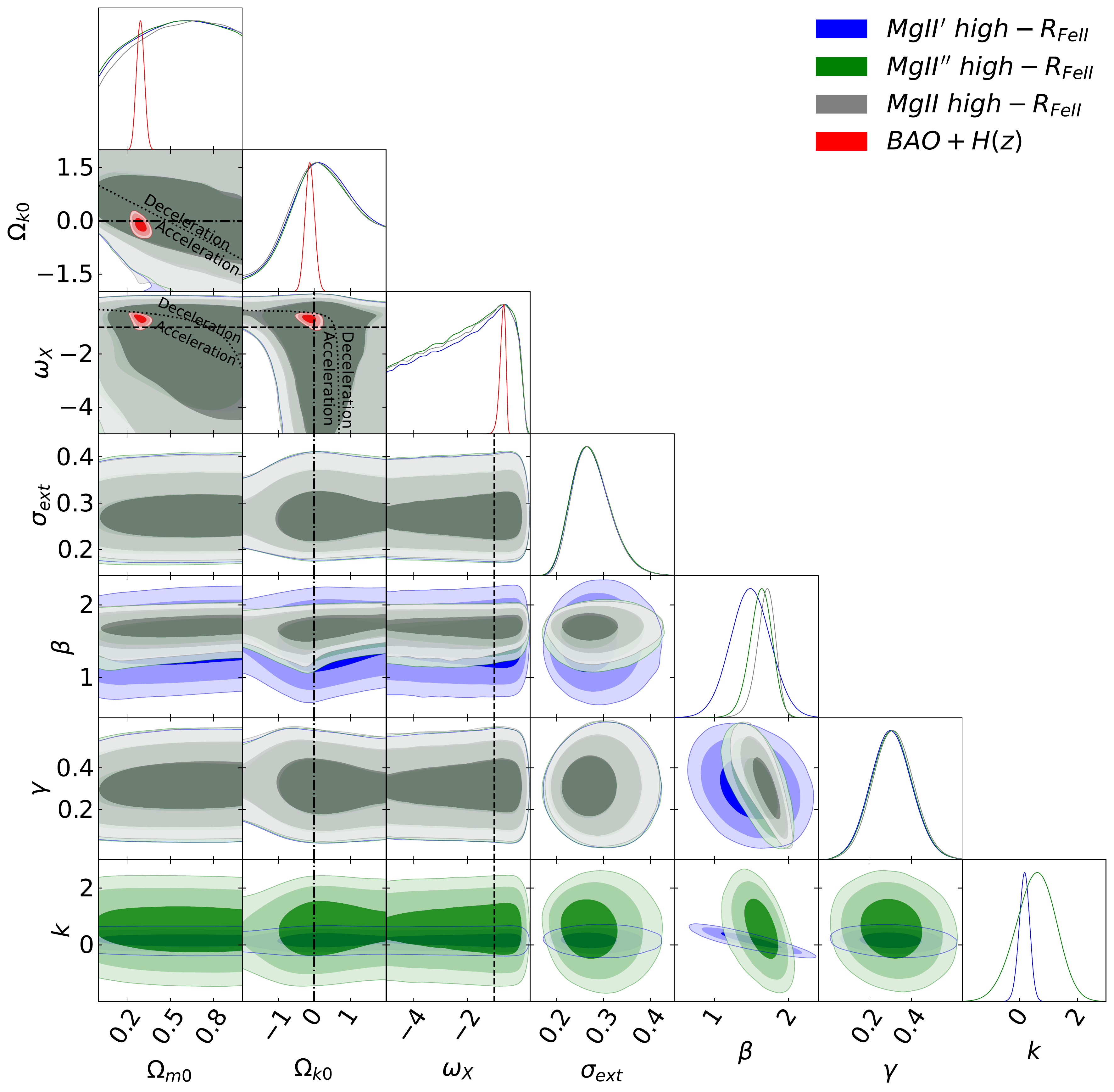}\par
    \includegraphics[width=\linewidth,height=7cm]{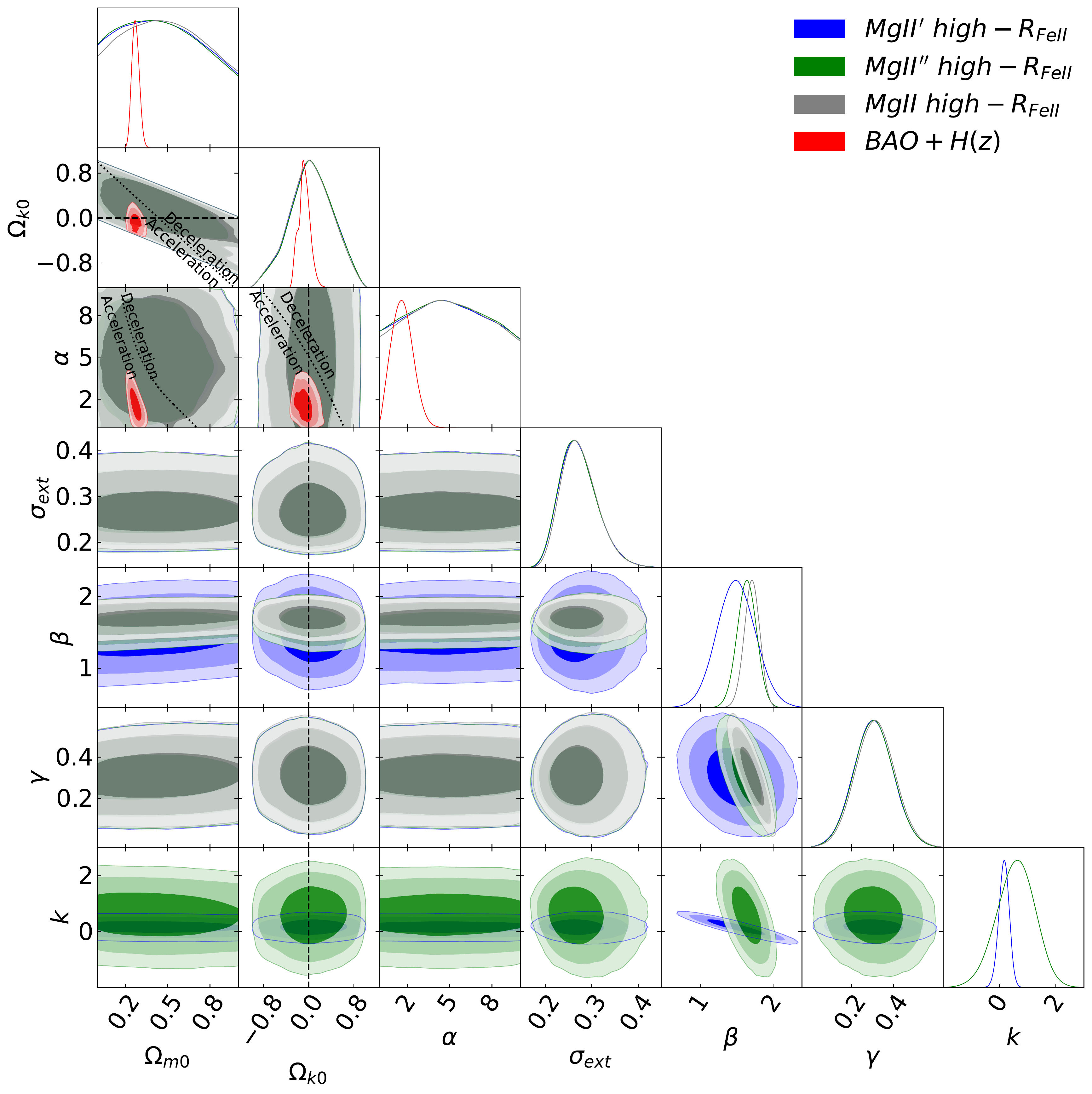}\par
\end{multicols}
\caption{One-dimensional likelihood distributions and two-dimensional likelihood contours at 1$\sigma$, 2$\sigma$, and 3$\sigma$ confidence levels using using \Mgii$^{\prime}$ high-\rfe\ (blue), \Mgii$^{\prime \prime}$ high-\rfe\ (green),  \Mgii\ high-\rfe\ (gray), and BAO + $H(z)$ (red) data for all free parameters. Left column shows the flat $\Lambda$CDM model, flat XCDM parametrization, and flat $\phi$CDM model respectively. The black dotted lines in all plots are the zero acceleration lines. The black dashed lines in the flat XCDM parametrization plots are the $\omega_X=-1$ lines. Right column shows the non-flat $\Lambda$CDM model, non-flat XCDM parametrization, and non-flat $\phi$CDM model respectively. Black dotted lines in all plots are the zero acceleration lines. Black dashed lines in the non-flat $\Lambda$CDM and $\phi$CDM model plots and black dotted-dashed lines in the non-flat XCDM parametrization plots correspond to $\Omega_{k0} = 0$. The black dashed lines in the non-flat XCDM parametrization plots are the $\omega_X=-1$ lines.}
\label{fig:6}
\end{figure*}

Results from the analyses of the QSO and BAO + $H(z)$ data sets are listed in Tables \ref{tab:1} and \ref{tab:2}. Unmarginalized best-fit parameter values are given in Table \ref{tab:1} and marginalized best-fit parameter values and error bars are listed in Table \ref{tab:2}. One-dimensional likelihoods and two-dimensional likelihood contours are shown in Figs.\ \ref{fig:1}--\ref{fig:6}. 

We use the BAO + $H(z)$ results to compare with the results from the QSO data sets. This comparison allows us to draw a qualitative conclusion about the consistency or inconsistency between the QSO results and the results from other well-established cosmological data. From Figs.\ \ref{fig:1}--\ref{fig:6} we see that the \Mgii\ QSO cosmological constraints are consistent with those derived from BAO + $H(z)$ data, which is also the case for the cosmological constraints derived from the full 78 source \Mgii\ data set \citep{khadka2021}. This, however, differs from what happens with the \hb\ sources whose cosmological constraints are $\sim 2\sigma$ inconsistent with the BAO + $H(z)$ cosmological constraints \citep{Khadkaetal2021c}.

From Table \ref{tab:2}, for the 2-parameter \Mgii\ QSO-66 data set, in all models, the measured values of $\beta$, $\gamma$, and $\sigma_{\rm ext}$ lie in the range $1.747^{+0.086}_{-0.061}$ to $1.782^{+0.056}_{-0.056}$, $0.239^{+0.055}_{-0.055}$ to $0.245^{+0.057}_{-0.057}$, and $0.285^{+0.024}_{-0.032}$ to $0.286^{+0.024}_{-0.032}$ respectively. For the 3-parameter linear-\rfe\ \Mgii$^{\prime}$ QSO-66 data set, in all models, the measured values of $\beta$, $\gamma$, $k$, and $\sigma_{\rm ext}$ lie in the range $1.690^{+0.110}_{-0.110}$ to $1.725^{+0.098}_{-0.098}$, $0.217^{+0.063}_{-0.063}$ to $0.223^{+0.066}_{-0.066}$, $0.069^{+0.099}_{-0.099}$ to $0.070^{+0.100}_{-0.100}$, and $0.286^{+0.024}_{-0.032}$ to $0.287^{+0.025}_{-0.032}$ respectively. For the 3-parameter log-\rfe\ \Mgii$^{\prime \prime}$ QSO-66 data set, in all models, the measured values of $\beta$, $\gamma$, $k$, and $\sigma_{\rm ext}$ lie in the range $1.787^{+0.091}_{-0.070}$ to $1.818^{+0.065}_{-0.065}$, $0.201^{+0.065}_{-0.065}$ to $0.207^{+0.068}_{-0.068}$, $0.170^{+0.170}_{-0.170}$ to $0.180^{+0.170}_{-0.170}$, and $0.284^{+0.024}_{-0.032}$ to $0.286^{+0.024}_{-0.032}$ respectively. In each of the three QSO-66 data sets the $R-L$ correlation parameter values are independent, within the error bars, of the cosmological model used in the analysis and so all three QSO-66 data sets are standardizable.\footnote{This is also true for all the high-\rfe\ and low-\rfe\ data subsets.} Importantly, there is no change in the intrinsic dispersion $\sigma_{\rm ext}$ when going from the 2-parameter $R-L$ relation to either of the 3-parameter $R-L$ relations. This differs mildly from what happens in the \hb\ case \citep{Khadkaetal2021c} where $\sigma_{\rm ext}$ is about 0.75$\sigma$ smaller in the 3-parameter linear-\rfe\ $R-L$ relation case compared to the 2-parameter one.\footnote{\citet{Khadkaetal2021c} did not study the log-\rfe\ 3-parameter relation in the \hb\ case.} 

From Table \ref{tab:2}, for the 2-parameter \Mgii\ low-\rfe\ data subset, in all models, the measured values of $\beta$, $\gamma$, and $\sigma_{\rm ext}$ lie in the range $1.776^{+0.092}_{-0.074}$ to $1.803^{+0.072}_{-0.072}$, $0.207^{+0.078}_{-0.078}$ to $0.212^{+0.079}_{-0.079}$, and $0.316^{+0.038}_{-0.054}$ to $0.318^{+0.036}_{-0.053}$ respectively. For the 2-parameter \Mgii\ high-\rfe\ data subset, in all models, the measured values of $\beta$, $\gamma$, and $\sigma_{\rm ext}$ lie in the range $1.660^{+0.150}_{-0.110}$ to $1.710^{+0.110}_{-0.110}$, $0.310^{+0.095}_{-0.095}$ to $0.320^{+0.098}_{-0.098}$, and $0.275^{+0.031}_{-0.046}$ to $0.276^{+0.031}_{-0.047}$ respectively. From Table \ref{tab:3}, for the 2-parameter \Mgii\ low-\rfe\ and high-\rfe\ data subsets, the difference in $\beta$ values, $\Delta\beta$, between different cosmological models lie in the range (0.64--0.77)$\sigma$ which is not statistically significant. The difference in $\gamma$ values, $\Delta\gamma$, between different cosmological models lie in the range (0.84--0.86)$\sigma$ which is not statistically significant. The difference in $\sigma_{\rm ext}$ values, $\Delta\sigma_{\rm ext}$ between different cosmological models lie in the range (0.65--0.69)$\sigma$ which is statistically insignificant. This is very different from what happens with the 2-parameter $R-L$ relation for \hb\ QSOs where $\Delta\beta \sim (3-4)\sigma$ \citep{Khadkaetal2021c}.

From Table \ref{tab:2}, for the 3-parameter linear-\rfe\ \Mgii$^{\prime}$ low-\rfe\ data subset, in all models, the measured values of $\beta$, $\gamma$, $k$, and $\sigma_{\rm ext}$ lie in the range $1.640^{+0.160}_{-0.160}$ to $1.660^{+0.160}_{-0.160}$, $0.162^{+0.072}_{-0.100}$ to $0.165^{+0.073}_{-0.100}$, $0.230^{+0.230}_{-0.230}$, and $0.315^{+0.036}_{-0.053}$ to $0.316^{+0.036}_{-0.054}$ respectively. For the 3-parameter linear-\rfe\ \Mgii$^{\prime}$ high-\rfe\ data subset, in all models, the measured values of $\beta$, $\gamma$, $k$, and $\sigma_{\rm ext}$ lie in the range $1.430^{+0.290}_{-0.290}$ to $1.490^{+0.270}_{-0.270}$, $0.305^{+0.098}_{-0.098}$ to $0.311^{+0.098}_{-0.098}$, $0.150^{+0.190}_{-0.190}$ to $0.160^{+0.190}_{-0.190}$, and $0.273^{+0.031}_{-0.047}$ to $0.275^{+0.032}_{-0.047}$ respectively. From Table \ref{tab:4}, for the 3-parameter linear-\rfe\ \Mgii$^{\prime}$ low-\rfe\ and high-\rfe\ data subsets, the difference in $\beta$ values, $\Delta\beta$, between different cosmological models lie in the range (0.54--0.63)$\sigma$ which is statistically insignificant. The difference in $\gamma$ values, $\Delta\gamma$, between different cosmological models lie in the range (1.17--1.19)$\sigma$ which is not statistically significant. The difference in $k$ values, $\Delta k$, between different cosmological models lie in the range (0.23--0.27)$\sigma$ which is statistically insignificant. The difference in $\sigma_{\rm ext}$ values, $\Delta\sigma_{\rm ext}$, between different cosmological models lie in the range (0.65--0.69)$\sigma$ which is statistically insignificant. Mostly these differences are insignificant but $\Delta \gamma$ are larger than in the 2-parameter $R-L$ relation case shown in Table \ref{tab:3}. This differs from what happens with the \hb\ sources, where $\Delta\gamma$ are smaller in the 3-parameter case compared to the 2-parameter case \citep{Khadkaetal2021c}. 

From Table \ref{tab:2}, for the 3-parameter log-\rfe\ \Mgii$^{\prime \prime}$ low-\rfe\ data subset, in all models, the measured values of $\beta$, $\gamma$, $k$, and $\sigma_{\rm ext}$ lie in the range $1.890^{+0.120}_{-0.094}$ to $1.910^{+0.110}_{-0.092}$, $0.142^{+0.063}_{-0.097}$ to $0.146^{+0.064}_{-0.098}$, $0.350^{+0.240}_{-0.240}$ to $0.360^{+0.240}_{-0.240}$, and $0.307^{+0.035}_{-0.051}$ to $0.309^{+0.035}_{-0.053}$ respectively. For the 3-parameter log-\rfe\ \Mgii$^{\prime \prime}$ high-\rfe\ data subset, in all models, the measured values of $\beta$, $\gamma$, $k$, and $\sigma_{\rm ext}$ lie in the range $1.580^{+0.170}_{-0.140}$ to $1.640^{+0.130}_{-0.130}$, $0.304^{+0.095}_{-0.095}$ to $0.312^{+0.098}_{-0.098}$, $0.550^{+0.730}_{-0.650}$ to $0.590^{+0.720}_{-0.650}$, and $0.272^{+0.032}_{-0.047}$ to $0.275^{+0.033}_{-0.048}$ respectively. From Table \ref{tab:5}, for the 3-parameter log-\rfe\ \Mgii$^{\prime \prime}$ low-\rfe\ and high-\rfe\ data subsets, the difference in $\beta$ values, $\Delta\beta$, between different cosmological models lie in the range (1.58--1.78)$\sigma$ which could be statistically significant. The difference in $\gamma$ values, $\Delta\gamma$, between different cosmological models lie in the range (1.40--1.44)$\sigma$ which could be statistically significant. The difference in $k$ values, $\Delta k$, between different cosmological models lie in the range (0.27--0.35)$\sigma$ which is statistically insignificant. The difference in $\sigma_{\rm ext}$ values, $\Delta\sigma_{\rm ext}$, between different cosmological models lie in the range (0.54--0.58)$\sigma$ which is statistically insignificant. $\Delta\beta$ values are (somewhat significantly) larger in the log-\rfe\ 3-parameter case compared to the linear-\rfe\ 3-parameter and the 2-parameter $R-L$ correlation cases shown in Tables \ref{tab:4} and \ref{tab:3}.

This difference in $\Delta\beta$ values is connected to the smaller and mostly negative range of log \rfe\ $(-1.0, 0.37)$ in comparison to the larger and positive range of \rfe\ $(0.1, 2.36)$. This results in the normalization, $\beta$, values being larger in the log-\rfe\ case compared to those in the linear-\rfe\ case. Another qualitative difference is that \Mgii$^{\prime\prime}$ high-\rfe\ sources always have positive log \rfe\ values, with mean value of $0.13$, while \Mgii$^{\prime\prime}$ low-\rfe\ sources have mostly negative log \rfe\ values, with mean value of $-0.21$. This apparently makes the inferred high- and low-\rfe\ $\beta$ values differ more in the log-\rfe\ case than in the linear-\rfe\ case.

From Table \ref{tab:1}, from the $\Delta AIC$ values, there is mostly only weak evidence (mostly) for or against the 2-parameter $R-L$ relation. The exceptions for the 66 source data set are the flat $\Lambda$CDM model and the non-flat XCDM parametrization where there is positive evidence for and against the 2-parameter $R-L$ relation relative to the 3-parameter linear-\rfe\ $R-L$ relations, respectively. From the $\Delta BIC$ values, in all 66 source cases, there is positive evidence in favor of the 2-parameter $R-L$ relation over the 3-parameter $R-L$ relations. The high- and low-\rfe\ data subset $\Delta BIC$ values indicate that the 2-parameter relation is weakly or positively favored over both the 3-parameter relations. The results for the 66 source data set analyses are dramatically different from those for the full \hb\ data set analyses \citep{Khadkaetal2021c} where the linear-\rfe\ 3-parameter $R-L$ is very strongly favored over the 2-parameter one. This difference is a consequence of the significant difference between low- and high-\rfe\ \hb\ sources relative to the small difference between these two subsets in the \Mgii\ case here. 

When considering the full \Mgii\ sample, using both the linear- and log-\rfe\ relations, $k$ is consistent with zero at less than $1\sigma$. It is only for the \Mgii$^{\prime\prime}$ low-\rfe\ case that $k$ is a little less than 1.5$\sigma$ away from zero. This is another indication that \rfe\ is of no significant relevance for current \Mgii\, data. This is another difference compared to the \hb\ case, where $k<0$ at more than 4$\sigma$ for the full sample in the linear-\rfe\ case.   

The \Mgii\ cosmological constraints obtained in this paper are listed in Table \ref{tab:2}. They are weak. Since we have found no reason to favor either of the 3-parameter $R-L$ relations over the 2-parameter one, the 2-parameter $R-L$ relation cosmological constraints derived in \citet{khadka2021} from the complete 78 source data set are the appropriate \Mgii\ cosmological constraints.

\onecolumn
\begin{landscape}
\addtolength{\tabcolsep}{-1pt}
\begin{longtable}{lccccccccccccccccccccc}
\caption{Unmarginalized best-fit parameters for all data sets.}
\label{tab:1}\\
\hline
Model & Data set & $\Omega_{b}h^2$ & $\Omega_{c}h^2$& $\Omega_{m0}$ & $\Omega_{k0}$ & $\omega_{X}$ & $\alpha$ & $H_0$\footnotesize{$^a$} & $\sigma_{\rm ext}$ & $\beta$ & $\gamma$ & $k$ & $dof$ & $-2\ln({\rm LF}_{\rm max})$ & $AIC$ & $BIC$ & $\Delta AIC$ & $\Delta BIC$\\
\hline
\endfirsthead
\hline
Model & Data set & $\Omega_{b}h^2$ & $\Omega_{c}h^2$& $\Omega_{m0}$ & $\Omega_{k0}$ & $\omega_{X}$ & $\alpha$ & $H_0$\footnotesize{$^a$} & $\sigma_{\rm ext}$ & $\beta$ & $\gamma$ & $k$ & $dof$ & $-2\ln({\rm LF}_{\rm max})$ & $AIC$ & $BIC$ & $\Delta AIC$ & $\Delta BIC$\\
\hline
\endhead
\hline
& \Mgii\ QSO-66 & - & -& 0.211 & - & - & - & - & 0.274 & 1.745 & 0.230 & - & 62 & 20.52 & 28.52 & 37.28 & - & -\\
Flat & \Mgii\ low-\rfe & - & -& 0.534 & - & - & - &- & 0.291 & 1.796 & 0.206 & - &29 & 14.30 & 22.30 & 28.29 & - & -\\
$\Lambda$CDM & \Mgii\ high-\rfe & - & - & 0.132 & - & - & - &- & 0.253 & 1.651 & 0.294 & - &29 & 4.64 & 12.64 & 18.63 & - & -\\
& \Mgii$^{\prime}$ QSO-66 & - & -& 0.120 & - & - & - & - & 0.272 & 1.684 & 0.202 & 0.076 & 61 & 19.90 & 29.90 & 40.85 & $1.38$ & $3.57$\\
& \Mgii$^{\prime}$ low-\rfe & - & -& 0.993 & - & - & - & - & 0.284 & 1.662 & 0.149 & 0.255 & 28 & 12.90 & 22.90 & 30.38 & $0.60$ & $2.09$\\
& \Mgii$^{\prime}$ high-\rfe & - & -& 0.027 & - & - & - & - & 0.243 & 1.375 & 0.278 & 0.194 & 28 & 3.12 & 13.12 & 20.60 & $0.48$ & $1.97$\\
& \Mgii$^{\prime \prime}$ QSO-66 & - & -& 0.088 & - & - & - & - & 0.270 & 1.817 & 0.270 & 0.182 & 61 & 19.26 & 26.29 & 40.21 & $-$2.23 & $2.93$\\
& \Mgii$^{\prime \prime}$ low-\rfe & - & -& 0.959 & - & - & - & - & 0.276 & 1.939 & 0.122 & 0.393 & 28 & 11.34 & 21.34 & 28.82 & $-$0.96 & $0.53$\\
& \Mgii$^{\prime \prime}$ high-\rfe & - & -& 0.030 & - & - & - & - & 0.241 & 1.546 & 0.272 & 0.791 & 28 & 3.10 & 13.10 & 20.58 & $0.46$ & $1.95$\\
& BAO + $H(z)$ & 0.024 & 0.119 & 0.298 & - & - & - &69.119&-&-&-& - &39 & 23.66&29.66&34.87 & - & -\\
\hline
& \Mgii\ QSO-66 & - & -& 0.347 & $-1.050$ & - &-&- & 0.271 & 1.683 & 0.295 & - &61 & 18.28 & 28.28 & 39.23 &-&-\\
Non-flat & \Mgii\ low-\rfe & - & - & 0.393 & $-$1.120 & - & - &- & 0.286 & 1.731 & 0.237 & - & 28 & 14.08 & 24.08 & 31.56 &-&-\\
$\Lambda$CDM & \Mgii\ high-\rfe &- & -& 0.254 & $-$0.853 & - & - &- & 0.242 & 1.561 & 0.368 & - &28 & 1.22 & 11.22 & 18.70 &-&-\\
& \Mgii$^{\prime}$ QSO-66 &- & -& 0.337 & $-$1.025 & - & - &- & 0.272 & 1.645 & 0.253 & 0.057 &  60 & 18.16 & 30.16 & 43.30 & $1.88$ & $4.07$\\
& \Mgii$^{\prime}$ low-\rfe &- & -& 0.737 & $-$1.773 & - & - &- & 0.281 & 1.632 & 0.195 & 0.230 &  27 & 12.78 & 24.78 & 33.76 & $0.70$ & $2.20$\\
& \Mgii$^{\prime}$ high-\rfe &- & -& 0.237 & $-$0.800 & - & - &- & 0.239 & 1.392 & 0.392 & 0.094 &  27 & 0.84 & 12.84 & 21.82 & $1.62$ & $3.12$\\
& \Mgii$^{\prime \prime}$ QSO-66 &- & -& 0.291 & $-$0.933 & - & - &- & 0.271 & 1.722 & 0.251 & 0.108 &  60 & 17.56 & 29.56 & 42.70 & $1.28$ & $3.47$\\
& \Mgii$^{\prime \prime}$ low-\rfe &- & -& 0.615 & $-$1.587 & - & - &- & 0.280 & 1.883 & 0.164 & 0.367 &  27 & 11.22 & 23.22 & 32.20 & $-$0.86 & $0.64$\\
& \Mgii$^{\prime \prime}$ high-\rfe &- & -& 0.214 & $-$0.750 & - & - &- & 0.230 & 1.525 & 0.343 & 0.426 &  27 & 0.90 & 12.90 & 21.88 & $1.68$ & $3.18$\\
& BAO + $H(z)$ & 0.025 & 0.114 & 0.294 & 0.021 & - & - &68.701&-&-&-&- &38&23.60&31.60&38.55 &-&-\\
\hline
& \Mgii\ QSO-66 &- & -& 0.002 & - & $-$4.549 &-&- & 0.276 & 1.517 & 0.178 & - &61 & 19.58 & 29.58 & 40.53 &-&-\\
Flat & \Mgii\ low-\rfe &- & -& 0.431 & - & $-1.454$ & - &- & 0.291 & 1.780 & 0.205 & - &28 & 14.30 & 24.30 & 31.78 &-&-\\
XCDM & \Mgii\ high-\rfe &- & -& 0.000 & - & $-4.910$ & - &- & 0.233 & 1.287 & 0.234 &- & 28 & 2.46 & 12.46 & 19.94 &-&-\\
& \Mgii$^{\prime}$ QSO-66 &- & -& 0.015 & - & $-$3.639 & - & - & 0.265 & 1.520 & 0.178 & 0.094 & 60 & 18.96 & 30.96 & 44.10 & $1.38$ & $3.57$\\
& \Mgii$^{\prime}$ low-\rfe &- & -& 0.280 & - & 0.091 &-&- & 0.285 & 1.656 & 0.143 & 0.272 & 27 & 12.90 & 24.90 & 33.88 & $0.60$ & $2.10$\\
& \Mgii$^{\prime}$ high-\rfe &- & -& 0.002 & - & $-$4.373 &-&- & 0.223 & 0.980 & 0.228 & 0.297 & 27 & 0.40 & 12.40 & 21.38 & $-$0.06 & $1.44$\\
& \Mgii$^{\prime \prime}$ QSO-66 &- & -& 0.003 & - & $-$4.524 &-&- & 0.268 & 1.559 & 0.170 & 0.151 & 60 & 18.34 & 30.34 & 43.48 & $0.76$ & $2.95$\\
& \Mgii$^{\prime \prime}$ low-\rfe &- & -& 0.770 & - & 0.058 &-&- & 0.276 & 1.939 & 0.128 & 0.375 & 27 & 11.34 & 23.34 & 32.32 & $-$0.96 & $0.54$\\
& \Mgii$^{\prime \prime}$ high-\rfe &- & -& 0.002 & - & $-$4.655 &-&- & 0.244 & 1.102 & 0.274 & 1.136 & 27 & $-0.12$ & 11.88 & 20.86 & $-$0.58 & $0.92$\\
& BAO + $H(z)$ & 0.031 & 0.088 & 0.280 & - & $-$0.691 & - &65.036& - & - & -&- &38&19.66&27.66&34.61 &-&-\\
\hline
& \Mgii\ QSO-66 &- & -& 0.129 & $-$0.248 & $-$2.424 &-&- & 0.264 & 1.552 & 0.261 & - &60 & 16.38 & 28.38 & 41.52 &-&-\\
Non-flat & \Mgii\ low-\rfe &- & - & 0.154 & $-$0.353 & $-$1.771 & - &- & 0.293 & 1.697 & 0.187 & - &27 & 13.88 & 25.88 & 34.86 &-&-\\
XCDM & \Mgii\ high-\rfe &- & -& 0.073 & $-0.152$ & $-$2.501 & - &- & 0.213 & 1.298 & 0.372 & - &27 & $-$3.34 & 8.66 & 17.64 &-&-\\
& \Mgii$^{\prime}$ QSO-66 &- & -& 0.110 & $-$0.231 & $-$2.246 &-&- & 0.260& 1.565 & 0.268 & $-0.004$ & 59 & 15.96 & 29.96 & 45.29 & $1.58$ & $3.77$\\
& \Mgii$^{\prime}$ low-\rfe &- & -& 0.634 & $-$1.833 & $-$0.776 &-&- & 0.285 & 1.628 & 0.167 & 0.261 & 26 & 12.78 & 26.78 & 37.26 & $0.90$ & $2.40$\\
& \Mgii$^{\prime}$ high-\rfe &- & -& 0.054 & $-$0.133 & $-$2.083 &-&- & 0.213 & 1.163 & 0.316 & 0.167 & 26 & $-2.56$ & 11.44 & 21.92 & $2.78$ & $4.28$\\
& \Mgii$^{\prime \prime}$ QSO-66 &- & -& 0.146 & $-$0.326 & $-$1.886 &-&- & 0.282& 1.620 & 0.254 & 0.065 & 59 & 16.42 & 30.42 & 45.75 & $2.04$ & $4.23$\\
& \Mgii$^{\prime \prime}$ low-\rfe &- & -& 0.991 & $-$1.250 & $-$4.657 &-&- & 0.271 & 1.865 & 0.166 & 0.424 & 26 & 11.22 & 25.22 & 35.70 & $-$0.66 & $0.84$\\
& \Mgii$^{\prime \prime}$ high-\rfe &- & -& 0.039 & $-$0.084 & $-$2.682 &-&- & 0.215 & 1.248 & 0.377 & 0.164 & 26 & $-5.28$ & 8.72 & 19.20 & $0.06$ & $1.56$\\
& BAO + $H(z)$ & 0.030 & 0.094 & 0.291 & $-$0.147 & $-$0.641 & - &65.204& - & - & -&- &37&18.34&28.34&37.03 &-&-\\
\hline
& \Mgii\ QSO-66 &- & -& 0.123 & - & - &0.174&- & 0.273 & 1.739 & 0.229 & - &61 & 20.56 & 30.56 & 41.51 &-&-\\
Flat & \Mgii\ low-\rfe &- & -& 0.866 & - & - & 5.741 &- & 0.291 & 1.816 & 0.214 & - &28 & 14.30 & 24.30 & 31.78 &-&-\\
$\phi$CDM & Mg II high-\rfe &- & -& 0.096 & - & - & 0.113 &- & 0.250 & 1.652 & 0.291 & - &28 & 4.64 & 14.64 & 22.12 &-&-\\
& \Mgii$^{\prime}$ QSO-66 &- & -& 0.143 & - & - & 0.217 &- & 0.272 & 1.681 & 0.209 & 0.078 & 60 & 19.92 & 31.92 & 45.06 & $1.36$ & $3.55$\\
& \Mgii$^{\prime}$ low-\rfe &- & -& 0.957 & - & - & 0.814 &- & 0.281 & 1.660 & 0.150 & 0.262 & 27 & 12.90 & 24.90 & 33.88 & $0.60$ & $2.10$\\
& \Mgii$^{\prime}$ high-\rfe &- & -& 0.039 & - & - & 0.196 &- & 0.243 & 1.336 & 0.282 & 0.231 & 27 & 3.20 & 15.20 & 24.18 & $0.60$ & $2.06$\\
& \Mgii$^{\prime \prime}$ QSO-66 &- & -& 0.197 & - & - & 0.089 &- & 0.275 & 1.786 & 0.198 & 0.182 & 60 & 19.28 & 31.28 & 44.42 & $0.72$ & $2.91$\\
& \Mgii$^{\prime \prime}$ low-\rfe &- & -& 0.969 & - & - & 6.156 &- & 0.278 & 1.937 & 0.126 & 0.385 & 27 & 11.34 & 23.34 & 32.32 & $-$0.96 & $0.54$\\
& \Mgii$^{\prime \prime}$ high-\rfe &- & -& 0.028 & - & - & 0.397 &- & 0.238 & 1.564 & 0.271 & 0.776 & 27 & 3.20 & 15.20 & 24.18 & $0.56$ & $2.06$\\
& BAO + $H(z)$ & 0.033 & 0.080 & 0.265 & - & - & 1.445 &65.272& - & - & -&- &38&19.56&27.56&34.51&-&-\\
\hline
& \Mgii\ QSO-66 &- & -& 0.342 & $-$0.308 & - & 0.194 &- & 0.275 & 1.746 & 0.238 & - &60 & 20.44 & 32.44 & 45.58 &-&-\\
Non-flat & \Mgii\ low-\rfe &- & - & 0.987 & $-$0.854 & - & 0.393 & - & 0.291 & 1.809 & 0.210 & - & 27 & 14.28 & 26.28 & 35.26 &-&-\\
$\phi$CDM & \Mgii\ high-\rfe & - & - & 0.278 & $-$0.267 & - & 0.136 &- & 0.245 & 1.649 & 0.316 & - &27 & 4.58 & 16.58 & 25.56 &-&-\\
& \Mgii$^{\prime}$ QSO-66 &- & -& 0.487 & $-$0.365 & - & 0.404 &- & 0.271 & 1.718 & 0.219 & 0.057 & 59 & 19.94 & 33.94 & 49.27 & $1.50$ & $3.69$\\
& \Mgii$^{\prime}$ low-\rfe &- & -& 0.970 & $-$0.898 & - & 5.136 &- & 0.282 & 1.672 & 0.154 & 0.255 & 26 & 12.88 & 26.88 & 37.36 & $0.60$ & $2.10$\\
& \Mgii$^{\prime}$ high-\rfe &- & -& 0.064 & 0.042 & - & 0.082 &- & 0.243 & 1.431 & 0.298 & 0.155 & 26 & 3.32 & 17.32 & 27.80 & $0.74$ & $2.24$\\
& \Mgii$^{\prime \prime}$ QSO-66 &- & -& 0.401 & $-$0.191 & - & 0.331 &- & 0.268 & 1.804 & 0.202 & 0.188 & 59 & 19.24 & 33.24 & 48.57 & $0.80$ & $2.99$\\
& \Mgii$^{\prime \prime}$ low-\rfe &- & -& 0.857 & $-$0.708 & - & 3.576 &- & 0.280 & 1.942 & 0.124& 0.384 & 26 & 11.32 & 25.32 & 35.80 & $-$0.96 & $0.53$\\
& \Mgii$^{\prime \prime}$ high-\rfe &- & -& 0.065 & $-$0.021 & - & 0.315 &- & 0.248 & 1.545 & 0.284 & 0.727 & 26 & 3.30 & 17.30 & 27.78 & $0.72$ & $2.22$\\
& BAO + $H(z)$ & 0.035 & 0.078 & 0.261 & $-$0.155 & - & 2.042 &65.720& - & - & -&- &37&18.16&28.16&36.85 &-&-\\
\hline
\end{longtable}
\footnotesize{\hspace{-0.6cm}$^a$ ${\rm km}\hspace{1mm}{\rm s}^{-1}{\rm Mpc}^{-1}$. $H_0$ is set to $70$ ${\rm km}\hspace{1mm}{\rm s}^{-1}{\rm Mpc}^{-1}$ for the QSO-only analyses.}

\addtolength{\tabcolsep}{-1pt}
\begin{longtable}{lccccccccccccccc}
\caption{Marginalized one-dimensional best-fit parameters with 1$\sigma$ confidence intervals or 2$\sigma$ limits for all data sets.}
\label{tab:2}\\
\hline
Model & Data & \ \ \ \ \ \ $\Omega_{b}h^2$ \ \ \ \ \ \ & \ \ \ \ \ \ $\Omega_{c}h^2$ \ \ \ \ \ \ & \ \ \ \ \ \ $\om$ \ \ \ \ \ \ & \ \ \ \ \ \ $\ol$\footnotesize{$^a$} \ \ \ \ \ \ & \ \ \ \ \ \ $\ok$ \ \ \ \ \ \ & \ \ \ \ \ \ $\omega_{X}$ \ \ \ \ \ \ & \ \ \ \ \ \ $\alpha$ \ \ \ \ \ \ & \ \ \ \ \ \ $H_0$\footnotesize{$^b$} \ \ \ \ \ \ & \ \ \ \ \ \ $\sigma_{\rm ext}$ \ \ \ \ \ \ & \ \ \ \ \ \ $\beta$ \ \ \ \ \ \ & \ \ \ \ \ \ $\gamma$ \ \ \ \ \ \ & \ \ \ \ \ \ $k$ \ \ \ \ \ \ \\
\hline
\endfirsthead
\hline
Model & Data & \ \ \ \ \ \ $\Omega_{b}h^2$ \ \ \ \ \ \ & \ \ \ \ \ \ $\Omega_{c}h^2$ \ \ \ \ \ \ & \ \ \ \ \ \ $\om$ \ \ \ \ \ \ & \ \ \ \ \ \ $\ol$\footnotesize{$^a$} \ \ \ \ \ \ & \ \ \ \ \ \ $\ok$ \ \ \ \ \ \ & \ \ \ \ \ \ $\omega_{X}$ \ \ \ \ \ \ & \ \ \ \ \ \ $\alpha$ \ \ \ \ \ \ & \ \ \ \ \ \ $H_0$\footnotesize{$^b$} \ \ \ \ \ \ & \ \ \ \ \ \ $\sigma_{\rm ext}$ \ \ \ \ \ \ & \ \ \ \ \ \ $\beta$ \ \ \ \ \ \ & \ \ \ \ \ \ $\gamma$ \ \ \ \ \ \ & \ \ \ \ \ \ $k$ \ \ \ \ \ \ \\
\hline
\endhead
\hline
& \Mgii\ QSO-66 &-&-& --- & $0.490^{+0.280}_{-0.280}$ & - & - & - &-& $0.285^{+0.024}_{-0.032}$ & $1.766^{-0.062}_{-0.062}$ & $0.239^{+0.055}_{-0.055}$&-\\
Flat & \Mgii\ low-\rfe &-&-& --- & $0.490^{+0.260}_{-0.480}$ & - & - & - &-& $0.318^{+0.036}_{-0.054}$ & $1.790^{-0.077}_{-0.077}$ & $0.207^{+0.078}_{-0.078}$&-\\
\lcdm\ & \Mgii\ high-\rfe &-&-& --- & $0.500^{+0.360}_{-0.310}$ & - & - &-&-& $0.276^{+0.031}_{-0.047}$ & $1.680^{-0.120}_{-0.120}$ & $0.310^{+0.095}_{-0.095}$&-\\
& \Mgii$^{\prime}$ QSO-66 &-&-& --- & $0.500^{+0.280}_{-0.280}$ & - & - & - &-& $0.287^{+0.025}_{-0.032}$ & $1.710^{-0.100}_{-0.100}$ & $0.217^{+0.063}_{-0.063}$&$0.070^{+0.100}_{-0.100}$\\
& \Mgii$^{\prime}$ low-\rfe &-&-& --- & $0.490^{+0.230}_{-0.480}$ & - & - & - &-& $0.316^{+0.036}_{-0.054}$ & $1.650^{-0.160}_{-0.160}$ & $0.162^{+0.072}_{-0.100}$&$0.230^{+0.240}_{-0.240}$\\
& \Mgii$^{\prime}$ high-\rfe &-&-& --- & $0.520^{+0.470}_{-0.230}$ & - & - & - &-& $0.274^{+0.032}_{-0.048}$ & $1.460^{-0.270}_{-0.270}$ & $0.303^{+0.095}_{-0.095}$&$0.160^{+0.190}_{-0.190}$\\
& \Mgii$^{\prime \prime}$ QSO-66 &-&-& --- & $0.490^{+0.280}_{-0.280}$ & - & - & - &-& $0.285^{+0.025}_{-0.032}$ & $1.804^{+0.071}_{-0.071}$ & $0.201^{+0.065}_{-0.065}$&$0.180^{+0.170}_{-0.170}$\\
& \Mgii$^{\prime \prime}$ low-\rfe &-&-& --- & $0.490^{+0.300}_{-0.480}$ & - & - & - &-& $0.309^{+0.036}_{-0.053}$ & $1.900^{-0.110}_{-0.094}$ & $0.143^{+0.064}_{-0.097}$&$0.350^{+0.240}_{-0.240}$\\
& \Mgii$^{\prime \prime}$ high-\rfe &-&-& --- & $0.520^{+0.470}_{-0.190}$ & - & - & - &-& $0.273^{+0.032}_{-0.048}$ & $1.610^{-0.140}_{-0.140}$ & $0.304^{+0.095}_{-0.095}$&$0.590^{+0.720}_{-0.650}$\\
& BAO + $H(z)$ & $0.024^{+0.003}_{-0.003}$ & $0.119^{+0.008}_{-0.008}$ & $0.299^{+0.015}_{-0.017}$ & - & - & - & - &$69.30^{+1.80}_{-1.80}$&-&-&-&-\\
\hline
& \Mgii\ QSO-66 &-&-& --- & $< 1.770$ & $0.100^{+1.100}_{-1.200}$ & - &-&-& $0.285^{+0.024}_{-0.032}$ & $1.768^{+0.069}_{-0.061}$ & $0.245^{+0.057}_{-0.057}$&-\\
Non-flat & \Mgii\ low-\rfe &-&-& --- & $< 1.500$ & $0.180^{+1.300}_{-0.980}$ & - &-&-& $0.318^{+0.036}_{-0.054}$ & $1.795^{+0.077}_{-0.077}$ & $0.212^{+0.079}_{-0.079}$&-\\
\lcdm\ & \Mgii\ high-\rfe &-&-& $0.550^{+0.430}_{-0.160}$ & $0.40^{+1.400}_{-0.710}$ & $0.000^{+1.000}_{-1.300}$ & - &-&-& $0.275^{+0.031}_{-0.046}$ & $1.680^{+0.130}_{-0.110}$ & $0.320^{+0.098}_{-0.098}$&-\\
&\Mgii$^{\prime}$ QSO-66 &-&-& --- & $< 1.640$ & $0.100^{+1.000}_{-1.000}$  &-&-&-& $0.286^{+0.025}_{-0.032}$ & $1.720^{+0.100}_{-0.100}$ & $0.223^{+0.066}_{-0.066}$ & $0.070^{+0.100}_{-0.100}$\\
&\Mgii$^{\prime}$ low-\rfe &-&-& --- & $< 1.500$ & $0.180^{+1.300}_{-0.980}$  &-&-&-& $0.316^{+0.036}_{-0.053}$ & $1.660^{+0.160}_{-0.160}$ & $0.165^{+0.073}_{-0.100}$ & $0.230^{+0.230}_{-0.230}$\\
&\Mgii$^{\prime}$ high-\rfe &-&-& $0.540^{+0.420}_{-0.190}$ & $< 1.650$ & $0.100^{+1.000}_{-1.000}$  &-&-&-& $0.273^{+0.032}_{-0.047}$ & $1.480^{+0.270}_{-0.270}$ & $0.311^{+0.098}_{-0.098}$ & $0.150^{+0.190}_{-0.170}$\\
&\Mgii$^{\prime \prime}$ QSO-66 &-&-& --- & $< 2.000$ & $0.100^{+1.000}_{-1.000}$  &-&-&-& $0.285^{+0.024}_{-0.032}$ & $1.805^{+0.078}_{-0.068}$ & $0.207^{+0.068}_{-0.068}$ & $0.170^{+0.170}_{-0.170}$\\
&\Mgii$^{\prime \prime}$ low-\rfe &-&-& --- & $< 2.000$ & $0.170^{+1.300}_{-0.970}$  &-&-&-& $0.308^{+0.035}_{-0.052}$ & $1.910^{+0.110}_{-0.092}$ & $0.146^{+0.064}_{-0.098}$ & $0.350^{+0.240}_{-0.240}$\\
&\Mgii$^{\prime \prime}$ high-\rfe &-&-& $0.540^{+0.420}_{-0.190}$ & $< 1.900$ & $0.100^{+1.000}_{-1.000}$  &-&-&-& $0.273^{+0.032}_{-0.047}$ & $1.620^{+0.140}_{-0.140}$ & $0.312^{+0.098}_{-0.098}$ & $0.550^{+0.730}_{-0.650}$\\
& BAO + $H(z)$ & $0.025^{+0.004}_{-0.004}$ & $0.113^{+0.019}_{-0.019}$ & $0.292^{+0.023}_{-0.023}$ & $0.667^{+0.093}_{+0.081}$ & $-0.014^{+0.075}_{-0.075}$ & - & - &$68.70^{+2.30}_{-2.30}$&-&-&-&-\\
\hline
& \Mgii\ QSO-66 &-&-& $0.500^{+0.280}_{-0.280}$ & - & - & $< 0.000$ & - &-& $0.286^{+0.024}_{-0.032}$ & $1.747^{+0.086}_{-0.061}$ & $0.240^{+0.055}_{-0.055}$&-\\
Flat & \Mgii\ low-\rfe &-&-& $0.520^{+0.390}_{0.250}$ & - & - & $< 0.000$ & - &-& $0.318^{+0.036}_{-0.053}$ & $1.776^{+0.092}_{-0.074}$ & $0.208^{+0.078}_{-0.078}$&-\\
XCDM & \Mgii\ high-\rfe &-&-& --- & - & - & $< -0.163$ & - &-& $0.276^{+0.031}_{-0.047}$ & $1.660^{+0.150}_{-0.110}$ & $0.312^{+0.096}_{-0.096}$&-\\
& \Mgii$^{\prime}$ QSO-66 &-&-& --- & - & - & $< -0.153$ & - &-& $0.287^{+0.025}_{-0.032}$ & $1.690^{+0.110}_{-0.110}$ & $0.218^{+0.064}_{-0.064}$ & $0.070^{+0.100}_{-0.100}$\\
& \Mgii$^{\prime}$ low-\rfe &-&-& $0.520^{+0.410}_{-0.230}$ & - & - & $< -0.100$ & - &-& $0.316^{+0.036}_{-0.054}$ & $1.640^{+0.160}_{-0.160}$ & $0.162^{+0.072}_{-0.100}$ & $0.230^{+0.230}_{-0.230}$\\
& \Mgii$^{\prime}$ high-\rfe &-&-& --- & - & - & $< -0.177$ & - &-& $0.274^{+0.032}_{-0.048}$ & $1.430^{+0.290}_{-0.290}$ & $0.305^{+0.096}_{-0.096}$ & $0.160^{+0.190}_{-0.190}$\\
& \Mgii$^{\prime \prime}$ QSO-66 &-&-& $0.500^{+0.280}_{-0.280}$ & - & - & $< 0.145$ & - &-& $0.285^{+0.024}_{-0.032}$ & $1.787^{+0.091}_{-0.070}$ & $0.203^{+0.065}_{-0.065}$ & $0.180^{+0.170}_{-0.170}$\\
& \Mgii$^{\prime \prime}$ low-\rfe &-&-& $0.520^{+0.390}_{-0.250}$ & - & - & $< 0.100$ & - &-& $0.308^{+0.035}_{-0.053}$ & $1.890^{+0.120}_{-0.094}$ & $0.142^{+0.063}_{-0.097}$ & $0.360^{+0.240}_{-0.240}$\\
& \Mgii$^{\prime \prime}$ high-\rfe &-&-& --- & - & - & $< -0.183$ & - &-& $0.273^{+0.032}_{-0.048}$ & $1.580^{+0.170}_{-0.140}$ & $0.306^{+0.095}_{-0.095}$ & $0.580^{+0.720}_{-0.650}$\\
& BAO + $H(z)$ & $0.030^{+0.005}_{-0.005}$ & $0.093^{+0.019}_{-0.017}$ & $0.282^{+0.021}_{-0.021}$ & - & - & $-0.744^{+0.140}_{-0.097}$ & - &$65.80^{+2.20}_{-2.50}$& - & - & -&-\\
\hline
& \Mgii\ QSO-66 &-&- & --- & - & $0.260^{+0.920}_{-0.920}$ & $< 0.000$ & - &-& $0.286^{+0.024}_{-0.032}$ & $1.772^{+0.083}_{-0.060}$ & $0.242^{+0.057}_{-0.057}$&-\\
Non-flat & \Mgii\ low-\rfe &-&-& --- & - & $0.280^{+0.930}_{-0.930}$ & $< 0.100$ & - &-& $0.318^{+0.036}_{-0.054}$ & $1.796^{+0.089}_{-0.071}$ & $0.210^{+0.080}_{-0.080}$&-\\
XCDM & \Mgii\ high-\rfe &-&-& $0.540^{+0.460}_{-0.160}$ & - & $0.260^{+0.910}_{-0.910}$ & $< 0.000$ & - &-& $0.276^{+0.031}_{-0.047}$ & $1.690^{+0.140}_{-0.110}$ & $0.315^{+0.097}_{-0.097}$&-\\
& \Mgii$^{\prime}$ QSO-66 &-&- & --- & - & $0.270^{+0.920}_{-0.920}$ & $< -0.138$ & - &-& $0.287^{+0.025}_{-0.032}$ & $1.720^{+0.110}_{-0.110}$ & $0.220^{+0.065}_{-0.065}$ & $0.070^{+0.100}_{-0.100}$\\
& \Mgii$^{\prime}$ low-\rfe &-&- & --- & - & $0.270^{+0.940}_{-0.940}$ & $< 0.100$ & - &-& $0.316^{+0.036}_{-0.054}$ & $1.650^{+0.072}_{-0.100}$ & $0.163^{+0.160}_{-0.160}$ & $0.230^{+0.230}_{-0.230}$\\
& \Mgii$^{\prime}$ high-\rfe &-&- & $0.530^{+0.450}_{-0.180}$ & - & $0.290^{+0.910}_{-0.910}$ & $< 0.000$ & - &-& $0.275^{+0.032}_{-0.047}$ & $1.480^{+0.270}_{-0.270}$ & $0.305^{+0.098}_{-0.098}$ & $0.160^{+0.190}_{-0.190}$\\
& \Mgii$^{\prime \prime}$ QSO-66 &-&- & --- & - & $0.280^{+0.930}_{-0.930}$ & $< -0.134$ & - &-& $0.286^{+0.024}_{-0.032}$ & $1.810^{+0.087}_{-0.065}$ & $0.203^{+0.067}_{-0.067}$ & $0.180^{+0.170}_{-0.170}$\\
& \Mgii$^{\prime \prime}$ low-\rfe &-&- & --- & - & $0.270^{+0.940}_{-0.940}$ & $< 0.100$ & - &-& $0.309^{+0.035}_{-0.053}$ & $1.910^{+0.110}_{-0.090}$ & $0.145^{+0.063}_{-0.099}$ & $0.360^{+0.240}_{-0.240}$\\
& \Mgii$^{\prime \prime}$ high-\rfe &-&- & $0.520^{+0.430}_{-0.200}$ & - & $0.300^{+0.900}_{-0.900}$ & $< 0.000$ & - &-& $0.275^{+0.033}_{-0.048}$ & $1.620^{+0.160}_{-0.140}$ & $0.308^{+0.098}_{-0.098}$ & $0.550^{+0.730}_{-0.650}$\\
& BAO + $H(z)$ & $0.029^{+0.005}_{-0.005}$ & $0.099^{+0.021}_{-0.021}$ & $0.293^{+0.027}_{-0.027}$ & - & $-0.120^{+0.130}_{-0.130}$ & $-0.693^{+0.130}_{-0.077}$ & - &$65.90^{+2.40}_{-2.40}$& - & - & -&-\\
\hline
& \Mgii\ QSO-66 &-&-& --- & - & - & - & --- &-& $0.285^{+0.024}_{-0.031}$ & $1.781^{+0.057}_{-0.057}$ & $0.241^{+0.054}_{-0.054}$&-\\
Flat & \Mgii\ low-\rfe &-&-& --- & - & - & - & --- &-& $0.316^{+0.035}_{-0.053}$ & $1.802^{+0.072}_{-0.072}$ & $0.210^{+0.077}_{-0.077}$&-\\
$\phi$CDM & \Mgii\ high-\rfe &-&-& --- & - & - & - & --- &-& $0.275^{+0.030}_{-0.046}$ & $1.700^{+0.110}_{-0.110}$ & $0.313^{+0.095}_{-0.095}$&-\\
& \Mgii$^{\prime}$ QSO-66 &-&-& --- & - & - & - & --- &-& $0.286^{+0.024}_{-0.032}$ & $1.725^{+0.098}_{-0.098}$ & $0.219^{+0.063}_{-0.063}$ & $0.069^{+0.099}_{-0.099}$\\
& \Mgii$^{\prime}$ low-\rfe &-&-& --- & - & - & - & --- &-& $0.315^{+0.036}_{-0.053}$ & $1.660^{+0.160}_{-0.160}$ & $0.163^{+0.073}_{-0.100}$ & $0.230^{+0.230}_{-0.230}$\\
& \Mgii$^{\prime}$ high-\rfe &-&-& --- & - & - & - & --- &-& $0.273^{+0.031}_{-0.047}$ & $1.490^{+0.270}_{-0.270}$ & $0.305^{+0.095}_{-0.095}$ & $0.160^{+0.180}_{-0.180}$\\
& \Mgii$^{\prime \prime}$ QSO-66 &-&-& --- & - & - & - & --- &-& $0.284^{+0.024}_{-0.032}$ & $1.817^{+0.066}_{-0.066}$ & $0.203^{+0.065}_{-0.065}$ & $0.180^{+0.170}_{-0.170}$\\
& \Mgii$^{\prime \prime}$ low-\rfe &-&-& --- & - & - & - & --- &-& $0.307^{+0.035}_{-0.051}$ & $1.910^{+0.100}_{-0.089}$ & $0.143^{+0.063}_{-0.096}$ & $0.360^{+0.240}_{-0.240}$\\
& \Mgii$^{\prime \prime}$ high-\rfe &-&-& --- & - & - & - & --- &-& $0.273^{+0.031}_{-0.047}$ & $1.630^{+0.130}_{-0.130}$ & $0.307^{+0.095}_{-0.095}$ & $0.570^{+0.710}_{-0.630}$\\
& BAO + $H(z)$ & $0.032^{+0.006}_{-0.003}$ & $0.081^{+0.017}_{-0.017}$ & $0.266^{+0.023}_{-0.023}$ & - & - & - & $1.530^{+0.620}_{-0.850}$ &$65.10^{+2.10}_{-2.10}$& - & - & -&-\\
\hline
& \Mgii\ QSO-66 &-&-& --- & - & $0.050^{+0.370}_{-0.370}$ & - & --- &-& $0.285^{+0.025}_{-0.032}$ & $1.782^{+0.056}_{-0.056}$ & $0.241^{+0.054}_{-0.054}$&-\\
Non-flat & \Mgii\ low-\rfe &-&-& --- & - & $0.040^{+0.370}_{-0.370}$ & - & --- &-& $0.316^{+0.038}_{-0.054}$ & $1.803^{+0.072}_{-0.072}$ & $0.210^{+0.077}_{-0.077}$&-\\
$\phi$CDM & \Mgii\ high-\rfe &-&-& $0.480^{+0.260}_{-0.370}$ & - & $0.040^{+0.370}_{-0.370}$ & - & --- &-& $0.275^{+0.033}_{-0.047}$ & $1.710^{+0.110}_{-0.110}$ & $0.313^{+0.095}_{-0.095}$&-\\
& \Mgii$^{\prime}$ QSO-66 &-&-& $0.480^{+0.260}_{-0.370}$ & - & $0.040^{+0.370}_{-0.370}$ & - & --- &-& $0.286^{+0.024}_{-0.032}$ & $1.725^{+0.098}_{-0.098}$ & $0.219^{+0.063}_{-0.063}$ & $0.069^{+0.099}_{-0.099}$\\
& \Mgii$^{\prime}$ low-\rfe &-&-& --- & - & $0.040^{+0.380}_{-0.380}$ & - & --- &-& $0.315^{+0.036}_{-0.053}$ & $1.660^{+0.160}_{-0.160}$ & $0.162^{+0.074}_{-0.100}$ & $0.230^{+0.230}_{-0.230}$\\
& \Mgii$^{\prime}$ high-\rfe &-&-& $0.470^{+0.220}_{-0.410}$ & - & $0.060^{+0.370}_{-0.370}$ & - & --- &-& $0.273^{+0.032}_{-0.047}$ & $1.490^{+0.270}_{-0.270}$ & $0.305^{+0.095}_{-0.095}$ & $0.160^{+0.180}_{-0.180}$\\
& \Mgii$^{\prime \prime}$ QSO-66 &-&-& --- & - & $0.040^{+0.370}_{-0.370}$ & - & --- &-& $0.284^{+0.024}_{-0.032}$ & $1.818^{+0.065}_{-0.065}$ & $0.203^{+0.065}_{-0.065}$ & $0.180^{+0.170}_{-0.170}$\\
& \Mgii$^{\prime \prime}$ low-\rfe &-&-& --- & - & $0.040^{+0.380}_{-0.380}$ & - & --- &-& $0.307^{+0.035}_{-0.053}$ & $1.910^{+0.100}_{-0.091}$ & $0.143^{+0.065}_{-0.097}$ & $0.360^{+0.240}_{-0.240}$\\
& \Mgii$^{\prime \prime}$ high-\rfe &-&-& --- & - & $0.060^{+0.370}_{-0.370}$ & - & --- &-& $0.272^{+0.032}_{-0.047}$ & $1.640^{+0.130}_{-0.130}$ & $0.306^{+0.095}_{-0.095}$ & $0.580^{+0.710}_{-0.640}$\\
& BAO + $H(z)$ & $0.032^{+0.006}_{-0.004}$ & $0.085^{+0.017}_{-0.021}$ & $0.271^{+0.024}_{-0.028}$ & - & $-0.080^{+0.100}_{-0.100}$ & - & $1.660^{+0.670}_{-0.830}$ &$65.50^{+2.50}_{-2.50}$& - & - & -&-\\
\hline
\end{longtable}
\footnotesize{$\hspace{-0.6cm}^a$ In our analyses $\Omega_{\Lambda}$ is a derived parameter and in each case $\Omega_{\Lambda}$ chains are derived using the current energy budget equation $\Omega_{\Lambda}= 1-\Omega_{m0}-\Omega_{k0}$ (where $\Omega_{k0}=0$ in the flat $\Lambda$CDM model).}\\
\footnotesize{$^b$ ${\rm km}\hspace{1mm}{\rm s}^{-1}{\rm Mpc}^{-1}$. $H_0$ is set to $70$ ${\rm km}\hspace{1mm}{\rm s}^{-1}{\rm Mpc}^{-1}$ for the QSO-only analyses.}
\end{landscape}
\clearpage
\twocolumn 

\begin{table}
\begin{threeparttable}
\caption{Two-parameter $R-L$ relation parameters (and $\sigma_{\rm ext}$) differences between the \Mgii\ low-\rfe\ and high-\rfe\ data sets.}
\label{tab:3}
\setlength{\tabcolsep}{14.5pt}
\begin{tabular}{lccc}
\hline
Model & $\Delta \sigma_{\rm ext}$  & $\Delta \gamma$  & $\Delta \beta$\\
\hline
Flat \lcdm\  & $0.67\sigma$ & $0.84\sigma$ & $0.77\sigma$\\
Non-flat \lcdm\ & $0.69\sigma$ & $0.86\sigma$ & $0.76\sigma$ \\
Flat XCDM  & $0.68\sigma$ & $0.85\sigma$ & $0.69\sigma$\\
Non-flat XCDM  & $0.67\sigma$ & $0.84\sigma$ & $0.64\sigma$\\
Flat $\phi$CDM  & $0.67\sigma$ & $0.84\sigma$ & $0.76\sigma$\\
Non-flat $\phi$CDM & $0.65\sigma$ & $0.84\sigma$ & $0.71\sigma$\\
\hline
\end{tabular}
\end{threeparttable}
\end{table}

\begin{table}
\begin{threeparttable}
\caption{Three-parameter linear-\rfe\ $R-L$ relation parameters (and $\sigma_{\rm ext}$) differences between the \Mgii$^{\prime}$ low-\rfe\ and high-\rfe\ data sets.}
\label{tab:4}
\setlength{\tabcolsep}{10pt}
\begin{tabular}{lcccc}
\hline
Model & $\Delta \sigma_{\rm ext}$  & $\Delta \gamma$  & $\Delta \beta$ & $\Delta k$\\
\hline
Flat \lcdm\  & $0.67\sigma$ & $1.18\sigma$ & $0.61\sigma$ & $0.23\sigma$\\
Non-flat \lcdm\ & $0.69\sigma$ & $1.19\sigma$ & $0.57\sigma$ & $0.27\sigma$ \\
Flat XCDM  & $0.67\sigma$ & $1.19\sigma$ & $0.63\sigma$ & $0.23\sigma$\\
Non-flat XCDM  & $0.65\sigma$ & $1.17\sigma$ & $0.54\sigma$ & $0.23\sigma$\\
Flat $\phi$CDM  & $0.68\sigma$ & $1.19\sigma$ & $0.54\sigma$ & $0.24\sigma$\\
Non-flat $\phi$CDM & $0.69\sigma$ & $1.19\sigma$ & $0.54\sigma$ & $0.24\sigma$\\
\hline
\end{tabular}
\end{threeparttable}
\end{table}

\begin{table}
\begin{threeparttable}
\caption{Three-parameter log-\rfe\ $R-L$ relation parameters (and $\sigma_{\rm ext}$) differences between the \Mgii$^{\prime\prime}$ low-\rfe\ and high-\rfe\ data sets.}
\label{tab:5}
\setlength{\tabcolsep}{10pt}
\begin{tabular}{lcccc}
\hline
Model & $\Delta \sigma_{\rm ext}$  & $\Delta \gamma$  & $\Delta \beta$ & $\Delta k$\\
\hline
Flat \lcdm\  & $0.58\sigma$ & $1.41\sigma$ & $1.72\sigma$ & $0.35\sigma$\\
Non-flat \lcdm\ & $0.58\sigma$ & $1.41\sigma$ & $1.72\sigma$ & $0.35\sigma$ \\
Flat XCDM  & $0.57\sigma$ & $1.44\sigma$ & $1.60\sigma$ & $0.32\sigma$\\
Non-flat XCDM  & $0.54\sigma$ & $1.40\sigma$ & $1.58\sigma$ & $0.27\sigma$\\
Flat $\phi$CDM  & $0.57\sigma$ & $1.44\sigma$ & $1.78\sigma$ & $0.31\sigma$\\
Non-flat $\phi$CDM & $0.57\sigma$ & $1.42\sigma$ & $1.70\sigma$ & $0.32\sigma$\\
\hline
\end{tabular}
\end{threeparttable}
\end{table}

\section{Discussion}
\label{sec_discussion}

In this paper we analyzed a sample of 66 \Mgii\ quasars with inferred UV \rfe\ ratios (or relative \Feii\ strengths) that are meant to serve as accretion-rate proxies \citep{2019ApJ...882...79P,2020ApJ...900...64S}. We derived cosmological constraints using six cosmological models (flat and non-flat $\Lambda$CDM, $\phi$CDM, and XCDM), and 2- and 3-parameter $R-L$ relations, while considering the full \Mgii\ sample as well as two equally-sized subsamples of low- and high-\rfe\ sources, with 33 \Mgii\ sources in each subsample. We included the relative UV \Feii\ strength in an attempt to correct for the accretion-rate effect. The cosmological constraints are in all cases consistent among the different samples as well as with constraints from better-established cosmological probes (BAO + $H(z)$ data). The inclusion of the third parameter \rfe\ in the $R-L$ relation does not result in the widely-expected reduction of the intrinsic scatter $\sigma_{\rm ext}$. In most cases the 2-parameter relation is mildly to positively preferred over the 3-parameter $R-L$ relation, see Table~\ref{tab:1}.

Previously, \citet{Khadkaetal2021c} found for H$\beta$ quasars that including the optical \rfe\ parameter in a 3-parameter $R-L$ relation also did not result in the widely-expected reduction of the intrinsic scatter. Moreover, in contrast to \Mgii\ quasars, low- and high-\rfe\ H$\beta$ sources obey different $R-L$ relations, with the 3-parameter $R-L$ relation being strongly favored over the 2-parameter one for the full sample of 118 sources, and with the inferred cosmological constraints being in $\sim 2\sigma$ tension with those found using better established cosmological probes.

There are several potential reasons for these differences between standardizable \Mgii\ quasars and H$\beta$ sources that may or may not be standardizable: 
\begin{itemize}
    \item[(i)] For the sample of 66 \Mgii\ sources, low- and high-\rfe\ sources have consistent $R-L$ relations and intrinsic scatter. In fact,  high-\rfe\ \Mgii\ sources appear to have somewhat smaller scatter in comparison with that of the full sample as well as that of the low-\rfe\ sources \citep[also see][]{Mary2020}. This is unlike H$\beta$ sources where high-accretors exhibit a larger scatter with respect to low-\rfe\, sources. \citep{MartinezAldama2019,duwang_2019,Khadkaetal2021c, 2022FrASS...950409P},
    \item[(ii)] The sample of 66 \Mgii\ quasars used here is about a factor of 2 smaller than the H$\beta$ sample of 118 sources. The majority of the \Mgii\ sources, 57, are from the SDSS-reverberation-mapping (SDSS-RM) program \citep{Homayouni2020} and cover a rather narrow range in monochromatic luminosity and rest-frame time-delay. The H$\beta$ sample of 118 sources is not only larger, but also more heterogeneous in terms of the time-delay determination methods.
    \item[(iii)] \Mgii\ sources have a larger median redshift in comparison with the H$\beta$ quasars, $z=0.990$ vs.\ $z=0.157$, hence their redshifts are not significantly affected by peculiar velocities. 
    \item[(iv)] The UV region is not significantly affected by host starlight, which is the case for low-luminous H$\beta$ sources \citep{2013ApJ...767..149B}. On the other hand, UV continuum emission may be prone to dust-extinction within their host galaxies and the intergalactic medium, but this effect is systematic for all \Mgii\ sources. 
    \item[(v)] Broad H$\beta$ and \Mgii\ lines are excited by different mechanisms: H$\beta$ is a recombination line, while \Mgii\ is a resonance line \citep{2019ApJ...875..133P,2020ApJ...900...64S}. Different photoionization mechanisms may account for some of the differences, such as the $R-L$ relation slope, though a detailed analysis is beyond the scope of the current study. The \Mgii\ line is $\sim 20\%$ narrower than the H$\beta$ line \citep{2013A&A...555A..89M} which hints at its position being farther away from the supermassive black hole. In addition, \citet{2020MNRAS.493.5773Y} found that the \Mgii\ line responds to the variable continuum with a smaller amplitude, while the \Mgii\ line width does not significantly change with luminosity (lack of ``breathing''), as  \citet{guo2020} found. 
    \item[(vi)] The smaller slope of the \Mgii-based $R-L$ relation in comparison with both the H$\beta$-based $R-L$ correlation slope \citep{MartinezAldama2019,Khadkaetal2021c} and the slope predicted by simple photoionization theory \citep[$\gamma_{\rm ion}$=0.5;][]{2021bhns.confE...1K,2022FrASS...950409P} is either caused by the source selection effect (cumulation of SDSS-RM sources at intermediate luminosities) or by the different photoionization mechanism and/or distribution of \Mgii-emitting gas \citep[e.g. lack of the intrinsic ``breathing'' effect for \Mgii-emitting material;][]{guo2020,2020ApJ...903...51W}.
\end{itemize}

The common feature for both \Mgii\ and H$\beta$ sources is that the inclusion of the third parameter, \rfe, in a 3-parameter $R-L$ relation does not result in the widely-expected significant reduction of the intrinsic scatter \citep[this paper and][]{Khadkaetal2021c}. This may be due to the fact that the correlation between the Eddington ratio and the \rfe\, parameter is generally not very strong \citep{Michal2021} and/or the accretion-rate effect is not entirely responsible for the intrinsic scatter in the $R-L$ relation.\footnote{\citet{duwang_2019} report a significant scatter reduction for their sample of H$\beta$ sources, going from $\sigma_{\rm ext}\sim 0.28$ to $\sigma_{\rm ext}=0.196$ after the inclusion of the optical \rfe\ parameter. This is, however, based on an approximate computation that only accounts for the variation in the $R-L$ relation parameters and the intrinsic dispersion, while holding fixed all cosmological-model parameters (they adopted the flat $\Lambda$CDM model with $H_0=67\,{\rm km\,s^{-1}\,Mpc^{-1}}$, $\Omega_{\Lambda}=0.68$, and $\Omega_{m0}=0.32$), unlike the more complete computations done in \citet{Khadkaetal2021c} and this paper. In addition, their sample size was smaller (75 vs.\ 118 sources) in comparison with \citet{Khadkaetal2021c}.} The scatter is partially also driven by the time-delay uncertainties since the determination of the broad-line time-delay requires long enough monitoring to be reliable. Adding just a few extra years of monitoring can change the rest-frame time-delay by as much as a factor of two, as was the case for one of the \Mgii\ quasars CTS C30.10 \citep{2019ApJ...880...46C,2022arXiv220111062P}. This can be mitigated by continued monitoring of already reverberation-mapped sources and by enlarging the sample using extensive photometric and spectroscopic surveys of higher-redshift quasars.

\section{Conclusion}
\label{con}

We have investigated two 3-parameter $R-L$ relations for \Mgii\ QSO sources. In our previous work with Mg II QSOs, we only studied the 2-parameter $R-L$ relation \citep{khadka2021} and found that the \Mgii\ QSO cosmological constraints were weak but consistent with cosmological constraints derived using better-established cosmological probes. Some discrepancies in the $R-L$ relations of H$\beta$ QSOs \citep{Khadkaetal2021c}, and the $\sim 2\sigma$ discrepancy between the \hb\ cosmological constraints and those derived using better-established cosmological probes, motivated our current paper. 

We find here that for \Mgii\ data and the 2-parameter $R-L$ relation, the low-\rfe\ and high-\rfe\ data subsets obey the same $R-L$ relation within the error bars. This is strikingly different compared to what happens in the H$\beta$ QSO case \citep{Khadkaetal2021c} and is a positive result for the on-going development of \Mgii\ QSOs as cosmological probes. Our analyses with the 3-parameter $R-L$ relations here show that the inclusion of the third parameter $k$ increases the $R-L$ relation parameter differences between the low-\rfe\ and high-\rfe\ data subsets. This increase is not very significant for the linear-\rfe\ 3-parameter $R-L$ relation in eq.\ (\ref{eq:corr}) with $q=$\rfe\, but is more significant for the log-\rfe\ 3-parameter $R-L$ relation in eq.\ (\ref{eq:corr}) with $q=\log{}$\rfe. However, the motivation for extending the 2-parameter $R-L$ relation to a 3-parameter one is the attempt to reduce the intrinsic dispersion of the $R-L$ relation; our analyses here shows no evidence for this \citep[and this is similar to what happens in the \hb\ case,][]{Khadkaetal2021c}.\footnote{In \citet{Khadkaetal2021c} we found that for the full \hb\ data set the 3-parameter $R-L$ relation provided a significantly better fit to data than did the 2-parameter $R-L$ relation, unlike what we have found for \Mgii\ data here. This is because the \hb\ low-\rfe\ and high-\rfe\ data subsets obey very different 2-parameter $R-L$ relations, unlike the case for \Mgii\ data here.}

Unlike current \hb\ QSOs, current \Mgii\ QSOs consistently obey the 2-parameter $R-L$ relation and so are standardizable and can be used as cosmological probes. It is important to continue to check whether this remains true for future \Mgii\ QSOs. If it does, and if there is a significant increase in the quantity and quality of reverberation-measured \Mgii\ time-lag QSOs over an extended redshift range, this will allow for tighter \Mgii\ cosmological constraints as well as a more definitive study of the adequacy of the 2-parameter $R-L$ relation for \Mgii\ sources.

\section{ACKNOWLEDGEMENTS}

This research was supported in part by US DOE grant DE-SC0011840, by the Polish Funding Agency National Science Centre, project 2017/26/A/ST9/00756 (Maestro 9), by GAČR EXPRO grant 21-13491X, by Millenium Nucleus NCN$19\_058$ (TITANs), and by the Conselho Nacional de Desenvolvimento Científico e Tecnológico (CNPq) Fellowship (164753/2020-6). Part of the computation for this project was performed on the Beocat Research Cluster at Kansas State University.

\section*{Data availability}

The data analysed in this article are listed in Table \ref{tab:MgQSOdata} of this paper.



\bibliographystyle{mnras}
\bibliography{mybibfile}




\onecolumn
\begin{appendix}
\section{\Mgii\ QSO data}
\label{sec:appendix}
\addtolength{\tabcolsep}{0pt}
\LTcapwidth=\linewidth
\begin{longtable}{lccccccc}
\caption{Reverberation-measured \Mgii\ QSO data. For each source, columns list: QSO name, redshift, continuum flux density at 3000\,\AA, measured rest-frame time-delay, the ratio of the UV \Feii\  flux density with respect to the MgII flux density \rfe, the original source reference, and the \rfe\ class where number 1 corresponds to low-\rfe\ sources (\rfe$<1.0467$) and number 2 to high-\rfe\ sources (\rfe$>1.0467$). The first 65 sources were analyzed by \citet{Mary2020}, the last source HE 0435-4312 is from \citet{Michal2021}. The references are designated as follows: (a) stands for sources taken from \citet{Homayouni2020} (in total 57 source, where the object number corresponds to the RMID number from the original catalogue); (b) \citet{2016ApJ...818...30S}; (c) \citet{shen_2019}; (d) \citet{2019ApJ...880...46C}; (e) \citet{2022arXiv220111062P}; (f) \citet{Michal2020}; (g) monochromatic luminosity inferred following \citet{2015AcA....65..251K}; (h) \citet{Michal2021}.  }
\label{tab:MgQSOdata}\\
\hline\hline
Object &  $z$ &  $\log \left(F_{3000}/{\rm erg}\,{\rm s^{-1}}{\rm cm^{-2}}\right)$  &  $\log \left(L_{3000}/{\rm erg}\,{\rm s^{-1}}\right)$  &  $\tau$ (day) & \rfe & Ref. & \rfe\ class\\
\hline
\endfirsthead
\caption{continued.}\\
\hline\hline
Object &  $z$ &  $\log \left(F_{3000}/{\rm erg}\,{\rm s^{-1}}{\rm cm^{-2}}\right)$  &  $\log \left(L_{3000}/{\rm erg}\,{\rm s^{-1}}\right)$  &  $\tau$ (day) & \rfe & Ref. & \rfe\ class\\
\hline
\endhead
\hline
\endfoot
018& 0.848 & $-13.1081 \pm 0.0009  $&  $44.4331 \pm 0.0009 $ & $125.9^{+6.8}_{-7.0}$ & $0.4716\pm0.0194$ & (a) & 1\\
028& 1.392& $-12.4342 \pm 0.0004 $& $45.6391 \pm    0.0004 $  & $65.7^{+24.8}_{-14.2}$ &$1.0340\pm0.0134$ & (a) & 1\\
038& 1.383& $-12.3956 \pm  0.0003$ & $ 45.6707 \pm  0.0003 $     &         $120.7^{+27.9}_{-28.7}$&$1.2057\pm0.0152$ & (a) & 2\\
044& 1.233& $-13.0287 \pm 0.0013 $ & $ 44.9143 \pm    0.0013$  & $65.8^{+18.8}_{-4.8}$&$1.0594\pm0.0644$ & (a) & 2\\
102& 0.861& $-12.5513 \pm 0.0005$ & $45.0061 \pm    0.0005 $  & $86.9^{+16.2}_{-13.3}$&$1.4152\pm0.0259$ & (a) & 2\\
114& 1.226& $-11.8682 \pm 0.0003 $& $46.0687 \pm    0.0003 $   & $186.6^{+20.3}_{-15.4}$&$1.3508\pm0.0399$ & (a) & 2\\
118 & 0.715& $ -12.2374 \pm     0.0006 $&  $ 45.1217 \pm 0.0006$ &   $102.2^{+27.0}_{-19.5}$&$1.2126\pm0.0270$ & (a) & 2\\
123& 0.891& $-12.9147 \pm 0.0009$& $ 44.6794 \pm    0.0009 $ & $81.6^{+28.0}_{-26.6}$&$1.1069\pm0.325$ & (a) & 2\\
135& 1.315& $-12.7843 \pm 0.0005$& $45.2279 \pm 0.0005$ & $93.0^{+9.6}_{-9.8}$&$1.0785\pm0.0097$ & (a) & 2\\
158& 1.478& $-13.2009 \pm 0.0012$& $44.9367 \pm 0.0012$ & $119.1^{+4.0}_{-11.8}$ &$0.9489\pm0.0211$ & (a) & 1\\
159& 1.587& $ -12.7115  \pm  0.0006$& $45.5023 \pm  0.0006$ & $324.2^{+25.3}_{-19.4}$&$1.0098\pm0.0180$ & (a) & 1\\
160& 0.36 & $-12.8476 \pm 0.0013$& $43.7965  \pm   0.0013 $ & $106.5^{+18.2}_{-16.6}$&$0.2622\pm0.0055$ & (a) & 1\\
170& 1.163& $ -12.6724 \pm 0.0005$& $ 45.2078 \pm 0.0005$ & $98.5^{+6.7}_{-17.7}$&$0.7566\pm0.0109$ & (a) & 1\\
185& 0.987& $-12.8520 \pm 0.0094$& $44.8519 \pm  0.0094$ & $387.9^{+3.3}_{-3.0}$&$1.2979\pm0.3264$ & (a) & 2\\
191& 0.442& $-13.0675 \pm 0.0012$& $43.7869 \pm  0.0012$ & $93.9^{+24.3}_{-29.1}$&$0.6649\pm0.0339$ & (a) & 1\\
228& 1.264& $-13.2425 \pm 0.0011 $ & $44.7272 \pm 0.0011$ & $37.9^{+14.4}_{-9.1}$&$1.0697\pm0.0420$ & (a) & 2\\
232& 0.808& $-13.2316 \pm 0.0014$& $44.2579 \pm 0.0014$ & $273.8^{+5.1}_{-4.1}$&$1.0113\pm0.0901$ & (a) & 1\\
240& 0.762& $-13.3448 \pm    0.0021 $& $44.0821 \pm 0.0021$ & $17.2^{+3.5}_{-2.8}$&$0.6603\pm0.0276$ & (a) & 1\\
260& 0.995& $-12.4492 \pm 0.0004$& $45.2633 \pm 0.0004 $ & $94.9^{+18.7}_{-17.2}$&$0.9968\pm0.0182$ & (a) & 1\\
280& 1.366& $-12.5521 \pm 0.0003$& $45.5009 \pm 0.0003$ & $99.1^{+3.3}_{-9.5}$&$1.0222\pm0.0145$ & (a) & 1\\
285& 1.034& $-13.2726 \pm 0.0020 $& $44.4812 \pm 0.0020 $ & $138.5^{+15.2}_{-21.1}$&$0.5753\pm0.0189$ & (a) & 1\\
291& 0.532& $-13.2325 \pm 0.0016$& $43.8146 \pm  0.0016$ & $39.7^{+4.2}_{-2.6}$&$0.1551\pm0.0118$ & (a) & 1\\
294& 1.215& $-12.4673 \pm 0.0004 $& $45.4599 \pm 0.0004$ & $71.8^{+17.8}_{-9.5}$&$1.3498\pm0.0247$ & (a) & 2\\
301& 0.548& $-12.8427 \pm 0.0011$& $44.2355  \pm   0.0011$ & $136.3^{+17.0}_{-16.9}$&$0.7894\pm0.0270$ & (a) & 1\\
303& 0.821& $-13.2805 \pm   0.0013$& $44.2261 \pm 0.0013$ & $57.7^{+10.5}_{-8.3}$&$0.8923\pm0.0170$ & (a) & 1\\
329& 0.721& $-11.9688 \pm 0.0007$& $45.3992 \pm 0.0007$ & $87.5^{+23.8}_{-14.0}$&$1.5348\pm0.0333$ & (a) & 2\\
338& 0.418& $-13.0371 \pm 0.0013$& $43.7598 \pm 0.0013$ & $22.1^{+8.8}_{-6.2}$&$0.1975\pm0.0190$ & (a) & 1\\
419& 1.272& $-12.9275 \pm 0.0011 $& $45.0489 \pm 0.0011$ & $95.5^{+15.2}_{-15.5}$&$1.1103\pm0.0257$ & (a) & 2\\
422& 1.074& $-13.0888 \pm 0.0011$& $44.7058 \pm 0.0011$ & $109.3^{+25.4}_{-29.6}$&$0.3254\pm0.0073$ & (a) & 1\\
440& 0.754& $-12.4820 \pm 0.0004 $ & $44.9337 \pm  0.0004$ & $114.6^{+7.4}_{-10.8}$&$0.9773\pm0.0139$ & (a) & 1\\
441& 1.397& $-12.6008 \pm 0.0004$ & $45.4763 \pm 0.0004$ & $127.7^{+5.7}_{-7.3}$&$1.3191\pm0.0424$ & (a) & 2\\
449& 1.218& $-12.9573 \pm 0.0013$& $44.9725 \pm 0.0013 $ & $119.8^{+14.7}_{-24.4}$&$1.1464\pm0.0336$ & (a) & 2\\
457& 0.604& $-13.4931 \pm 0.0029 $& $43.6874 \pm 0.0029$ & $20.50^{+7.7}_{-5.3}$&$0.2002\pm0.0283$ & (a) & 1\\
459& 1.156& $-12.8865 \pm 0.0011$& $44.9872 \pm 0.0011$ & $122.8^{+5.1}_{-5.7}$&$1.0102\pm0.0346$ & (a) & 1\\
469& 1.004& $-12.1473 \pm  0.0002$& $45.5749 \pm 0.0002 $ & $224.1^{+27.9}_{-74.3}$&$1.2398\pm0.0295$ & (a) & 2\\
492& 0.964& $ -12.3949 \pm 0.0004$& $45.2837 \pm 0.0004$ & $92.0^{+16.3}_{-12.7}$&$1.0328\pm0.0167$ & (a) & 1\\
493& 1.592& $ -12.2609 \pm 0.0004 $& $ 45.9563 \pm 0.0004$ & $315.6^{+30.7}_{-35.7}$&$1.3944\pm0.0360$ & (a) & 2\\
501& 1.155& $-12.9605 \pm 0.0009$& $ 44.9123 \pm 0.0009$ &  $44.9^{+11.7}_{-10.4}$&$1.0066\pm0.0444$ & (a) & 1\\
505& 1.144& $-13.0664 \pm 0.0011$& $44.7961 \pm 0.0011 $ & $94.7^{+10.8}_{-16.7}$&$0.8675\pm0.0209$ & (a) & 1\\
522& 1.384& $ -12.9840 \pm 0.0006 $& $ 45.0831 \pm 0.0006$ & $115.8^{+11.3}_{-16.0}$&$1.2665\pm0.0320$ & (a) & 2\\
556& 1.494& $-12.6554 \pm 0.0005$& $45.4937 \pm 0.0005$ & $98.7^{+13.9}_{-10.8}$&$1.2118\pm0.0164$ & (a) & 2\\
588& 0.998& $-12.1342 \pm 0.0002$& $45.5816 \pm 0.0002$ & $74.3^{+23.0}_{-18.2}$&$1.0015\pm0.0125$ & (a) &  1\\
593& 0.992& $-12.7386 \pm 0.0006$& $44.9707 \pm 0.0006$ & $80.1^{+21.4}_{-20.8}$&$1.0595\pm0.0176$ & (a) & 2\\
622& 0.572& $-12.6271 \pm 0.0005$& $44.4961 \pm 0.0005$ & $61.7^{+6.0}_{-4.3}$&$1.7362\pm0.0368$ & (a) & 2\\
645& 0.474& $-12.7685 \pm 0.0009$& $44.1583 \pm 0.0009$ & $30.2^{+26.8}_{-8.9}$&$1.3497\pm0.0440$ & (a) & 2\\
649& 0.85 & $-13.0933 \pm 0.0013$& $44.4504 \pm 0.0013$ & $165.5^{+22.2}_{-25.1}$&$2.3486\pm0.3300$ & (a) & 2\\
651& 1.486& $ -12.9595 \pm 0.0011$& $45.1839 \pm 0.0011$ & $76.5^{+18.0}_{-15.6}$&$0.8697\pm0.0216$ & (a) & 1\\
675& 0.919& $-12.5484 \pm 0.0005$& $45.0789 \pm 0.0005$ & $139.8^{+12.0}_{-22.6}$&$1.2550\pm0.0146$ & (a) & 2\\
678& 1.463& $-12.8283 \pm 0.0007 $& $45.2983 \pm 0.0007$ & $82.9^{+11.9}_{-10.2}$&$1.1805\pm0.0254$ & (a) & 2\\
709& 1.251& $-12.9770 \pm 0.0010$& $44.9816 \pm 0.0010$ & $85.4^{+17.7}_{-19.3}$&$1.3036\pm0.0475$ & (a) & 2\\
714& 0.921& $-12.8414 \pm 0.0012$& $44.7882 \pm 0.0012$ & $320.1^{+11.3}_{-11.2}$&$0.5856\pm0.0216$ & (a) & 1\\
756& 0.852& $-13.1449 \pm 0.0023 $& $44.4013 \pm 0.0023$ & $315.3^{+20.5}_{-16.4}$&$1.2265\pm0.0345$ & (a) & 2\\
761& 0.771& $-12.6558 \pm 0.0024$& $44.7837 \pm 0.0024$ & $102.1^{+8.2}_{-7.4}$&$1.4789\pm0.0420$ & (a) & 2\\
771& 1.492& $-12.4571 \pm 0.0004$& $45.6906 \pm 0.0004$ & $31.3^{+8.1}_{-4.6}$&$0.7654\pm0.0084$ & (a) & 1\\
774& 1.686& $-12.5769 \pm  0.0004$& $45.7017 \pm 0.0004$ & $58.9^{+13.7}_{-10.1}$&$1.1880\pm0.0306$ & (a) & 2\\
792& 0.526& $-13.4922 \pm 0.0030$& $43.5431 \pm 0.0030$ & $111.4^{+29.5}_{-20.0}$&$0.6261\pm0.1034$ & (a) & 1\\
848& 0.757& $-13.3426 \pm 0.0015$& $44.0773 \pm 0.0015$ & $65.1^{+29.4}_{-16.3}$&$0.8990\pm0.0584$ & (a) & 1\\
J141214.20$+$532546.7 &  0.45810 & $-12.2526 \pm 0.0004$  & $44.6388 \pm 0.0004$ & $36.7^{+10.4}_{-4.8}$ & $1.650\pm 0.201$    & (b), (c) & 2\\
J141018.04$+$532937.5& 0.4696& $-13.1883\pm0.0051$& $43.7288 \pm 0.0051$ & $32.3^{+12.9}_{-5.3}$&$0.900\pm0.085$ &(b), (c) &  1\\
J141417.13$+$515722.6& 0.6037& $-13.4925 \pm 0.0029$& $ 43.6874 \pm 0.0029$ & $29.1^{+3.6}_{-8.8}$&$0.200\pm0.028$ & (b), (c) & 1\\
J142049.28$+$521053.3& 0.751 & $-12.7205\pm0.0009$& $44.6909  \pm   0.0009$ & $34.0^{+6.7}_{-12.0}$&$1.450\pm0.034$ & (b), (c) & 2\\
J141650.93$+$535157.0& 0.5266& $-13.2587 \pm 0.0020$& $43.7778 \pm 0.0020$ & $25.1^{+2.0}_{-2.6}$&$0.100\pm0.011$ & (b), (c) & 1\\
J141644.17$+$532556.1& 0.4253& $-12.8668 \pm 0.0010$& $43.9479 \pm 0.0010 $ & $17.2^{+2.7}_{-2.7}$&$1.450\pm0.068$ & (b), (c) & 2\\
CTS C30.10& 0.9005& $ -11.5825 \pm    0.0260$& $46.0230 \pm 0.0260 $ & $275.5^{+12.4}_{-19.5}$ &$1.600\pm0.004$ & (d), (e) &  2\\
HE0413-4031& 1.3765& $-11.3302 \pm    0.0824 $& $46.7310  \pm   0.0810$ &  $302.9^{+23.7}_{-19.1}$&$0.800\pm0.020$ & (f), (g) &  1\\
HE0435-4312& 1.2231& $-11.5853  \pm   0.0361$& $46.3490 \pm 0.0360 $ & $296^{+13.0}_{-14.0}$&$2.36\pm0.03$ & (h), (g) & 2\\
\hline
\end{longtable}

\end{appendix}



\bsp	
\label{lastpage}
\end{document}